\documentclass[journal]{IEEEtran}
\usepackage[utf8]{inputenc} 
\usepackage[numbers,compress]{natbib}
\usepackage[cmex10]{amsmath}
\usepackage{amsfonts,mathrsfs,mathdots,mathtools,amsthm,amssymb,centernot}
\usepackage[cal=euler,
    bb=ams]{mathalfa}
\usepackage{graphicx}
\usepackage{textcomp}
\usepackage{array}
\usepackage[dvipsnames, table]{xcolor}
\usepackage{dsfont}
\usepackage{pifont}
\usepackage{siunitx}
\usepackage{soul}
\usepackage{bm}	
\usepackage[inline]{enumitem}
\setlist[enumerate,1]{label = (\roman*),ref = \roman*}
\usepackage{nicefrac}
\usepackage{subcaption}
\usepackage{colortbl}
\usepackage{booktabs}
\usepackage{diagbox}
\usepackage{url}
\usepackage{ifthen}
\usepackage{balance}
\usepackage{dirtytalk}

\usepackage{hyperref}  
\hypersetup{colorlinks,
            linkcolor=blue,
            citecolor=blue,
            urlcolor=blue,
            anchorcolor=blue,
            filecolor=blue,
            linktocpage,
            plainpages=false,
            breaklinks=true}
\definecolor{inner}{HTML}{ffffea}
\definecolor{middle}{HTML}{fff0d4}
\definecolor{outer}{HTML}{fbe0e0}

\DeclarePairedDelimiter\abs{\lvert}{\rvert}%
\DeclarePairedDelimiter\norm{\lVert}{\rVert}%

\DeclareMathOperator*{\argmax}{arg\,max}
\DeclareMathOperator*{\argmin}{arg\,min}

\DeclareMathOperator*{\esssup}{ess\,sup}
\DeclareMathOperator*{\essinf}{ess\,inf}

\theoremstyle{definition}
\newtheorem{definition}{Definition}
\newtheorem{lemma}{Lemma}
\newtheorem{theorem}{Theorem}

\newtheorem{corollary}{Corollary}
\newtheorem{proposition}{Proposition}
\newtheorem{example}{Example}
\newtheorem{remark}{Remark}

\newcommand{\bE}{\ensuremath{\mathbb{E}}}

\newcommand{\bP}{\ensuremath{\mathbb{P}}}
\newcommand{\bQ}{\ensuremath{\mathbb{Q}}}
\newcommand{\bR}{\ensuremath{\mathbb{R}}}

\newcommand{\bZ}{\ensuremath{\mathbb{Z}}}

\newcommand{\cA}{\ensuremath{\mathcal{A}}}
\newcommand{\cB}{\ensuremath{\mathcal{B}}}
\newcommand{\cC}{\ensuremath{\mathcal{C}}}
\newcommand{\cD}{\ensuremath{\mathcal{D}}}
\newcommand{\cE}{\ensuremath{\mathcal{E}}}
\newcommand{\cF}{\ensuremath{\mathcal{F}}}
\newcommand{\cG}{\ensuremath{\mathcal{G}}}
\newcommand{\cH}{\ensuremath{\mathcal{H}}}

\newcommand{\cL}{\ensuremath{\mathcal{L}}}

\newcommand{\cP}{\ensuremath{\mathcal{P}}}

\newcommand{\cS}{\ensuremath{\mathcal{S}}}

\newcommand{\cU}{\ensuremath{\mathcal{U}}}

\newcommand{\cW}{\ensuremath{\mathcal{W}}}
\newcommand{\cX}{\ensuremath{\mathcal{X}}}
\newcommand{\cY}{\ensuremath{\mathcal{Y}}}
\newcommand{\cZ}{\ensuremath{\mathcal{Z}}}

\newcommand{\sfc}{\ensuremath{\mathsf{c}}}

\newcommand{\sfr}{\ensuremath{\mathsf{r}}}

\newcommand{\sfN}{\ensuremath{\mathsf{N}}}

\newcommand{\sfU}{\ensuremath{\mathsf{U}}}

\newcommand{\dinf}{\ensuremath{D_\infty}}
\newcommand{\ind}{\ensuremath{\mathbf{1}}}
\newcommand{\nll}{\centernot{\ll}}
\newcommand{\Ren}{R\'enyi }

\mathtoolsset{showonlyrefs}
\allowdisplaybreaks
\begin{document}
\title{Information Density Bounds for Privacy} 

\author{Sara~Saeidian,~\IEEEmembership{Member,~IEEE,}
        Leonhard Grosse,~\IEEEmembership{Member,~IEEE,}
        Parastoo~Sadeghi,~\IEEEmembership{Senior~Member,~IEEE,}
        Mikael~Skoglund,~\IEEEmembership{Fellow,~IEEE,}
        Tobias~J.~Oechtering,~\IEEEmembership{Senior~Member,~IEEE}
\thanks{This article was presented in part at the 2024 IEEE International Symposium on Information Theory.}   
\thanks{This work was supported by the Swedish Research Council (VR) under grants 2023-04787, 2023-03684 and 2024-06615, and the Digital Futures center within the collaborative project DataLEASH in Action.}
\thanks{Sara~Saeidian is with the Division of Information Science and Engineering, School of Electrical Engineering and Computer Science, KTH Royal Institute of Technology, 100 44 Stockholm, Sweden, and also with Inria Saclay, 91120 Palaiseau, France  (email: saeidian@kth.se).}
\thanks{Leonhard~Grosse, Mikael~Skoglund, and Tobias~J.~Oechtering are with the Division of Information Science and Engineering, School of Electrical Engineering and Computer Science, KTH Royal Institute of Technology, 100 44 Stockholm, Sweden (email: \{lgrosse, skoglund, oech\}@kth.se). }
\thanks{Parastoo Sadeghi is with the School of Engineering and Technology, University of New South Wales, Canberra, ACT 2601, Australia (e-mail: p.sadeghi@unsw.edu.au). The work of P. Sadeghi was partly supported by the Australian Research Council, under Future Fellowship Scheme, grant number FT19100429 and partly by the Digital Futures Visiting Fellowship.}
}

\maketitle

\begin{abstract}
This paper explores the implications of guaranteeing privacy by imposing a lower bound on the information density between the private and the public data. We introduce a novel and operationally meaningful privacy measure called \emph{pointwise maximal cost} (PMC) and demonstrate that imposing an upper bound on PMC is equivalent to enforcing a lower bound on the information density. PMC quantifies the information leakage about a secret to adversaries who aim to minimize non-negative cost functions after observing the outcome of a privacy mechanism. When restricted to finite alphabets, PMC can equivalently be defined as the information leakage to adversaries aiming to minimize the probability of incorrectly guessing randomized functions of the secret. We study the properties of PMC and apply it to standard privacy mechanisms to demonstrate its practical relevance. Through a detailed examination, we connect PMC with other privacy measures that impose upper or lower bounds on the information density. These are pointwise maximal leakage (PML), local differential privacy (LDP), and (asymmetric) local information privacy. In particular, we show that a mechanism satisfies LDP if and only if it has both bounded PMC and bounded PML. Overall, our work fills a conceptual and operational gap in the taxonomy of privacy measures, bridges existing disconnects between different frameworks, and offers insights for selecting a suitable notion of privacy in a given application.
\end{abstract}

\begin{IEEEkeywords}
Privacy, information density, pointwise maximal leakage, pointwise maximal cost, local differential privacy, information privacy, gain function, cost function.
\end{IEEEkeywords}

\section{Introduction}
Recent advancements in machine learning and artificial intelligence have driven progress across numerous domains of human life such as healthcare, finance, education, and transportation. However, as these technologies become omnipresent in our daily lives, the need for reliable and robust privacy protection has become more urgent than ever. Research suggests that as machine learning models become more sophisticated, the potential for privacy breaches escalates~\cite{rigaki2023survey}. This, in turn, underscores the need for new protective measures for safeguarding sensitive information and ensuring the ethical deployment of AI technologies.

Traditionally, privacy research has focused on developing techniques to enable the release of tabular data immune to specific types of attacks, such as re-identification or attribute disclosure. These methods, which fall under the umbrella term \emph{syntactic privacy}, include $k$-anonymity~\cite{sweeney2002k}, $l$-diversity~\cite{machanavajjhala2007diversity}, and $t$-closeness~\cite{li2006t}. While these techniques are still widely used both in the public sector and by private companies, recent research has shifted towards methods that define and study privacy as an intrinsic property of the data release mechanism. This shift was synthesized in the framework of \emph{differential privacy} (DP)~\cite{DPoriginalpaper}. DP is a mathematical framework with provable privacy guarantees. Its goal is to enable the theoretical analysis and implementation of statistical methods on aggregated datasets while protecting the privacy of each individual data entry. 

The main principle behind DP is that two datasets differing in a single entry should be hard to distinguish based on the statistics released from each. This is achieved by controlling a key quantity known as the \emph{privacy loss random variable}~\cite{dwork2014algorithmic}. Let $x_1$ and $x_2$ be two databases differing in a single entry, i.e., \emph{neighboring} databases. The privacy loss random variable associated with a data release mechanism $M$ compares the probability of producing an outcome $Y$ from database $x_1$ versus database $x_2$, that is, 
\begin{equation}
\label{eq:privacy_loss_dp}
    \cL_{x_1, x_2} (Y) \coloneqq \log \frac{\bP[M(x_1) = Y]}{\bP[M(x_2) = Y]}. 
\end{equation}
Various definitions of DP emerge by imposing different restrictions on the privacy loss random variable. For example, \emph{pure DP} requires the privacy loss to be bounded for all pairs of neighboring databases~\cite{DPoriginalpaper}. Later, the concept of DP was generalized from databases to data pertaining to a single individual, leading to the development of \emph{local differential privacy} (LDP)~\cite{kasiviswanathan2011can,duchi2013LDPminmaxDEF}. LDP can be also formulated as a constraint on a privacy loss similar to~\eqref{eq:privacy_loss_dp}; the key difference is that the loss is restricted for all possible pairs of inputs.

It is natural to question whether the definition of privacy loss, and more broadly the principles underlying DP, can be justified beyond intuition. One commonly accepted explanation is that DP balances Type I and Type II error probabilities in a hypothesis test between two neighboring datasets~\cite{wassermanStatisticalFrameworkDifferential2010a,kairouz2015composition}. In general, the approach of defining a mathematical measure for privacy and justifying it retrospectively has inherent limitations since it is difficult to ensure that the resulting definition is neither too weak nor excessively pessimistic. In the context of DP, for instance, opinions are notably divided. Some argue that DP is overly conservative, while others believe it may fail to provide sufficient protection, especially in scenarios involving correlated datasets~\cite{tschantz2020differential}. 

The limitations of ex-post justified privacy measures suggest the need for alternatives that are \emph{operationally meaningful}. Such measures need not rely on natural language descriptions or intuition to be interpreted. This motivation has led to the development of a novel privacy measure called \emph{pointwise maximal leakage} (PML)~\cite{saeidian2023pointwise,saeidian2023pointwisegeneral}. Unlike most measures, PML arises from a broad and encompassing threat model, first introduced in~\cite{alvim2012measuring}. In this model, adversaries aim to maximize arbitrary non-negative gain functions encoding their objectives. This flexibility allows PML to address a wide array of attack scenarios, including membership inference, reconstruction, and attribute disclosure attacks~\cite{saeidian2023pointwise}.

Similar to the role of the privacy loss random variable in DP, PML is a random variable that quantifies the information leakage about a private quantity $X$ (modeling any type of sensitive information) to a public quantity $Y$. PML can be expressed as 
\begin{equation*}
    \ell_{P_{XY}}(X \to Y) = \esssup_{P_X} \; i_{P_{XY}}(X;Y),
\end{equation*}
where 
\begin{equation*}
    i_{P_{XY}} = \log \frac{d P_{X Y}}{d (P_X \times P_Y)},
\end{equation*}
is the \emph{information density} between $X$ and $Y$. Thus, to restrict the information leakage between $X$ and $Y$, a natural approach is to impose an upper bound on PML. This, in turn, imposes an upper bound on the information density between $X$ and $Y$. 

PML is not the only privacy measure related to information density. In fact, information theory has a long tradition of addressing privacy and security problems by enforcing various constraints on the information density and related quantities, especially mutual information. Other notable examples of privacy measures that restrict the information density are \emph{local information privacy} (LIP)~\cite{6483382,jiang2018context,jiang2020LIP,jiang2021LIPcontextaware} and \emph{asymmetric local information privacy} (ALIP)~\cite{zarrabian2023lift,zarrabian2022asymmetric}. These measures are defined by axiomatically imposing lower and upper bounds on $i_{P_{XY}}$. While LIP imposes bounds symmetrically, ALIP allows for different upper and lower bounds.

Now, imposing an upper bound on the information density is operationally justified through PML, and its implications have been rigorously examined in previous works~\cite{saeidian2023pointwise,saeidian2023pointwisegeneral,saeidian2023inferential}. On the other hand, the lower bound on the information density is not as well understood. Definitions such as LIP and ALIP in effect use the lower bound to remain consistent with and provide (local) DP guarantees. However, beyond compliance with DP-based definitions, the consequences of imposing a lower bound on the information density or its operational significance have remained largely unexplored in the literature.

\subsection{Motivations}
There are two key motivations behind this paper. First, we seek to provide an in-depth understanding of the implications of defining privacy through a lower bound on the information density. To this end, we introduce an operationally meaningful privacy measure, which we call \emph{pointwise maximal cost} (PMC). We demonstrate that imposing an upper bound on PMC is equivalent to imposing a lower bound on the information density. PMC may be thought of as a complementary definition to PML. More precisely, while PML considers \emph{opportunistic} adversaries aiming to maximize gain functions, PMC focuses on \emph{risk-averse} adversaries aiming to minimize arbitrary non-negative cost functions. The following simple example illustrates how the threat associated with opportunistic and risk-averse adversaries is quantified differently in a membership inference scenario~\cite{dwork2017exposed}.
\begin{example}
Consider a membership inference scenario, where $X$ denotes a database and the adversary aims to determine whether or not a particular individual, Alice, is present in $X$. Let $\cX$ denote the set of all possible databases, $\cX_1 \subset \cX$ denote the set of databases that include Alice's data, and $\cX_0 = \cX \setminus \cX_1$ denote the set of databases that do not include Alice's data. Thus, each database $x \in \cX$ satisfies $x \in \cX_i$ for some $i \in \{0,1\}$, indicating Alice’s membership status.

In this context, an \emph{opportunistic} adversary may be modeled via the gain function $g(w, x) = \ind[w = i]$ for $x \in \cX_i$, where $w \in \{0,1\}$ is the adversary's guess, and $\ind[\cdot]$ denotes the indicator function. Here, the adversary receives a gain of $1$ for correctly guessing the membership and no gain otherwise. The adversary would then seek a strategy for producing guesses that maximizes this gain.

On the other hand, a \emph{risk-averse} adversary for the membership inference attack may be modeled via a cost function $c(w, x) = \ind[w \neq i]$ for $x \in \cX_i$ and $w \in \{0,1\}$. This adversary would seek a strategy for producing guesses that minimizes the cost, reflecting a preference to avoid incorrect guesses.
\end{example}

Our second goal is to explore the connections and the interplay between the upper and lower bounds of information density. Specifically, we investigate whether a PML guarantee inherently provides a PMC guarantee, and vice versa. We address these questions both assuming finite alphabets and more general probability spaces. This exploration leads to insightful results connecting PML, PMC, LIP, ALIP, and LDP. In doing so, our work significantly closes the gap in the known relations between various privacy frameworks. 

\subsection{Contributions and Summary of the Main Results}
Section~\ref{sec:operational} is dedicated to defining pointwise maximal cost and drawing connections to similar definitions. We begin by assuming finite alphabets and define PMC as the multiplicative decrease in the probability of incorrectly guessing the value of an arbitrary randomized function of $X$ upon observing an outcome $y$ of the privacy mechanism. We call this threat model the \emph{randomized function model} of PMC, and prove in Theorem~\ref{thm:randomized_function} that PMC, denoted by $\Lambda_{P_{XY}}(X \to y)$, can be expressed as
\begin{equation*}
    \Lambda_{P_{XY}}(X \to y) = \log \; \max_{x \in \cX} \; \frac{P_{X}(x)}{P_{X \mid Y=y}(x)}.
\end{equation*}

Next, we argue that PMC can alternatively be defined as the multiplicative decrease in the expected value of an arbitrary non-negative \emph{cost} function upon observing an outcome $y$ of the privacy mechanism. This setup is referred to as the \emph{cost function model}. Essentially, we prove an equivalence between the randomized function model and the cost function model of PMC (Theorem~\ref{thm:equivalence}). This equivalence signifies a parallelism with PML since a similar equivalence between the gain function model and the randomized function model of PML was previously established in~\cite[Thm. 2]{saeidian2023pointwise}. Table~\ref{tab:pml_pmc_summary} provides a comparative summary of PMC and PML, with the relevant definitions presented side by side to highlight their similarities and differences.

\begin{table*}[tp]
\caption{Comparative summary of PMC and PML.}
\centering
\renewcommand{\arraystretch}{1.4}
\begin{tabular}{|p{0.47\linewidth} p{0.47\linewidth}|}
\hline
\multicolumn{1}{|c}{\textbf{\normalsize Pointwise Maximal Cost}}  & \multicolumn{1}{c|}{\textbf{\normalsize Pointwise Maximal Leakage}}\\ \hline
\begin{itemize}[leftmargin=*]
\item Quantifies the threat of risk-averse adversaries
\item Definition via cost functions (Def. \ref{def:general_leakage})
\begin{equation}
    \Lambda_{P_{XY}}(X \to y) \coloneqq \log \sup_{\substack{(\cW, \cS_\cW), \\ c \in \cC}} \; \frac{\inf\limits_{P_W} \bE \Big[c(X,W) \Big]}{\inf\limits_{P_{\hat W \mid Y}} \bE \Big[c(X, \hat W) \mid Y=y \Big]}
\end{equation}

\item Definition via randomized functions (Def. \ref{def:randomized_function})
\begin{equation*}
    \Lambda_{P_{XY}}(X \to y) \coloneqq \sup_{U: U-X-Y} \log \frac{\inf\limits_{P_{\tilde U}}\; \bP \Big[U \neq \tilde{U} \Big]}{\inf\limits_{P_{\hat U \mid Y}} \bP \Big[U \neq \hat{U} \mid Y=y \Big]}
\end{equation*}

\item When $\cX$ is finite, cost and randomized function definitions are equivalent (Thm. \ref{thm:equivalence}).  

\item Computable form (Thm.~\ref{thm:general_leakage})
\begin{equation*}
    \Lambda_{P_{XY}}(X \to y) = \dinf(P_{X} \Vert P_{X \mid Y=y})
\end{equation*} 

\item Lower bounds the information density 
\begin{equation}
    \Lambda_{P_{XY}}(X \to y) \leq \varepsilon \iff -\varepsilon \leq i_{P_{XY}}(X;y) \;\; \text{a.s.}
\end{equation}

\item $\Lambda_{P_{XY}}(X \to y)$ can be infinite, even when $\cX$ is finite.
\end{itemize}
&
\begin{itemize}[leftmargin=*]
\item Quantifies the threat of opportunistic adversaries
\item Definition via gain functions \cite[Def. 3]{saeidian2023pointwisegeneral}
\begin{equation}
    \ell_{P_{XY}}(X \to y) \coloneqq \log \sup_{\substack{(\cW, \cS_\cW), \\ g \in \cG}} \frac{\sup\limits_{P_{\hat W \mid Y}} \bE \big[g(X,\hat W) \mid Y=y \big]}{\sup\limits_{P_W} \bE[g(X,W)]}
\end{equation}

\item Definition via randomized functions \cite[Def. 1]{saeidian2023pointwise}
\begin{equation*}
    \ell_{P_{XY}}(X\to y) = \sup_{U: U-X-Y} \log \frac{\sup\limits_{P_{\hat U \mid Y}} \bP \left[U=\hat U \mid Y=y \right]}{\sup\limits_{P_{\tilde U}}\; \bP \Big[U = \tilde{U} \Big]}
\end{equation*} 

\item When $\cX$ is finite, gain and randomized function definitions are equivalent \cite[Thm. 2]{saeidian2023pointwise}.

\item Computable form \cite[Thm. 3]{saeidian2023pointwisegeneral}
\begin{equation*}
    \ell_{P_{XY}}(X \to y) = \dinf(P_{X \mid Y=y} \Vert P_{X} )
\end{equation*}

\item Upper bounds the information density 
\begin{equation}
    \ell_{P_{XY}}(X \to y) \leq \varepsilon \iff i_{P_{XY}}(X;y) \leq \varepsilon \; \; \text{a.s.}
\end{equation}

\item $\ell_{P_{XY}}(X \to y) \leq \log \frac{1}{p_{\min}}$ when $\cX$ is finite.
\end{itemize}
\\ \hline  
\end{tabular}
\label{tab:pml_pmc_summary}
\end{table*}

In Sections~\ref{ssec:mcl_and_mrc} and~\ref{ssec:guesswork}, we explore the connections between PMC and other information measures designed with risk-averse adversaries in mind. These are \emph{maximal cost leakage}~\cite[Def. 11]{IssaMaxL}, \emph{maximal realizable cost}~\cite[Def. 12]{IssaMaxL}, and \emph{maximal guesswork leakage}~\cite{kurri2024maximal}. By carefully examining these definitions and their formulations, we observe that they are often more cumbersome and complex without strengthening the privacy model. For instance, both maximal cost leakage and maximal realizable cost involve a supremum over a Markov chain, which we demonstrate can be omitted. By contrast, we show that PMC offers a streamlined yet powerful and flexible formulation that effectively encompasses these alternative definitions. It is important to note that our discussions here are restricted to finite random variables, as the other measures have only been defined in this context.

In Section~\ref{ssec:general_alphabet}, we extend the cost function model to random variables on standard Borel spaces. Here, PMC is defined as the multiplicative decrease in the expected value of an arbitrary non-negative measurable cost function upon observing an outcome $y$ of the privacy mechanism. By doing so, we obtain a general form of PMC which can be written as
\begin{equation*}
    \Lambda_{P_{XY}}(X \to y) = \esssup_{P_X} \Big(- i_{P_{XY}}(X;y) \Big).
\end{equation*}
The above expression illustrates that imposing an upper bound on PMC is equivalent to imposing a lower bound on the information density. 

Moving beyond definitions, in Section~\ref{sec:properties} we explore the properties of PMC and examine its application to several well-known examples. We demonstrate that PMC meets the usual criteria expected of an information measure, such as non-negativity and satisfying data-processing inequalities (Lemma~\ref{lemma:properties}). We then use PMC to examine some central examples in privacy, including the randomized response mechanism~\cite{warnerRRoriginal, extremalmechanismLong}, the Laplace mechanism~\cite{DPoriginalpaper} used to release the sample mean of a dataset, and the Gaussian mechanism perturbing the value of a bounded random variable, among others.

In Section~\ref{sec:bounds}, we explore the interactions between PMC and PML guarantees and their relationships with other definitions, specifically ALIP, LIP, and LDP. We begin with finite random variables in Section~\ref{ssec:connections_finite} and demonstrate that if a mechanism satisfies a PML guarantee with a sufficiently small parameter, then it also satisfies a PMC guarantee. Conversely, we show that a PMC guarantee always implies a PML guarantee (Theorem~\ref{thm:liftasym}). This result is then leveraged to establish connections between PML, PMC, LIP, ALIP, and LDP and derive bounds between these measures (Corollary~\ref{cor:relations}).

We further show that a mechanism satisfies LDP if and only if it has bounded PMC for all priors with fixed support (Theorem~\ref{thm:ldp_pmc_sup}). This result mirrors a known characterization of LDP in terms of PML (see Theorem~\ref{thm:ldp_pml_sup}) and also offers a novel interpretation of LDP through risk-averse adversaries. Table~\ref{tab:relations} summarizes key relationships between these privacy measures for finite random variables, including both results from prior work and new connections established in this paper.

Next, in Section~\ref{ssec:connections_general}, we extend our analysis to random variables on standard Borel spaces. While in finite spaces a PMC guarantee always implies a PML guarantee, we show that this implication breaks down in more general settings. We construct two illustrative examples: one involving countably infinite random variables and another involving absolutely continuous ones. In both cases, the mechanisms have bounded PMC but unbounded PML. Lastly, we prove that a mechanism satisfies LDP if and only if it has both bounded PMC and bounded PML (Theorem~\ref{thm:general_ldp_pml_pmc}). This shows that LDP is equivalent to simultaneously restricting the threat imposed by risk-averse and opportunistic adversaries. Overall, our findings in Section~\ref{sec:bounds} provide a clear and unified understanding of the relationships among privacy measures that restrict the information density.

\subsection{Other Related Works} 
Operationally meaningful privacy measures first emerged within the area of \emph{quantitative information flow}~\cite{alvim2020science} through the concept of \emph{min-entropy leakage} \cite{smith2009foundations, braun2009quantitative}, also known as \emph{multiplicative Bayes leakage}. Min-entropy leakage assumes that an adversary attempts to guess the value of the secret $X$ in one try. Various extensions of this adversarial model have since been explored~\cite{alvim2012measuring, espinoza2013min, alvim2014additive, alvim2020science}. Among these works, the \emph{$g$-leakage} framework of ~\citet{alvim2012measuring} has been particularly influential. This framework generalizes min-entropy leakage by allowing adversaries to construct guesses of $X$ that maximize arbitrary non-negative gain functions. 

\citet{IssaMaxL} introduced the threat model approach to privacy into the domain of information theory through the development of \emph{maximal leakage}. This measure focuses on adversaries aiming to guess the values of arbitrary (randomized) functions of $X$. \citet{IssaMaxL} showed that maximal leakage is a highly robust measure since many variations of their original threat model still yield the same quantity as maximal leakage. \citet{liao2019tunable} further extended maximal leakage by introducing a tunable loss function called \emph{$\alpha$-loss}. Then, they considered the class of adversaries attempting to minimize the $\alpha$-loss, which gave rise to the notions of \emph{$\alpha$-leakage} and \emph{maximal $\alpha$-leakage}. More recent developments in this area include pointwise maximal leakage, defined as the pointwise adaptation of $g$-leakage and maximal leakage~\cite{saeidian2023pointwise}, \emph{maximal $(\alpha, \beta)$-leakage} \cite{gilani2024unifying}, and \emph{maximal $g$-leakage under multiple guesses} \cite{kurri2023operational}. 

The broader area of information theory has also contributed significantly to the study of privacy measures. Here, mutual information is notably the most extensively examined privacy measure~\cite{asoodeh2014notes, asoodeh2016information, asoodeh2015maximal, wang2016relation, makhdoumi2014information, liao2017hypothesis, rassouli2021perfect}. Beyond mutual information, generalized information measures such as $f$-information~\cite{diaz2019robustness} have also been used to assess privacy. Specific instances of $f$-information utilized as privacy measures include $\chi^2$-information~\cite{du2017principal, wang2019privacy} (associated with $\chi^2$-divergence), and total variation privacy~\cite{rassouli2019TVDprivacy} (associated with total variation distance). Additionally, some works have used per-letter $f$-divergences (as opposed to the average-case divergence in $f$-information) as privacy measures~\cite{zamani2021data, zamani2021design, zamani2023privacy}. Other research has examined the probability of correctly guessing~\cite{asoodeh2018estimation} or guesswork~\cite{massey1994guessing,malone2004guesswork} for guaranteeing privacy. The design of privacy mechanisms under various privacy constraints has also received considerable attention. These include mechanisms for local differential privacy~\cite{kalantari2018hamming, goseling2022robust, lopuhaa2024mechanisms}, maximal leakage~\cite{wu2020optimal, saeidian2021hamming}, total variation privacy~\cite{rassouli2019TVDprivacy}, and local information privacy~\cite{hsu2019information, sadeghi2021properties, jiang2021LIPcontextaware}, among others.

Finally, we note that privacy is often one of several competing goals in reliable and trustworthy learning systems, alongside objectives such as good generalization capabilities~\cite{shalev2014understanding}, fairness~\cite{dwork2012fairness}, and adversarial robustness~\cite{madry2018towards}. These goals have been jointly studied in various works. For instance, differential privacy has been linked to generalization in adaptive learning~\cite{dwork2015preserving, bassily2016algorithmic}. Recent work has also explored the interplay between privacy and adversarial robustness~\cite{zhang2025learning}.

\begin{table*}[t]
\caption{Summary of key implications and equivalences between LDP, PML, PMC, LIP, and ALIP for finite alphabets.}
\centering
\renewcommand{\arraystretch}{1.6}
\begin{tabular}{>{\raggedright\arraybackslash}p{0.5\linewidth} >{\raggedright\arraybackslash}p{0.2\linewidth}}
\toprule
\multicolumn{1}{c}{\textbf{Implications}} & \textbf{References} \\
\midrule

$\varepsilon$-LDP $\Rightarrow$ $-\log \left(p_{\min} + e^{-\varepsilon} (1 - p_{\min}) \right)$-PML & \cite[Thm. 1]{jiang2021LIPcontextaware} \\

$\varepsilon$-LDP $\Rightarrow$ $\log \left(p_{\min} + e^\varepsilon (1 - p_{\min}) \right)$-PMC & Prop.~\ref{prop:ldp_pmc} \\ 

$\varepsilon$-PML with $\varepsilon < \log \frac{1}{1 - p_{\min}}$ $\Rightarrow$ $ \log \left( \frac{p_{\min}}{1 - e^{\varepsilon}(1 - p_{\min})} \right)$-PMC & Thm.~\ref{thm:liftasym} \\

$\varepsilon$-PMC $\Rightarrow$ $\log \left( \frac{1 - e^{-\varepsilon}(1 - p_{\min})}{p_{\min}} \right)$-PML & Thm.~\ref{thm:liftasym} \\
\addlinespace[0.6em]
\midrule
\multicolumn{1}{c}{\textbf{Equivalences}} & \textbf{References}\\
\midrule

$\varepsilon$-LIP $\Leftrightarrow$ $\varepsilon$-PMC and $\varepsilon$-PML
& Thm.~\ref{thm:randomized_function}, \cite[Thm. 1]{saeidian2023pointwise} \\

$(\varepsilon_1, \varepsilon_2)$-ALIP $\Leftrightarrow$ $\varepsilon_1$-PMC and $\varepsilon_2$-PML
& Thm.~\ref{thm:randomized_function}, \cite[Thm. 1]{saeidian2023pointwise} \\

$\varepsilon$-LDP $\Leftrightarrow$ $\varepsilon$-PML for all $P_X$
& \cite[Thm. 14]{IssaMaxL}, \cite[Thm. 3]{fernandes2024explaining} \\

$\varepsilon$-LDP $\Leftrightarrow$ $\varepsilon$-PMC for all $P_X$
& Thm.~\ref{thm:ldp_pmc_sup} \\
\bottomrule
\end{tabular}
\label{tab:relations}
\end{table*}



\section{Preliminaries}
\label{sec:background}
\subsection{Notation and Assumptions}
\label{ssec:notation}
The following notations are used throughout the paper: $\mathbb R_+ = [0,\infty)$ and $\bar{\mathbb R}_+ = [0,\infty]$ denote the non-negative real numbers and the extended non-negative real numbers. The set $[n] \coloneqq \{1, \ldots, n\}$ represents the first $n$ positive integers, $\log(\cdot)$ refers to the natural logarithm, and $\ind_{\cA}$ denotes the indicator function of the set $\cA$. We also adopt the convention that $\nicefrac{0}{0} = 1$. 

Let $(\Omega, \mathcal H, \mathbb P)$ be an abstract probability space fixed in the background, where $\Omega$ is the sample space, $\cH$ is the event space (i.e., a $\sigma$-algebra on $\Omega$), and $\bP$ is a probability measure on the measurable space $(\Omega, \cH)$. We use $\cH_+$ to denote the set of all functions that are measurable relative to $\cH$ and $\mathcal B_{\bar{\mathbb R}_+}$, where $\mathcal B_{\bar{\mathbb R}_+}$ denotes the Borel $\sigma$-algebra on $\bar{\bR}_+$. Given a function $h \in \mathcal H_+$, the essential supremum of $h$ with respect to $\mathbb P$ is $\esssup\limits_{\mathbb P} h = \sup \{c \in \bar \bR_+ \colon \mathbb P(\{h > c\}) > 0\}$ and its essential infimum is $\essinf\limits_{\mathbb P} h = \inf \{c \in \mathbb R_+ \colon \mathbb P(\{h < c\}) > 0\}$. Suppose $\lambda$ and $\rho$ are measures on $(\Omega, \cH)$ and assume that $\lambda$ is $\sigma$-finite. If $\rho$ is absolutely continuous with respect to $\lambda$, denoted by $\rho \ll \lambda$, then we write $p = \frac{d\rho}{d\lambda}$ to imply that
\begin{equation*}
    \int_\Omega h(\omega) \; \rho(d\omega) = \int_\Omega h(\omega) \, p(\omega) \; \lambda(d\omega),
\end{equation*}
for all $h \in \cH_+$, where $p \in \cH_+$ is the Radon-Nikodym derivative of $\rho$ with respect to $\lambda$. 

Let $\cX$ be a set and $\cS_\cX$ a $\sigma$-algebra on $\cX$. A mapping $X : \Omega \to \cX$ is called a random variable taking values in $(\cX, \cS_\cX)$ if $X$ is measurable relative to $\mathcal H$ and $\cS_\cX$. In this paper, we exclusively use $X$ to denote some data containing sensitive information, i.e., the \emph{secret}. We use $P_X$ to denote the distribution of $X$. Often, we assume that $\cX$ is a discrete set and use some notations specific to this case. In particular, (with a slight abuse of notation) we also use $P_X$ to denote the probability mass function (pmf) of $X$ and write $P_X(x) \coloneqq P_X(\{x\})$ for $x \in \cX$. Moreover, we assume that $P_X$ has full support on $\cX$, that is, $P_X(x) > 0$ for all $x \in \cX$, and define $p_\mathrm{min} \coloneqq \min\limits_{x \in \cX} P_X(x)$ for finite $\cX$.

Let $(\cY, \cS_\cY)$ be a measurable space. A mapping $P_{Y \mid X} : \cX \times \cS_\cY \to \mathbb [0,1]$ is called a transition probability kernel (or simply kernel) from $(\cX, \cS_\cX)$ into $(\cY, \cS_\cY)$ if the mapping $x \mapsto P_{Y \mid X=x}(\cB)$ is in $\cS_{\cX_+}$ for all $\cB \in \cS_\cY$, and $P_{Y \mid X=x}(\cdot)$ is a probability measure on $(\cY, \cS_\cY)$ for all $x \in \cX$. We sometimes use $P_{Y \mid X} (\cB \mid x)$ instead of $P_{Y \mid X=x} (\cB)$. This notation is less awkward when we do not want to specify the outcome of $X$ but leave it as a random variable. The kernel $P_{Y \mid X}$ induces a random variable $Y$ taking values in $(\cY, \cS_\cY)$ with distribution $P_Y$, where 
\begin{equation}
\label{eq:margin}
    P_Y(\cB) = \int_\cX P_{Y \mid X=x}(\cB) \; P_X(dx), \quad \cB \in \cS_\cY.
\end{equation}
We write $P_Y = P_{Y \mid X} \circ P_X$ to represent \emph{marginalization} over $X$ described by~\eqref{eq:margin}. In this paper, we exclusively use $Y$ to denote some publicly available data that contains information about $X$, and refer to $P_{Y \mid X}$ as the \emph{privacy mechanism}. When $\cY$ is a discrete set, we use a similar notation described above in the case of discrete $\cX$. For example, we write $P_{Y \mid X=x}(y) \coloneqq P_{Y \mid X=x}(\{y\})$ with $y \in \cY$. 

Let $P_{XY}$ be a probability measure on the product space $(\cX \times \cY, \cS_\cX \otimes \cS_\cY)$ with marginals $P_X$ and $P_Y$. Then, we write $P_{XY} = P_{X} \times P_{Y \mid X}$ to denote \emph{disintegration}, i.e., to imply that 
\begin{align*}
    \mathbb E[h] &= \int_{\cX \times \cY} h(x,y) \; P_{XY} (dx,dy)\\
    &= \int_\cX P_X(dx) \int_\cY h(x,y) \; P_{Y \mid X=x} (dy),
\end{align*}
for all $h \in (\cS_\cX \otimes \cS_\cY)_+$. Let $P_X \times P_Y$ denote the product measure of $P_X$ and $P_Y$. Throughout the paper it is assumed that $P_{XY} \ll P_X \times P_Y$ and the logarithm of the Radon-Nikodym derivative 
\begin{equation}
\label{eq:info_density}
    i_{P_{XY}} \coloneqq \log \; \frac{dP_{XY}}{d (P_X \times P_Y)}, 
\end{equation}
is referred to as \emph{information density} of $X$ and $Y$. We assume that all probability measures on $(\cX, \cS_\cX)$ are absolutely continuous with respect to a common $\sigma$-finite measure $\mu$ and all probability measures on $(\cY, \cS_\cY)$ are absolutely continuous with respect to a common $\sigma$-finite measure $\nu$. Additionally, we assume that $P_{XY} \ll \mu \times \nu$. 

Finally, all measurable spaces are assumed to be \emph{standard Borel}~\citep[Def. 8.35]{klenke2013probability}. Standard Borel spaces have several convenient properties. Most prominently, joint distributions on standard Borel spaces can always be disintegrated into a kernel and a marginal distribution~\citep[Thm. IV.2.18]{ccinlar2011probability}. That is, if $P_{XY}$ is a distribution on $(\cX \times \cY, \cS_\cX \otimes \cS_\cY)$ with marginal $P_X$ on $(\cX, \cS_\cX)$, then there exists a transition probability kernel from $(\cX, \cS_\cX)$ into $(\cY, \cS_\cY)$ such that $P_{XY} = P_X \times P_{Y \mid X}$. Another advantage is that if $P_{Y \mid X}$ and $Q_{Y \mid X}$ are both kernels and $P_{Y \mid X=x} \ll Q_{Y \mid X=x}$ for all $x \in \cX$, then we may invoke Doob's version of the Radon-Nikodym theorem to obtain a Radon-Nikodym derivative $\frac{dP_{Y \mid X}}{dQ_{Y \mid X}}$ which is jointly measurable in $(x,y)$~\citep[Thm. V.4.44]{ccinlar2011probability}. This has the effect that we may alternatively use $\log \frac{dP_{Y \mid X}}{dP_Y}$ or $\log \frac{dP_{X \mid Y}}{dP_X}$ as the information density. Note that all measurable spaces of practical interest are standard Borel, e.g., countable sets, $\bR^n$, or complete separable metric spaces endowed with Borel $\sigma$-algebras.


\subsection{Pointwise Maximal Leakage}
\label{ssec:background_pml}
Pointwise maximal leakage (PML) is a recently introduced privacy measure that is operationally meaningful, robust, and flexible. For secrets with finite alphabets, PML is defined by quantifying risks in two powerful adversarial threat models: the \emph{gain function model}, first introduced by \citet{alvim2012measuring}, and the \emph{randomized function model}, put forward by \citet{IssaMaxL}. It has been shown that when $X$ is a finite random variable, these two models are equivalent \cite[Thm. 2]{saeidian2023pointwise}. Furthermore, the gain function model can be extended to secrets taking values in standard Borel spaces \cite{saeidian2023pointwisegeneral}.

Before we define PML, let us first recall the definition of Rényi divergence of order infinity~\cite{renyi1961entropy, van2014renyi}, which is used to provide simplified expressions for PML, and later, PMC.   

\begin{definition}[{Rényi divergence of order $\infty$~\cite[Thm. 6]{van2014renyi}}] 
Let $\bP$ and $\bQ$ be probability measures on the measurable space $(\Omega, \mathcal H)$. Let $\lambda$ be a $\sigma$-finite measure satisfying $\bP \ll \lambda$ and $\bQ \ll \lambda$. The Rényi divergence of order $\infty$ of $\bP$ from $\bQ$ is defined as
\begin{equation}
\label{eq:renyi_div_inf}
    D_\infty(\bP \Vert \bQ) = \log \sup_{\cA \in \mathcal H} \frac{\bP(\cA)}{\bQ(\cA)} = \log \left(\esssup_{\bP} \frac{p}{q}\right),
\end{equation}
where $p = \frac{d\bP}{d\lambda}$ and $q = \frac{d\bQ}{d\lambda}$. 
\end{definition}

If $\bP \ll \bQ$, then the divergence can also be expressed as 
\begin{align*}
    D_\infty(\bP \Vert \bQ) = \log \left(\esssup_{\bP} \frac{d\bP}{d\bQ} \right) = \log \left(\esssup_{\bQ} \frac{d\bP}{d\bQ} \right).
\end{align*}
On the other hand, if $\bP \nll \bQ$, then $D_\infty(\bP \Vert \bQ) = \infty$. When the sample space $\Omega$ is countable we may write~\eqref{eq:renyi_div_inf} in the form
\begin{equation*}
    D_\infty(\bP \Vert \bQ) = \log \left(\sup_{\omega \in \Omega} \frac{\bP(\omega)}{\bQ(\omega)} \right).
\end{equation*}

Below, we first define PML in its most general form using the gain function model. Then, assuming that $X$ is a finite random variable, we also define PML using the randomized function model.

\begin{definition}[Gain function view of PML {\cite[Def. 3]{saeidian2023pointwisegeneral}}]
\label{def:general_pml}
Given a joint distribution $P_{XY}$ on the product measurable space $(\cX \times \cY, \cS_\cX \otimes \cS_\cY)$, we define the pointwise maximal leakage from $X$ to $y \in \cY$ as
\begin{equation}
\label{eq:general_pml}
    \ell_{P_{XY}}(X \to y) \coloneqq \log \sup_{\substack{(\cW, \cS_\cW), \\ g \in \cG}} \frac{\sup\limits_{P_{\hat W \mid Y}} \bE \left[g(X,\hat W) \mid Y=y \right]}{\sup\limits_{P_W} \, \bE[g(X,W)]},
\end{equation}

where the supremum in the numerator is over all transition probability kernels $P_{\hat W \mid Y}$ from $(\cY, \cS_\cY)$ into $(\cW, \cS_\cW)$, and $\cG$ denotes the set of all gain functions defined as 
\begin{equation*}
    \cG \coloneqq \left\{g \in (\cS_\cX \otimes \cS_\cW)_+ \Big\vert \sup_{P_W} \, \bE[g(X,W)] < \infty \right\}.
\end{equation*}
\end{definition}
Definition~\ref{def:general_pml} can be explained in the following way. Consider an adversary who aims to construct an estimate of $X$ in order to maximize a non-negative gain function $g$. The gain function $g$ encapsulates the adversary's goal and can be adjusted to represent a variety of privacy attacks. For example, if $X$ is a dataset, then $g$ might model membership inference attacks \cite{saeidian2023pointwise}. To measure the information leakage associated with a single released outcome $y \in \cY$, we compare the expected value of $g$ after observing $y$ (in the numerator of~\eqref{eq:general_pml}) with the expected value of $g$ before observing $y$ (in the denominator). The posterior expected gain is calculated using the optimal estimation kernel $P_{\hat W \mid Y}$, while the prior expected gain is $\sup_{P_W} \, \bE[g(X,W)]$. Then, to ensure that the privacy measure resulting from this definition encompasses various attacks, the ratio of the posterior-to-prior expected gain is maximized over all possible measurable spaces $(\cW,\cS_\cW)$ and all gain functions in $\cG$.

Next, we consider finite $X$'s and define PML through the randomized function model. 
\begin{definition}[Randomized function view of PML {\cite[Def. 1]{saeidian2023pointwise}}]
\label{def:randomized_function_pml}
Suppose $(X,Y) \sim P_{XY}$, where $X$ is a finite random variable. The pointwise maximal leakage from $X$ to $y \in \cY$ is defined as
\begin{align}
\label{eq:pml_u_sup}
    \ell_{P_{XY}}(X\to y) &= \log \sup_{U: U-X-Y} \, \frac{\sup\limits_{P_{\hat U \mid Y}} \bP \left[U=\hat U \mid Y=y \right]}{\sup\limits_{P_{\tilde U}}\; \bP \Big[U = \tilde{U} \Big]},
\end{align}
where $U$, $\hat U$ and $\tilde U$ are random variables on a finite set $\cU$, $P_{U} = P_{U \mid X} \circ P_X$, and the Markov chain $U-X-Y-\hat U$ holds.
\end{definition}
Definition~\ref{def:randomized_function_pml} can be understood similarly to Definition~\ref{def:general_pml}. Let $U$ be a randomized function of $X$. To quantify the information leakage associated with a single outcome of the mechanism $y \in \cY$, we compare the probability of correctly guessing $U$ after observing $y$ with the \emph{a priori} probability of correctly guessing $U$ in a ratio. Then, to obtain a robust privacy measure, this posterior-to-prior ratio is maximized over all possible randomized functions of $X$, represented by the supremum over $U$'s satisfying the Markov chain $U-X-Y$. 

It has been shown that both Definitions~\ref{def:general_pml} and~\ref{def:randomized_function_pml} simplify to the following expression for PML~\cite{saeidian2023pointwise,saeidian2023pointwisegeneral}: 
\begin{equation*}
    \ell_{P_{XY}}(X \to y) = \dinf(P_{X \mid Y=y} \Vert P_X),  
\end{equation*}
where $P_{X \mid Y=y}$ denotes the posterior distribution of $X$ given $y \in \cY$. Observe that PML is a \emph{context-aware} privacy measure, that is, it depends on the prior distribution $P_X$.

Viewing $\ell_{P_{XY}}(X \to Y)$ as a function of the random variable $Y$, PML itself becomes a random variable whose various statistical properties can be examined and constrained to provide privacy guarantees. A stringent privacy guarantee can thus be defined by requiring $\ell_{P_{XY}}(X \to Y)$ to be essentially bounded. Formally, we say that a privacy mechanism $P_{Y \mid X}$ satisfies \emph{$\varepsilon$-PML} if $\ell_{P_{XY}}(X \to Y) \leq \varepsilon$ (almost surely) for some $\varepsilon \geq 0$. Building on this, \cite{grosse2024extremal} investigates the design of privacy mechanisms under the $\varepsilon$-PML constraint for secrets with finite alphabets. Specifically, the authors formulate a linear program for maximizing a class of convex utility functions called \emph{sub-convex} functions~\cite[Def. 5]{grosse2024extremal}. The optimal solutions to this linear program are referred to as \emph{PML extremal mechanisms}. The paper also introduces the concept of \emph{PML privacy regions} as a partitioning of the space of the privacy parameter $\varepsilon \in [0,\infty)$ into disjoint intervals. The privacy regions specify an upper bound on the number $(x,y) \in \cX \times \cY$ pairs with $P_{Y \mid X=x}(y)=0$. The first region, termed the \emph{high-privacy regime}, corresponds to the interval $0 \leq \varepsilon < \log\frac{1}{1- p_{\mathrm{min}}}$. In this region, no zero probability assignments are allowed in the mechanism. It is then proved that the PML extremal mechanism $P^*_{Y \mid X}$ in the high-privacy regime has the following form: 
\begin{equation}
     P^*_{Y \mid X=i}(j) = \begin{cases}
        1 - e^{\varepsilon}(1-P_X(i)) & \text{if}\; i=j, \\
        e^{\varepsilon}P_X(j) & \text{if}\; i \neq j,
    \end{cases}
\end{equation}
where $i,j \in [\abs{\cX}]$~\cite[Thm. 3]{grosse2024extremal}.

\subsection{Local Information Privacy and Asymmetric Local Information Privacy}
Besides PML, several other context-aware privacy measures have been defined in the literature. Two notable examples are \emph{local information privacy} (LIP) and \emph{asymmetric local information privacy} (ALIP). These measures are defined by axiomatically imposing lower and upper bounds on the information density between $X$ and $Y$. First, we define LIP which imposes symmetric lower and upper bounds.

\begin{definition}[Local information privacy {{\cite{jiang2018context,6483382}}}]
Given $\varepsilon \geq 0$, a privacy mechanism $P_{Y\mid X}$ is said to satisfy $\varepsilon$-LIP if $\abs{i_{P_{XY}}(X;Y)} \leq \varepsilon$ almost surely.
\end{definition}

It was observed in \cite{zarrabian2023lift} that the distribution of $i(X;Y)$ exhibits an asymmetry in the sense that it has a skewed long tail for negative values, and a much shorter range and a sharp fall-off for positive values. This asymmetry, which the authors call the \emph{lift-asymmetry}, motivated their definition of ALIP.

\begin{definition}[Asymmetric local information privacy \cite{zarrabian2023lift}]
Given $\varepsilon_l, \varepsilon_u \geq 0$, a privacy mechanism $P_{Y \mid X}$ is said to satisfy $(\varepsilon_l,\varepsilon_u)$-ALIP if $-\varepsilon_l \leq i_{P_{XY}}(X;Y) \leq \varepsilon_u$ almost surely.
\end{definition}

It follows directly from the above definitions that $\varepsilon_u$-LIP and $(\varepsilon_l,\varepsilon_u)$-ALIP both imply $\varepsilon_u$-PML. However, it is not immediately clear whether PML also provides LIP or ALIP guarantees, as PML does not directly impose a lower bound on the information density. As we will see, in some cases PML does implicitly impose a lower bound on the information density, thereby providing LIP and ALIP guarantees. We discuss this subject in Section~\ref{sec:bounds}.

\subsection{Local Differential Privacy}

Differential privacy in its original form assumes that all data is collected in a database managed by a trusted data curator. Here, we focus on the \emph{local} model of privacy offering an alternative that removes the need for a central database. In this model, each user's data is perturbed before collection to ensure that only the user has access to their original, unaltered information. This approach gives rise to the concept of \emph{local differential privacy} (LDP)~\citep{kasiviswanathan2011can, duchi2013LDPminmaxDEF}.

\begin{definition}[Local differential privacy]
\label{def:ldp}
Let $X$ and $X'$ be independent random variables distributed according to $P_X$. Given $\varepsilon \geq 0$, a privacy mechanism $P_{Y \mid X}$ satisfies $\varepsilon$-LDP if 
\begin{equation}
    \bP \left(\left \{\sup_{\cE \in \cS_\cY} \, \log \, \frac{P_{Y \mid X}(\cE \mid X)}{P_{Y \mid X}(\cE \mid X')} \leq \varepsilon \right \} \right) =1 ,
\end{equation}
where the probability is over the draws of $X$ and $X'$. Alternatively, $P_{Y \mid X}$ satisfies $\varepsilon$-LDP if  
\begin{equation*}
    \esssup_{P_X \times P_X} \; \dinf \Big(P_{Y \mid X} (\cdot \mid X) \big \Vert P_{Y \mid X} (\cdot \mid X') \Big) \leq \varepsilon. 
\end{equation*}
\end{definition}


\begin{remark}
It is usually said that local differential privacy is a \emph{context-free} definition independent of the distribution $P_X$. That is, it describes a property of the privacy mechanism $P_{Y \mid X}$. However, note that technically Definition~\ref{def:ldp} depends on $P_X$ through its support. 
\end{remark}

LDP has close connections to context-aware privacy measures. We state two results that illustrate these connections. The first, from~\cite{jiang2021LIPcontextaware}, demonstrates how LDP guarantees translate into PML, LIP, and ALIP guarantees assuming that $\cX$ and $\cY$ are finite given a fixed prior distribution $P_X$. Note that while~\cite[Thm.~1]{jiang2021LIPcontextaware} articulates the LIP result alone, its proof also provides guarantees in terms of ALIP and PML. Recall that $p_\mathrm{min} \coloneqq \min\limits_{x \in \cX} P_X(x)$. 
\begin{proposition}
\label{prop:ldp_pml_bound}
Let $X$ and $Y$ be finite random variables and suppose $X$ is distributed according to $P_X$. Given $\varepsilon \geq 0$, define 
\begin{gather}
    \varepsilon_1 =  \log \Big(p_\mathrm{min} + e^\varepsilon (1 - p_\mathrm{min}) \Big),\\
    \varepsilon_2 = - \log \Big(p_\mathrm{min} + e^{-\varepsilon} (1 - p_\mathrm{min}) \Big).
\end{gather}
If $P_{Y \mid X}$ satisfies $\varepsilon$-LDP, then it also satisfies $\varepsilon_1$-LIP, $(\varepsilon_1, \varepsilon_2)$-ALIP, and $\varepsilon_2$-PML.
\end{proposition}

The previous result is straightforward and intuitive, as it illustrates the transition from a context-free property of a mechanism to guarantees that hold under a specific prior distribution. However, the reverse direction is more intricate: it involves moving from a property tied to both the mechanism and a prior to a universal property that holds irrespective of the prior. Interestingly, \cite[Prop.~1]{jiang2021LIPcontextaware} showed that $\varepsilon$-LIP implies $2\varepsilon$-LDP, and \cite[Prop.~1]{zarrabian2023lift} established that $(\varepsilon_l, \varepsilon_u)$-ALIP yields $(\varepsilon_l + \varepsilon_u)$-LDP. In general, $\varepsilon$-PML may not imply any LDP guarantee, since a mechanism satisfying $\varepsilon$-PML may assign $P_{Y \mid X=x}(y)=0$ for some $(x,y) \in \cX \times \cY$. Nonetheless, as we will explore in Section~\ref{sec:bounds}, in the high-privacy regime PML guarantees can indeed translate into valid LDP guarantees. 

The second result, from~\cite{IssaMaxL} and~\cite{fernandes2024explaining}, provides an equivalent formulation of LDP via PML. Although not originally stated in this form, the equivalence follows directly from~\cite[Thm.~14]{IssaMaxL} and~\cite[Thm.~3]{fernandes2024explaining}, and leads to the following theorem.

\begin{theorem}
\label{thm:ldp_pml_sup}
Let $\cP_\cX$ denote the set of distributions with full support on $\cX$. For any $\varepsilon \geq 0$, the mechanism $P_{Y \mid X}$ satisfies $\varepsilon$-LDP, if and only if it satisfies $\varepsilon$-PML for all $P_X \in \cP_\cX$.
\end{theorem}
We establish a parallel result between LDP and PMC in Section~\ref{sec:bounds}.

\section{Quantifying Information Leaked to Risk-averse Adversaries}
\label{sec:operational}
The purpose of this section is to explore the implications of defining privacy by imposing a lower bound on the information density $i_{P_{XY}}$. We begin by assuming that $X$ is a finite random variable and demonstrate that given $y \in \cY$, $- \min_x i_{P_{XY}}(x;y)$ describes the information leaking to \emph{risk-averse} adversaries, as opposed to the \emph{opportunistic} adversaries of PML. In particular, we show that $- \min_x i_{P_{XY}}(x;y)$ is operationally meaningful in two adversarial threat models. The first setup called the \emph{randomized function} model, considers adversaries who wish to minimize the probability of incorrectly guessing the value of arbitrary (randomized) functions of $X$. The second setup, called the \emph{cost function} model, considers adversaries who wish to minimize arbitrary non-negative cost functions. We refer to the information leakage measure obtained in both threat models as \emph{pointwise maximal cost} (PMC). 

In Section~\ref{ssec:general_alphabet}, we extend the cost function definition of PMC to secrets on standard Borel spaces. When $\cX$ is finite, the cost function definition of PMC is closely related to three other informational leakage measures, namely \emph{maximal cost leakage}, \emph{maximal realizable cost}, and \emph{maximal guesswork leakage} defined in \cite{IssaMaxL} and \cite{kurri2024maximal}. We discuss their connections in Sections~\ref{ssec:mcl_and_mrc} and \ref{ssec:guesswork}. 

\subsection{Randomized Function Model}
Suppose $\cX$ is a finite set. Consider an adversary who attempts to guess the value of a (finite) randomized function of $X$, denoted by $U$. Upon observing $y \in \cY$, the adversary formulates a guess of $U$, denoted by $\hat U$ using a kernel $P_{\hat U \mid Y}$ that minimizes the probability of an incorrect guess. Essentially, this threat model is an adaptation of \cite{IssaMaxL} and \cite[Section II-A]{saeidian2023pointwise} to the risk-averse case, where the adversary seeks to minimize the probability of incorrectly guessing $U$ instead of maximizing the probability of correctly guessing it. To quantify the risk incurred by such an adversary, we compare the probability of incorrectly guessing $U$ with access to $y$ and the probability of incorrectly guessing $U$ without access. Consequently, we define 
\begin{equation}
\label{eq:U_def}
    \Lambda_{U}(X \to y) \coloneqq \log \; \frac{\inf_{P_{\tilde U}}\; \bP [U \neq \tilde{U}]}{\inf_{P_{\hat U \mid Y}} \; \bP [U \neq \hat{U} \mid Y=y]}. 
\end{equation}

In general, we may not know which function of $X$ the adversary is interested in, or different adversaries may be interested in different functions of $X$. Therefore, to achieve robustness in our model, we define the pointwise maximal cost (PMC) by maximizing $\Lambda_{U}(X \to y)$ over all possible $U$'s satisfying the Markov chain $U-X-Y$.

\begin{definition}[Randomized function view of PMC]
\label{def:randomized_function}
Suppose $X$ is a finite random variable and let $P_{XY}$ be the joint distribution of $X$ and $Y$. Given $y \in \cY$, the pointwise maximal cost from $X$ to $y$ is 
\begin{align}
\label{eq:lambda_sup_U}
    &\Lambda_{P_{XY}}(X \to y) \coloneqq \sup_{U: U-X-Y} \; \Lambda_{U}(X \to y) \\
    &=\sup_{U: U-X-Y} \; \log \; \frac{\inf_{P_{\tilde U}}\; \bP \Big[U \neq \tilde{U} \Big]}{\inf_{P_{\hat U \mid Y}} \; \bP \Big[U \neq \hat{U} \mid Y=y \Big]},
\end{align}
where $U, \tilde U$, and $\hat U$ are random variables defined on a finite set $\cU$. 
\end{definition}

Below, we show how $\Lambda_{P_{XY}}(X \to y)$ relates to $\min_x i(x;y)$.
\begin{theorem}
\label{thm:randomized_function}
Let $P_{XY}$ denote the joint distribution of $X$ and $Y$. Given $y \in \cY$, PMC can be expressed as 
\begin{equation*}
    \Lambda_{P_{XY}}(X \to y) = \dinf(P_X \Vert P_{X \mid Y=y}), 
\end{equation*}
where $P_{X \mid Y=y}$ denotes the posterior distribution of $X$. Thus, for all $y \in \mathcal Y$, lower bounding $\min_{x \in \cX} i(x;y)$ is equivalent to upper bounding $\Lambda_{P_{XY}}(X \to y)$. That is, given $\varepsilon \geq 0$ it holds that 
\begin{equation}
    -\varepsilon \leq \min_{x \in \mathcal X}i(x;y) \iff \Lambda_{P_{XY}}(X \to y) \leq \varepsilon,
\end{equation}
for all $y \in \cY$. 
\end{theorem}
\begin{IEEEproof}
Fix an arbitrary $y \in \cY$. First, we show that $\Lambda_{P_{XY}}(X \to y) \leq \dinf(P_X \Vert P_{X \mid Y=y})$. Let $u_1 \in \argmax\limits_{u \in \cU} P_U(u)$ and $u_2 \in \argmax\limits_{u \in \cU} P_{U \mid Y=y}(u)$. Using the fact that 
\begin{equation*}
    \inf_{P_{\tilde U}}\; \bP [U \neq \tilde{U}] = 1 - \sup_{P_{\tilde U}}\; \bP [U = \tilde{U}] = 1 - \max_{u \in \cU} P_U(u),
\end{equation*}
and
\begin{align*}
    \inf_{P_{\hat U \mid Y}} \; \bP [U \neq \hat{U} \mid Y=y] &= 1 - \sup_{P_{\hat U \mid Y}} \; \bP [U \neq \hat{U} \mid Y=y]\\
    &= 1 - \max_{u \in \cU} P_{U \mid Y=y}(u),
\end{align*}
we write 
\begin{align*}
    \exp\big(\Lambda_{P_{XY}}(X \to y)\big) &= \sup_{U: U-X-Y} \; \frac{1 - \max_u P_U(u)}{1 - \max_u P_{U \mid Y=y}(u)}\\
    &= \sup_{U: U-X-Y} \; \frac{\sum_{u \neq u_1} P_U(u)}{\sum_{u \neq u_2} P_{U \mid Y=y}(u)}. 
\end{align*}
Since $P_U(u_1) = \max_u P_{U}(u) \geq P_U(u_2)$, we have  
\begin{align*}
    &\frac{\sum_{u \neq u_1} P_U(u)}{\sum_{u \neq u_2} P_{U \mid Y=y}(u)} \leq \frac{\sum_{u \neq u_2} P_U(u)}{\sum_{u \neq u_2} P_{U \mid Y=y}(u)}\\
    &= \frac{\sum_{u \neq u_2} P_{U\mid Y=y}(u) \left(\frac{P_U(u)}{P_{U\mid Y=y}(u)}\right)}{\sum_{u \neq u_2} P_{U \mid Y=y}(u)}\\
    &\leq \max_{u \neq u_2} \frac{P_{U}(u)}{P_{U \mid Y=y}(u)}\\[0.5em]
    &= \max_{u \neq u_2} \frac{\sum\limits_{x \in \cX} P_{U \mid X=x}(u) P_X(x)}{\sum\limits_{x \in \cX} P_{U \mid X=x}(u) P_{X \mid Y=y}(x)}\\
    &= \max_{u \neq u_2} \frac{\sum\limits_{x \in \cX} P_{U \mid X=x}(u) P_{X\mid Y=y}(x) \left(\frac{P_X(x)}{P_{X \mid Y=y}(x)} \right)}{\sum\limits_{x \in \cX} P_{U \mid X=x}(u) P_{X \mid Y=y}(x)}\\
    &\leq \max_{x \in \cX} \frac{P_{X}(x)}{P_{X \mid Y=y}(x)}\\
    &= \exp \Big( \dinf(P_X \Vert P_{X \mid Y=y}) \Big). 
\end{align*}
We conclude that $\Lambda_{P_{XY}}(X \to y) \leq \dinf \Big(P_X \Vert P_{X \mid Y=y} \Big)$. 

Next, we show that $\Lambda_{P_{XY}}(X \to y) \geq \dinf(P_X \Vert P_{X \mid Y=y})$. Let $x^* \in \argmax_{x \in \cX} \frac{P_X(x)}{P_{X \mid Y=y}(x)}$, and define $V \coloneqq \ind_{\cX \setminus \{x^*\}}$. Then, $V$ is a binary random variable with distribution $P_{V}(0) = 1 - P_{V}(1) = P_X(x^*)$. Let $k$ be a large integer. We define a random variable $W$ with alphabet $\cW = \{1, \ldots, k+1\}$ and induced by the kernel 
\begin{equation*}
    P_{W \mid V=0}(w) = \begin{cases}
    \frac{1}{k} & \text{if} \; w \in \{1, \ldots, k\},\\
    0 & \text{if} \; w = k+1,\\
    \end{cases}
\end{equation*}
and 
\begin{equation*}
    P_{W \mid V=1}(w) = \begin{cases}
    0 & \text{if} \; w \in \{1, \ldots, k\},\\
    1 & \text{if} \; w = k+1.\\
    \end{cases}
\end{equation*}
By inspection, $P_W(w) = \frac{P_{V}(0)}{k} = \frac{P_X(x^*)}{k}$ for $w \in [k]$ and $P_W(k+1) = P_{V}(1)$. Thus, taking $k$ to be sufficiently large, we can ensure that $P_W(k+1) = \max_w P_W(w)$. Moreover, since the Markov chain $W-V-X-Y$ holds $P_{W \mid Y=y}(w) = \frac{P_{V \mid Y=y} (0)}{k} = \frac{P_{X \mid Y=y}(x^*)}{k}$ for $w \in [k]$ and $P_{W \mid Y=y}(k+1) = P_{V \mid Y=y}(1) = P_{X \mid Y=y}(\cX \setminus \{x^*\})$. Once again, taking $k$ to be sufficiently large we can ensure that $P_{W \mid Y=y}(k+1) = \max_w P_{W \mid Y=y}(w)$. 

Since $W - X - Y$ forms a Markov chain, we get 
\begin{align}
    \exp\Big(\Lambda_{P_{XY}}(X \to y)\Big) &= \sup_{U: U-X-Y} \; \frac{1 - \max_u P_U(u)}{1 - \max_u P_{U \mid Y=y}(u)}\\[0.5em]
    &\geq \frac{1 - \max_w P_W(w)}{1 - \max_w P_{W \mid Y=y}(w)}\\[0.5em]
    &= \frac{\sum_{w \neq k+1} P_W(w)}{\sum_{w \neq k+1} P_{W \mid Y=y}(w)}\\
    &= \frac{P_X(x^*)}{P_{X \mid Y=y}(x^*)}\\[0.5em]
    &= \max_{x \in \cX} \frac{P_X(x)}{P_{X \mid Y=y} (x)}\\[0.5em]
    &= \exp \left( \dinf \Big(P_X \Vert P_{X \mid Y=y}\Big) \right),
\end{align}
as desired.
\end{IEEEproof}

\begin{remark}[Closedness under pre-processing]
\label{rem:pre_proc}
Importantly, it follows directly from \eqref{eq:lambda_sup_U} that PMC is \emph{closed under pre-processing}, that is,
\begin{equation*}
    \Lambda_{P_{ZY}} (Z \to y) \leq \Lambda_{P_{XY}} (X \to y),
\end{equation*}
for all $Z$'s satisfying the Markov chain $Z-X-Y$ and all $y \in \cY$. Closedness under pre-processing implies that the amount of information leaking about (random) functions of $X$ can never exceed the amount of information leaking about $X$ itself.  
\end{remark}

\begin{remark}
\label{rem:pmc_unbounded}
It is worth emphasizing that even when $X$ and $Y$ are both finite random variables, PMC may be infinite. This occurs if $P_{Y \mid X=x}(y) =0$ for some $(x,y)$ pair with $P_X(x)P_Y(y) > 0$, in other words, if $P_X \times P_Y \nll P_{XY}$. This characteristic also distinguishes PMC from PML since PML is always bounded by $\log \frac{1}{p_{\min}}$ when $X$ is finite.
\end{remark}

\subsection{Cost Function Model}
\label{ssec:cost_function_model}
Once again, let $\cX$ be a finite set. Now, we consider adversaries aiming to minimize the expected value of non-negative \emph{cost functions} $c: \cX \times \cW \to \bR_+$, where $\cW$ is an arbitrary finite set representing the adversary's guessing space. In this scenario, the adversary observes $y \in \cY$ and selects $w \in \cW$ to minimize $\bE [c(X,w) \mid Y=y]$. Essentially, this threat model is an adaptation of \cite{alvim2012measuring} and \cite[Section II-B]{saeidian2023pointwise} to the risk-averse case, where the adversary seeks to minimize a cost function instead of maximizing a gain function. To quantify the risk associated with such an adversary, we compare the smallest posterior expected cost with the smallest prior expected cost in a ratio. Accordingly, we define 
\begin{equation}
\label{eq:cost_def}
    \Lambda_{c}(X \to y) \coloneqq \log \; \frac{\inf_{P_{W}} \; \bE \Big[c(X,W)\Big]}{\inf_{P_{\hat W \mid Y}} \; \bE \Big[c(X,\hat W) \mid Y=y \Big]}. 
\end{equation}
The expectation in the numerator of~\eqref{eq:cost_def} is taken with respect to $P_{XW} = P_X \times P_W$ since $X$ and $W$ are statistically independent. On the other hand, in the denominator of~\eqref{eq:cost_def}, the expectation is taken with respect to $P_{X \hat W \mid Y} = P_{X \mid Y} \times P_{\hat W \mid Y}$ since $X-Y-\hat W$ forms a Markov chain. 

Similar to the equivalence between the randomized function model and the gain function model established in \cite[Thm. 2]{saeidian2023pointwise} for PML, below we demonstrate that \eqref{eq:U_def} and \eqref{eq:cost_def} provide equivalent characterizations of PMC. This result is proved in Appendix~\ref{sec:proof_thm_equivalence}.

\begin{theorem}
\label{thm:equivalence}
Suppose $X$ is a finite random variable and let $y \in \cY$. For every randomized function of $X$, denoted by $U$, there exists a set $\cW_U$ and a cost function $c_{U} : \cX \times \cW_{U} \to \mathbb R_+$ such that $\Lambda_U(X \to y)  = \Lambda_{c_{U}} (X \to y)$. Conversely, for every cost function $c : \cX \times \cW \to \mathbb R_+$, there exists a randomized function of $X$, denoted by $U_c$, such that $\Lambda_c(X \to y) = \Lambda_{U_c}(X \to y)$.    
\end{theorem}

\begin{corollary}[Cost function view of PMC]
\label{cor:cost_function_model}
By Theorem~\ref{thm:equivalence}, PMC can alternatively be defined as 
\begin{equation}
    \Lambda_{P_{XY}}(X \to y) \coloneqq \sup_{c} \; \Lambda_c(X \to y),  
\end{equation}
where the supremum is over all non-negative cost functions with finite range. 
\end{corollary}

\subsection{Relationship to Maximal Cost Leakage and Maximal Realizable Cost}
\label{ssec:mcl_and_mrc}

In \cite{IssaMaxL}, Issa et al. defined two notions of information leakage that also correspond to risk-averse adversaries. The first notion, termed \emph{maximal cost leakage}, quantifies the average information leaking through a mechanism. The second notion, termed \emph{maximal realizable cost}, describes the largest amount of information leakage across all outcomes of a mechanism.

\begin{definition}[Maximal cost leakage~{\cite[Def. 11]{IssaMaxL}}] 
\label{def:mcl}
Given a joint distribution $P_{XY}$ on the finite set $\cX \times \cY$, the maximal cost leakage from $X$ to $Y$ is defined as 
\begin{equation}
\label{eq:mcl}
    \cL^{\sfc}(X \to Y) \coloneqq \!\!\sup_{\substack{U: U-X-Y, \\ \cW, c:\cU \times \cW \to \bR_+}} \; \log \; \frac{\inf\limits_{w \in \cW} \; \bE \Big[c(U,w) \Big]}{\inf\limits_{\hat w(\cdot)} \; \bE \Big[c(U, \hat w(Y)) \Big]}, 
\end{equation}
where $U$ takes values in an arbitrary finite set $\cU$, and the infimum in the denominator is over all functions $\hat w: \cY \to \cW$.  
\end{definition}

\citet[Thm. 15]{IssaMaxL} then showed that maximal cost leakage takes the simple form 
\begin{equation*}
    \cL^{\sfc}(X \to Y) = - \log \; \sum_{y \in \cY} \min_{x \in \cX} P_{Y \mid X=x}(y).
\end{equation*}

The general framework of maximal cost leakage closely resembles the cost function perspective of PMC outlined in Section~\ref{ssec:cost_function_model}. Similarly to PMC, Definition~\ref{def:mcl} compares the adversary's minimal prior cost with her minimal posterior cost. However, there are also two distinctions: Maximal cost leakage is defined for the average outcome of $Y$ and evaluates the costs of guessing randomized functions of $X$. In contrast, PMC is defined for each specific outcome $y \in \cY$ and evaluates the costs associated with guessing $X$ itself. 

Earlier in Remark~\ref{rem:pre_proc} we argued that PMC is closed under pre-processing; thus, it suffices to consider the costs of guessing $X$ itself. In a similar vein, below we demonstrate that when evaluating the information leaking to the average outcome of $Y$ as in maximal cost leakage, it suffices to analyze the costs associated with guessing $X$ alone. Specifically, we show that the supremum over $U$'s satisfying the Markov chain $U-X-Y$ in \eqref{eq:mcl} is superfluous. 

\begin{proposition}
\label{prop:mcl}
Given a joint distribution $P_{XY}$ on the finite set $\cX \times \cY$, the maximal cost leakage can alternatively be defined as 
\begin{equation*}
    \cL^{\sfc}(X \to Y) \coloneqq \sup_{\cW, c:\cX \times \cW \to \bR_+} \; \log \; \frac{\inf\limits_{w \in \cW} \; \bE \Big[c(X,w) \Big]}{\inf\limits_{\hat w(\cdot)} \; \bE \Big[c(X, \hat w(Y)) \Big]}. 
\end{equation*}
In other words, the supremum over $U$'s satisfying the Markov chain $U-X-Y$ in \eqref{eq:mcl} is superfluous. 
\end{proposition}

\begin{IEEEproof}
Let 
\begin{equation*}
    L_1 \coloneqq \sup_{\substack{U: U-X-Y, \\ \cW, c:\cU \times \cW \to \bR_+}} \; \log \; \frac{\inf\limits_{w \in \cW} \; \bE \Big[c(U,w) \Big]}{\inf\limits_{\hat w(\cdot)} \; \bE \Big[c(U, \hat w(Y)) \Big]}, 
\end{equation*}
and 
\begin{equation*}
    L_2 \coloneqq \sup_{\cW, c:\cX \times \cW \to \bR_+} \; \log \; \frac{\inf\limits_{w \in \cW} \; \bE \Big[c(X,w) \Big]}{\inf\limits_{\hat w(\cdot)} \; \bE \Big[c(X, \hat w(Y)) \Big]}. 
\end{equation*}
Our goal is to show that $L_1 = L_2$. Clearly, $L_1 \geq L_2$ so we show that $L_2 \geq L_1$.  

Fix a random variable $U$ satisfying the Markov chain $U-X-Y$ and let $\cU$ denote the alphabet of $U$. Let $c_1: \cU \times \cW \to \bR_+$ be a cost function, where $\cW$ is a finite set. Consider the cost function $c_2: \cX \times \cW \to \bR_+$ defined as 
\begin{equation*}
    c_2(x,w) = \sum_{u \in \cU} P_{U \mid X=x}(u) c_1(u,w) = \bE \Big[c_1(U,w) \mid X=x \Big].  
\end{equation*}
For all $w \in \cW$, we have 
\begin{align*}
    \bE \Big[c_1(U,w) \Big] = \bE_X \Big[ \bE\Big[c_1(U,w) \mid X \Big] \Big] =  \bE_X \Big[c_2(X,w) \Big]. 
\end{align*}
Similarly, given a function $\hat w: \cY \to \cW$ it holds that 
\begin{align*}
    &\bE \Big[c_1(U, \hat w(Y)) \Big] = \sum_{y \in \cY} P_Y(y) \, \bE\Big[c_1(U,\hat w(y)) \mid Y=y \Big]\\
    &=  \sum_{y \in \cY} \!P_Y(y) \sum_{x \in \cX} \!P_{X \mid Y=y}(x) \sum_{u \in \cU} \!P_{U \mid X=x, Y=y}(u)  \, c_1(u,\hat w(y))\\
    &= \sum_{y \in \cY} P_Y(y) \sum_{x \in \cX} P_{X \mid Y=y}(x) \sum_{u \in \cU} P_{U \mid X=x}(u)  \, c_1(u,\hat w(y))\\
    &= \sum_{y \in \cY} P_Y(y) \sum_{x \in \cX} P_{X \mid Y=y}(x) \, \bE \Big[c_1(U, \hat w(y)) \mid X=x \Big]\\
    &= \sum_{y \in \cY} P_Y(y) \sum_{x \in \cX} P_{X \mid Y=y}(x) \, c_2(x,\hat w(y))\\ 
    &= \bE \Big[c_2(X, \hat w(Y)) \Big],
\end{align*}
where the third equality is due to the Markov chain $U-X-Y$. Hence, for each random variable $U$ satisfying the Markov chain $U-X-Y$ and cost function $c_1: \cU \times \cW \to \bR_+$, there exists a cost function $c_2: \cX \times \cW \to \bR_+$ such that 
\begin{equation*}
    \frac{\inf\limits_{w \in \cW} \; \bE \Big[c_1(U,w) \Big]}{\inf\limits_{\hat w(\cdot)} \; \bE \Big[c_1(U, \hat w(Y)) \Big]} = \frac{\inf\limits_{w \in \cW} \; \bE \Big[c_2(X,w) \Big]}{\inf\limits_{\hat w(\cdot)} \; \bE \Big[c_2(X, \hat w(Y)) \Big]}.
\end{equation*}
This implies that $L_1 \leq L_2$, as desired.     
\end{IEEEproof}

It is also worth noting that maximal cost leakage does \emph{not} correspond to the expected value of the PMC. Using Corollary~\ref{cor:cost_function_model} we have
\begin{subequations}
\begin{align}
&\cL^{\sfc}(X \to Y)\\
&= \sup_{U : U-X-Y} \left(-\log \; \bE_{Y \sim P_Y} \left[\exp \Big(-\Lambda_{P_{UY}} (U \to Y) \Big) \right] \right)\\
&\leq \sup_{U : U-X-Y} \bE_{Y \sim P_Y} \left[\Lambda_{P_{UY}} (U \to Y) \right] \label{subeq:mcl_1}\\
&\leq \bE_{Y \sim P_Y} \left[\sup_{U : U-X-Y} \Lambda_{P_{UY}} (U \to Y) \right] \\  
&= \bE_{Y \sim P_Y} \left[\Lambda_{P_{XY}} (X \to Y) \right],  \label{subeq:mcl_2}
\end{align}
\end{subequations}
where~\eqref{subeq:mcl_1} is due to Jensen's inequality, and \eqref{subeq:mcl_2} follows from the closedness of PMC under pre-processing. Note that the strict concavity of $\log(\cdot)$ implies that Jensen's inequality is strict. Consequently, maximal cost leakage does not describe the expected value of PMC but underestimates it.

\begin{definition}[Maximal realizable cost~{\cite[Def. 12]{IssaMaxL}}]
Given a joint distribution $P_{XY}$ on the finite set $\cX \times \cY$, the maximal realizable cost from $X$ to $Y$ is defined as 
\begin{align}
\label{eq:mrc}
    \cL^{\sfr \sfc}&(X \to Y)\\
    &\coloneqq \!\!\sup_{\substack{U: U-X-Y, \\ {\cW}, c:\cU \times \cW \to \bR_+}} \; \log \; \frac{\inf\limits_{w \in \cW} \; \bE \Big[c(U,w) \Big]}{\min\limits_{y \in \cY} \; \inf\limits_{\hat w \in \cW} \; \bE \Big[c(U, \hat w) \mid Y=y \Big]}, 
\end{align}
where $U$ takes values in an arbitrary finite set $\cU$.
\end{definition}
Once again, the threat model of maximal realizable cost is similar to that of PMC. The difference is that maximal realizable cost is defined for the worst outcome of $Y$ and evaluates the costs of guessing randomized functions of $X$. 

\citet[Thm. 16]{IssaMaxL} showed that maximal realizable cost can be expressed as 
\begin{equation*}
    \cL^{\sfr \sfc}(X \to Y) = \dinf(P_{X} \times P_Y \Vert P_{XY}). 
\end{equation*}
Therefore, the maximal realizable cost is related to PMC by 
\begin{align*}
    \cL^{\sfr \sfc}(X \to Y) &= \max_{y \in \cY} \; \sup_{U:U-X-Y} \; \Lambda_{P_{UY}} (U \to y)\\
    &= \max_{y \in \cY} \; \Lambda_{P_{XY}} (X \to y),
\end{align*}
where the second equality is due to the closedness of PMC under pre-processing. This again implies that the supremum over $U$'s in \eqref{eq:mrc} is not required (cf. Propostion~\ref{prop:mcl}).

In general, defining an information measure for each outcome $y \in \cY$ as done by PMC has the advantage that $\Lambda_{P_{XY}} (X \to Y)$ may be viewed as a random variable that can be restricted in various ways. Hence, the resulting privacy analysis is not restricted to computing the average or maximum leakages but can be tailored to application-specific requirements, such as restricting the tail of $\Lambda_{P_{XY}}(X \to Y)$ or its different moments. Consequently, PMC is rendered a flexible tool that is applicable across a wider spectrum of scenarios. The flexibility of PMC will become more apparent when we discuss concrete examples in Section~\ref{ssec:examples}.

\subsection{Relationship to Maximal Guesswork Leakage} 
\label{ssec:guesswork}

A central example of a cost function is the expected number of attempts (or moments thereof) required to correctly guess a secret~\cite{massey1994guessing,malone2004guesswork}. In this context, a guessing function $G$ is a one-to-one mapping from $\cX$ to $[\abs\cX]$, and the \emph{guesswork} is the minimum expected number of attempts required to correctly identify $X$, i.e., $\min_G \bE[G(X)]$. Recently, guesswork has been used to define two information measures called \emph{maximal guesswork leakage} and \emph{pointwise maximal guesswork leakage}~\cite{kurri2024maximal}. Both measures describe adversaries aiming to minimize the guesswork associated with randomized functions of $X$ and compare the prior guesswork with the posterior guesswork. Per usual terminology, maximal guesswork leakage describes the average amount of information leaking through a mechanism, while pointwise maximal guesswork leakage is defined for each outcome $y \in \cY$. 

Here, we focus on pointwise maximal guesswork leakage and its connections to PMC. Similar discussions apply to maximal guesswork leakage.

\begin{definition}[Pointwise maximal guesswork leakage~{\cite[Def. 6]{kurri2024maximal}}]
Given a joint distribution $P_{XY}$ on the finite set $\cX \times \cY$, the pointwise maximal guesswork leakage from $X$ to $y \in \cY$ is defined as 
\begin{equation*}
    \cL^\mathsf{G-pw} (X \to y) \coloneqq \sup_{U:U-X-Y} \log \frac{\min_G \, \bE \Big[G(U) \Big]}{\min_{\hat G} \, \bE \Big[\hat G(U) \mid Y=y \Big]}, 
\end{equation*}    
where $U$ takes values in an arbitrary finite alphabet. 
\end{definition}

Pointwise maximal guesswork leakage is a special case of the cost function model of PMC. In particular, given a random variable $U$ satisfying the Markov chain $U-X-Y$ and a guessing function $G: \cU \to [\abs{\cU}]$, there exists a cost function $d: \cU \times \cW \to [\abs{\cU}]$ such that
\begin{gather*}
    \min_G \, \bE \Big[G(U) \Big] = \min_{w \in \cW} \bE \Big[d(U,w) \Big],\\
    \min_{\hat G} \, \bE \Big[\hat G(U) \mid Y=y \Big] = \min_{\hat w \in \cW} \bE \Big[d(U,\hat w) \mid Y=y\Big],
\end{gather*}
(see~\cite[Sec. III]{kurri2024maximal} for the exact construction). Furthermore, we showed in (the proof of) Proposition~\ref{prop:mcl} that for each $U$ satisfying the Markov chain $U-X-Y$ and each cost function $d: \cU \times \cW \to \bR_+$, there exists a cost function $c: \cX \times \cW \to \bR_+$ such that 
\begin{gather*}
    \bE \Big[d(U,w) \Big] =  \bE \Big[c(X,w) \Big],\\  
    \bE \Big[d(U,w) \mid Y=y \Big] =  \bE \Big[c(X,w) \mid Y=y \Big], 
\end{gather*}
for all $w \in \cW$. This argument shows that $\cL^\mathsf{G-pw} (X \to y) \leq \Lambda(X \to y)$. Interestingly, \citet{kurri2024maximal} proved that the inequality is in fact an equality, i.e., 
\begin{equation*}
    \cL^\mathsf{G-pw} (X \to y) = \Lambda(X \to y) = \dinf(P_X \Vert P_{X \mid Y=y}), \; y \in \cY. 
\end{equation*}

\subsection{PMC on General Alphabets}
\label{ssec:general_alphabet}
Now, we extend the cost function model described earlier to a much broader class of secrets. Suppose $X$ is defined on a measurable space $(\cX, \cS_\cX)$, and $Y$ is defined on a measurable space $(\cY, \cS_\cY)$. Let $P_{XY}$, denote the joint distribution of $X$ and $Y$.

\begin{definition}
\label{def:general_leakage}
Given a joint distribution $P_{XY}$ on the product space $(\cX \times \cY, \cS_\cX \otimes \cS_\cY)$ the pointwise maximal cost from $X$ to $y \in \cY$ is defined as
\begin{equation}
\label{eq:general_leakage}
    \Lambda_{P_{XY}}(X \to y) \coloneqq \log \!\sup_{\substack{(\cW, \cS_\cW), \\ c \in \cC}} \frac{\inf_{P_W} \; \bE \Big[c(X,W) \Big]}{\inf_{P_{\hat W \mid Y}} \bE \Big[c(X, \hat W) \mid Y=y \Big]},
\end{equation}
where the infimum in the denominator is over all transition probability kernels from $(\cY, \cS_\cY)$ into $(\cW, \cS_\cW)$, and $\cC$ denotes the set of all cost functions defined as 
\begin{equation*}
    \cC \coloneqq \left\{c \in (\cS_\cX \otimes \cS_\cW)_+ \Big\vert \inf_{P_W} \; \bE \Big[c(X,W) \Big] < \infty \right\}.
\end{equation*}
\end{definition}

Definition \ref{def:general_leakage} describes a threat model similar to the cost function model in Section \ref{ssec:cost_function_model}. It represents an adversary who aims to construct an estimate of $X$, denoted by $W$, minimizing the expected value of a cost function. Here, $W$ is a random variable on a measurable space $(\mathcal{W}, \cS_\cW)$, and cost functions are selected from the collection $\mathcal{C}$. To quantify the information revealed about $X$ by disclosing an outcome $y \in \mathcal{Y}$, we compare the minimum expected cost before observing $y$ in the numerator of \eqref{eq:general_leakage} with the minimum expected cost after observing $y$ in the denominator. We then take the supremum of this ratio across all measurable spaces $(\mathcal{W}, \mathcal{S}_\mathcal{W})$ and all $c \in \mathcal{C}$ to obtain a general and robust definition that does not depend on specific instances of $c$. Note that the condition $\inf_{P_W} \; \bE \Big[c(X,W) \Big] < \infty$  ensures that \eqref{eq:general_leakage} is well-defined since we exclude cost functions that result in infinite minimum costs both in the numerator and the denominator. 

Before simplifying \eqref{eq:general_leakage}, let us give two concrete examples of privacy attacks encompassed by Definition~\ref{def:general_leakage}. 

\begin{example}[Minimizing $L_p$-distance]
A key objective in statistics is to produce an estimate of a random quantity with the smallest mean squared error. More generally, suppose $X$ is a real-valued random variable with distribution $P_X$, let $p \geq 1$, and suppose $\bE[\abs{X}^p] < \infty$. The $L_p$-norm of $X$ is 
\begin{equation*}
    \norm{X}_p = \left( \int_{\bar \bR} \abs{x}^p \, P_X(dx) \right)^{\frac{1}{p}}.
\end{equation*}
The $L_p$-norm defines a metric space through the metric $d(X,W) \coloneqq \norm{X-W}_p$, where $W$ is also a real-valued random variable. 

In this context, an adversary may aim to construct an estimate of $X$, denoted by $W$, minimizing the $L_p$-distance to $X$. Using the notation of Definition \ref{def:general_leakage}, this attack can be modeled by taking $\cW = \bar \bR$ and $\cS_\cW = \cB_{\bar \bR}$. Then, the cost function $c_d(X,W) = \abs{X - W}^p$, $c_d \in \cC$ can be employed to capture the adversary's objective. 
\end{example}

\begin{example}[Minimizing $\alpha$-loss]
Let $\cX$ be a finite set with cardinality $k$, and let $\Delta^{k-1}$ denote the set of all distributions on $\cX$, i.e., the $(k-1)$-dimensional probability simplex. \citet{liao2019tunable} introduced a class of loss functions called \emph{$\alpha$-loss} with $\alpha \in [1, \infty]$ that interpolate between the log-loss at $\alpha =1$ and the 0-1 loss at $\alpha=\infty$. More precisely, they defined the $\alpha$-loss $c_\alpha : \cX \times \Delta^{k-1} \to \bar \bR_+$ by
\begin{equation*}
    c_\alpha(x,Q) \coloneqq \begin{cases}
        \frac{\alpha}{\alpha -1} \left(1-Q(x)^{\frac{\alpha-1}{\alpha}}\right) & \mathrm{if} \; \alpha \in (1, \infty),\\
        \log \frac{1}{Q(x)} & \mathrm{if} \; \alpha=1,\\ 
        1-Q(x) & \mathrm{if} \; \alpha=\infty, 
    \end{cases}
\end{equation*}
where $Q \in \Delta^{k-1}$ is a distribution on $\cX$. Then, they used the $\alpha$-loss to define a class of tunable information leakage measure called (\emph{maximal}) \emph{$\alpha$-leakage}~\cite[Def. 5 and 6]{liao2019tunable}.\footnote{Note that (maximal) $\alpha$-leakage assumes that the adversary maximizes an objective function such that the optimal action corresponds to minimizing the $\alpha$-loss. In contrast, here we assume that the adversary directly minimizes the $\alpha$-loss.} 

Notably, adversaries minimizing the expected value of an $\alpha$-loss are incorporated into the threat model of Definition~\ref{def:general_leakage}. To illustrate this, let $\cW = \Delta^{k-1}$, and let $\cS_\cW$ be the Borel $\sigma$-algebra on $\Delta^{k-1}$. Given $\alpha \in [1, \infty]$, we may use the cost function $c_\alpha(X,W) \in \cC$ to capture the adversary's objective. More generally, we can extend this model to encompass adversaries who minimize the $\alpha$-loss associated with randomized functions of $X$. 
\end{example}

Now, we show that PMC on general alphabets also reduces to $\dinf(P_X \Vert P_{X \mid Y=y})$. This result is proved in Appendix~\ref{sec:proof_thm_general_leakage}. 

\begin{theorem}
\label{thm:general_leakage}
Let $P_{XY}$ be the joint distribution of $X$ and $Y$. Given $y \in \cY$, PMC as described by Definition~\ref{def:general_leakage} simplifies to
\begin{equation*}
    \Lambda_{P_{XY}}(X \to y) = \dinf(P_{X} \Vert P_{X \mid Y=y}). 
\end{equation*}
\end{theorem}

Henceforth, we drop the subscript $P_{XY}$ in $\Lambda_{P_{XY}} (X \to y)$ and $i_{P_{XY}}$ when the joint distribution is clear from the context. 

\begin{remark}[Measurability of PMC] 
We discussed earlier that one of the advantages of PMC over similar existing notions is its flexibility, as $\Lambda(X \to Y)$ is a random variable whose various statistical properties can be examined and controlled. For this to work, the mapping $y \mapsto \Lambda_{P_{XY}}(X \to y)$ must be measurable. Note that the assumption that $(\cX, \cS_\cX)$ and $(\cY, \cS_\cY)$ are standard Borel spaces is sufficient for this purpose (see Section~\ref{ssec:notation}). Consequently, assuming that $P_X \times P_Y \ll P_{XY}$, we can express PMC by 
\begin{align*}
    \Lambda(X \to y) = \esssup_{P_X} \Big(- i(X;y) \Big).
\end{align*}

Let $\mu$ and $\nu$ be dominating $\sigma$-finite measures on $(\cX, \cS_\cX)$ and $(\cY, \cS_\cY)$, respectively. We can also express PMC using densities: 
\begin{align*}
    \Lambda(X \to y) &=\log \left(\esssup_{P_X} \frac{f_X(X)}{f_{X \mid Y} (X \mid y)}\right)\\[0.5em]
    &= \log \left(\esssup_{P_X} \frac{f_Y(y)}{f_{Y \mid X}(y \mid X)}\right),
\end{align*}
where $f_{X \mid Y} \in (\cS_\cX \otimes \cS_\cY)_+$ is the density of $P_{X \mid Y}$ with respect to $\mu$, and $f_X \in {\cS_\cX}_+$ is the density of $P_X$ with respect to $\mu$. The densities $f_{Y \mid X}$ and $f_Y$ are defined similarly.
\end{remark}

Having settled the measurability of PMC, we say that a privacy mechanism $P_{Y \mid X}$ satisfies $\varepsilon$-PMC if 
\begin{equation*}
    \bP \big\{ \Lambda(X \to Y) \leq \varepsilon \big \} = 1, 
\end{equation*}
where $\varepsilon \geq 0$. This represents the strongest type of guarantee that can be defined based on PMC, as it requires $\Lambda(X \to Y)$ to be essentially bounded.

\section{Properties of PMC and Examples}
\label{sec:properties}
This section focuses on two main topics. First, we explore the key properties of PMC, demonstrating that it satisfies the usual criteria expected from an information measure, such as non-negativity and data-processing inequalities. Next, we examine some prototypical examples in privacy, including the randomized response mechanism~\cite{warnerRRoriginal, extremalmechanismLong} applied to finite secrets, and the Laplace mechanism~\cite{DPoriginalpaper} for releasing the sample mean of a dataset. In all examples, we calculate, analyze, and discuss the PMC. 

\subsection{Properties of Pointwise Maximal Cost}
Before exploring the properties of PMC, we introduce a conditional version of PMC. This definition measures the amount of information leaked to an adversary who possesses some side information about the secret already before observing the mechanism's outcome. The side information is modeled as the outcome of a random variable $Z$ that is correlated with both $X$ and $Y$. Conditional PMC is particularly useful for deriving an upper bound on the leakage when multiple observations are available.

\begin{definition}[Conditional PMC]
\label{def:cond_pmc}
Let $Z$ be a random variable on a measurable space $(\cZ, \cS_\cZ)$. Given a transition probability kernel $P_{XY \mid Z}$ from $(\cZ, \cS_\cZ)$ into $(\cX \times \cY, \cS_\cX \otimes \cS_\cY)$, the conditional pointwise maximal cost from $X$ to $y \in \cY$ given $z \in \cZ$ is defined as
\begin{align}
    &\Lambda_{P_{XY \mid Z}}(X \to y \mid z)\\
    &\coloneqq \log \sup_{\substack{(\cW, \cS_\cW), \\ c \in \cC_z}} \;  \frac{\inf_{P_{W \mid Z}} \; \bE \Big[c(X,W) \mid Z=z \Big]}{ \inf_{P_{\hat W \mid Y,Z}} \bE \Big[c(X,\hat W) \mid Y=y,Z=z \Big]},
\end{align}
where $\cC_z$ denotes the set of all cost functions defined as 
\begin{equation*}
    \cC_z \coloneqq \left\{c \in (\cS_\cX \otimes \cS_\cW)_+ \Big\vert \inf_{P_{W \mid Z}} \bE \Big[c(X,W) \mid Z=z \Big] < \infty \right\}.
\end{equation*}
\end{definition}

Technically, Definition~\ref{def:cond_pmc} does not introduce a new concept; rather it is an adaptation of Definition~\ref{def:general_leakage} where all distributions are conditioned on $Z=z$. Thus, conditional PMC describes a similar threat model to PMC, with the key difference being the additional side information $Z=z$. Writing $P_{XY \mid Z} = P_{Y \mid X,Z} \times P_{X \mid Z}$, the outcome $z \in \cZ$ essentially parameterizes the distribution of $X$, denoted by $P_{X \mid Z}$, and the information release mechanism $P_{Y \mid X,Z}$. As such, by applying Theorem~\ref{thm:general_leakage}, we directly obtain a simplified form for the conditional PMC.

\begin{corollary}
Let $P_{XY \mid Z}$ be a transition probability kernel from $(\cZ, \cS_\cZ)$ into $(\cX \times \cY, \cS_\cX \otimes \cS_\cY)$. The conditional pointwise maximal cost from $X$ to $y \in \cY$ given $z \in \cZ$ can be expressed as 
\begin{equation*}
    \Lambda_{P_{XY \mid Z}}(X \to y \mid z) = \dinf(P_{X \mid Z=z} \Vert P_{X \mid Y=y, Z=z}). 
\end{equation*}
\end{corollary}
As a special case, if the Markov chain $Z-X-Y$ holds, then $\Lambda_{P_{XY \mid Z}}(X \to y \mid z) = \dinf(P_{X \mid Z=z} \Vert P_{X \mid Y=y})$. This scenario corresponds to the important case where the side information determines the adversary's prior knowledge about $X$ but does not influence the mechanism through which information is released.

Now, we discuss the properties of PMC. Most properties are consequences of the fact that PMC is a \Ren divergence between two distributions. The following lemma confirms that PMC satisfies all the usual properties expected from an information measure, e.g., non-negativity, data-processing inequalities, and additivity. 

\begin{lemma}[Properties of PMC]
\label{lemma:properties}
The pointwise maximal cost satisfies the following properties: 
\begin{enumerate}
    \item (Non-negativity). $\Lambda(X \to y) \geq 0$ for all $y \in \cY$. 
    \item (Independence). If $X$ and $Y$ are independent, then $\Lambda(X \to Y) = 0$ almost surely.
    \item (Additivity). If $P_{Y_1, \ldots, Y_k \mid X_1, \ldots, X_k} = \prod_{i=1}^k P_{Y_i \mid X_i}$ and $P_{X_1, \ldots, X_k} = \prod_{i=1}^k P_{X_i}$, then 
    \begin{equation*}
        \Lambda(X_1, \ldots, X_k \to Y_1, \ldots, Y_k) = \sum_{i=1}^k \Lambda(X_i \to Y_i).
    \end{equation*}
    \item (Concavity). Given $y \in \cY$, $\Lambda(X \to y)$ is concave in $P_X$. 
    \item (Pre-processing). Suppose the Markov chain $Z - X - Y$ holds. Then, for all $y \in \cY$ it holds that 
    \begin{equation*}
        \Lambda(Z \to y) \leq \Lambda(X \to y).
    \end{equation*}
    \item (Post-processing). Suppose the Markov chain $X - Y - Z$ holds. Then, we have 
    \begin{equation*}
        \esssup_{P_Z} \Lambda(X \to Z) \leq \esssup_{P_Y} \Lambda(X \to Y).
    \end{equation*}
    \item (Composition). Given a joint distribution $P_{X Y_1 Y_2}$, it holds that 
    \begin{multline*}
        \Lambda_{P_{X Y_1 Y_2}}(X \to y_1, y_2)\\
        \leq \Lambda_{P_{X Y_1}}(X \to y_1) + \Lambda_{P_{X Y_2 \mid Y_1}}(X \to y_2 \mid y_1), 
    \end{multline*}
    for all $y_1 \in \cY_1$ and $y_2 \in \cY_2$. 
\end{enumerate}
\end{lemma}
Lemma~\ref{lemma:properties} is proved in Appendix~\ref{sec:proof_lemma_properties}.

\subsection{Examples}
\label{ssec:examples}
In this section, we study PMC in several examples. By calculating PMC for commonly used mechanisms, such as the randomized response mechanism~\cite{warnerRRoriginal, extremalmechanismLong} and the Laplace mechanism~\cite{DPoriginalpaper,dwork2014algorithmic}, we demonstrate how PMC can be effectively used to quantify privacy.

\begin{figure*}
    \centering
    \begin{subfigure}[b]{0.45\textwidth}
        \centering
        \includegraphics[width=0.9\textwidth]{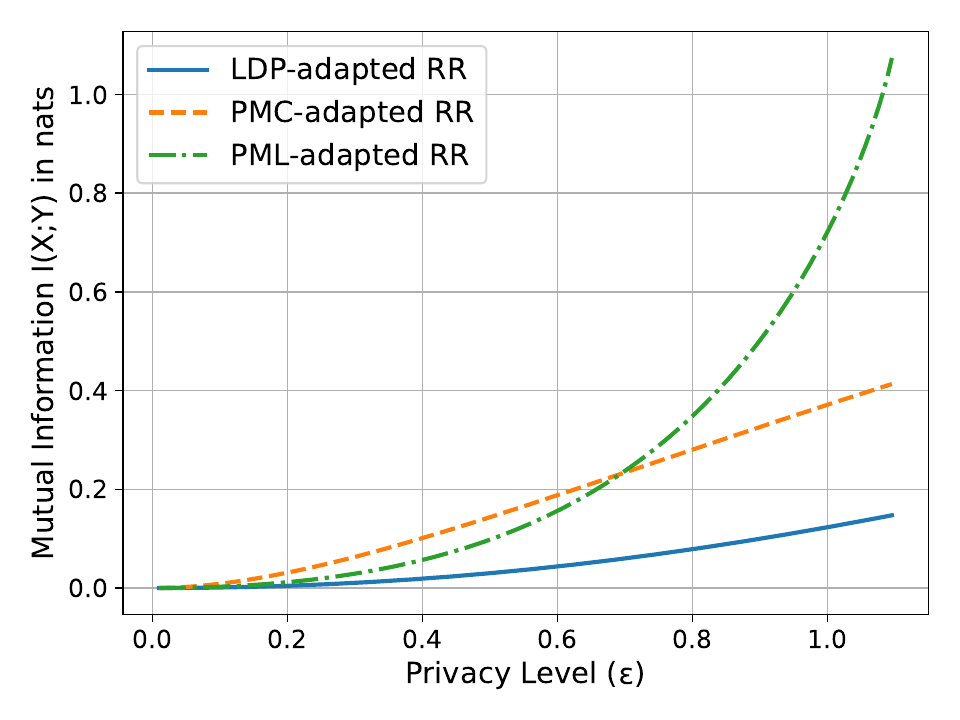}
        \caption{$n=3, P_X = (\frac{1}{3}, \frac{1}{3}, \frac{1}{3})$.}
    \end{subfigure}
    \hfill
    \begin{subfigure}[b]{0.45\textwidth}
        \centering
        \includegraphics[width=0.9\textwidth]{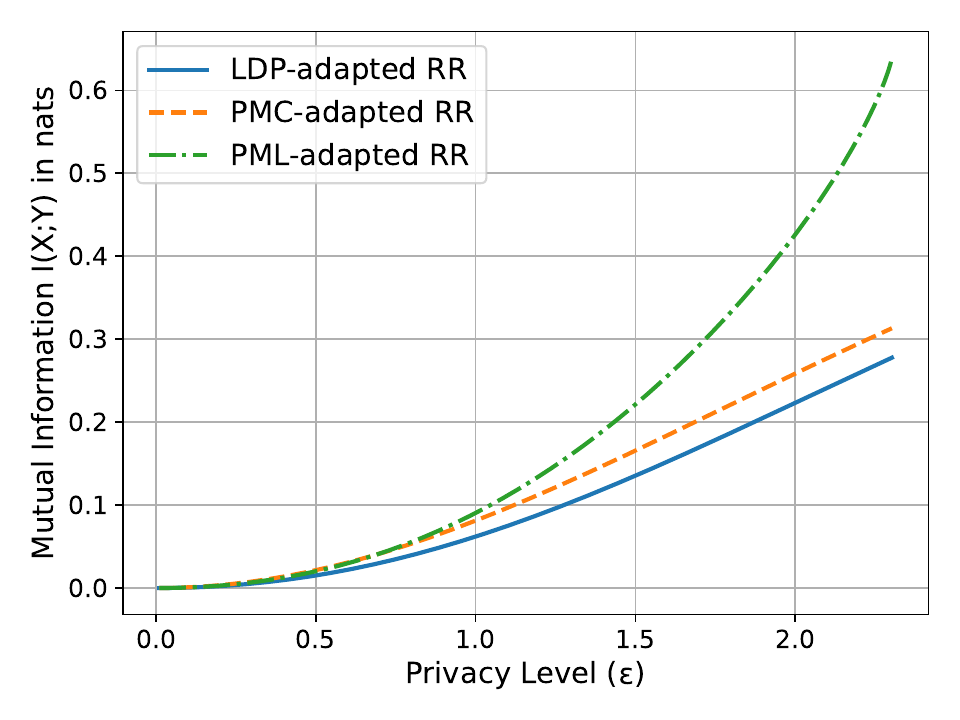}
        \caption{$n=3, P_X = (0.8, 0.1, 0.1)$.}
    \end{subfigure}
    \hfill
    \begin{subfigure}[b]{0.45\textwidth}
        \centering
        \includegraphics[width=0.9\textwidth]{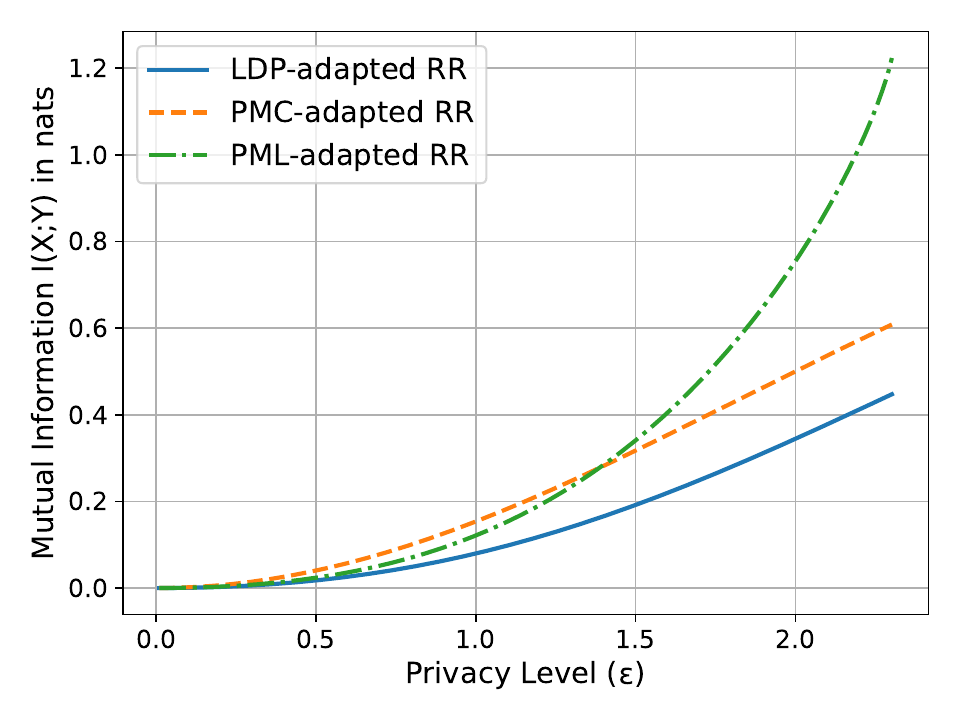}
        \caption{$n=5, P_X = (0.6, 0.1, 0.1, 0.1, 0.1)$.}
    \end{subfigure}
    \hfill
    \begin{subfigure}[b]{0.45\textwidth}
        \centering
        \includegraphics[width=0.9\textwidth]{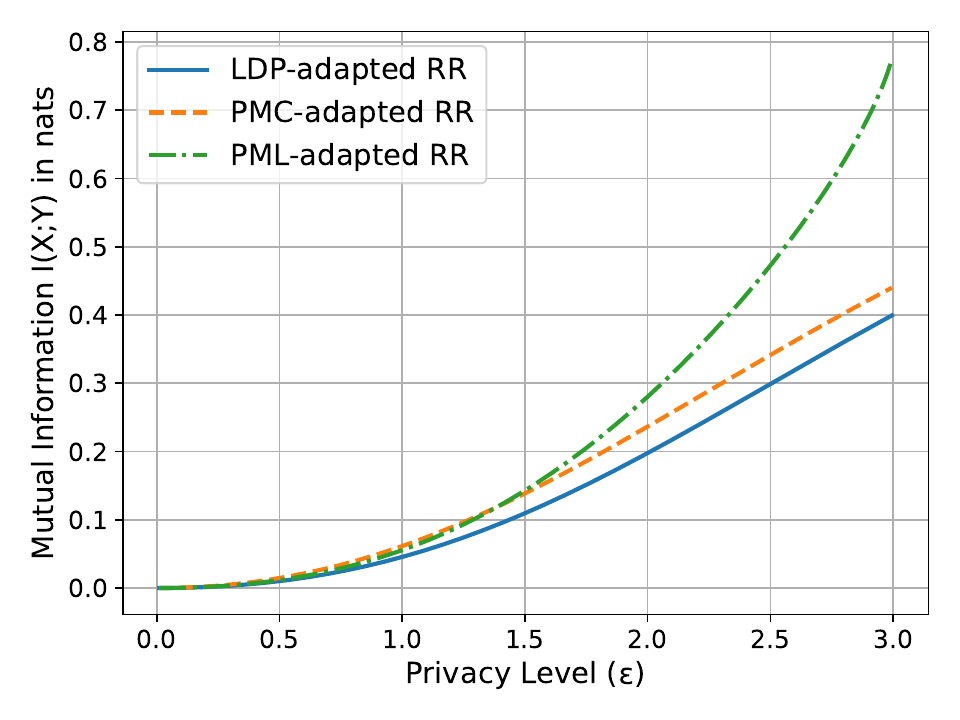}
        \caption{$n=5, P_X = (0.8, 0.05, 0.05, 0.05, 0.05)$.}
    \end{subfigure}
    \hfill
    \begin{subfigure}[b]{0.45\textwidth}
        \centering
        \includegraphics[width=0.9\textwidth]{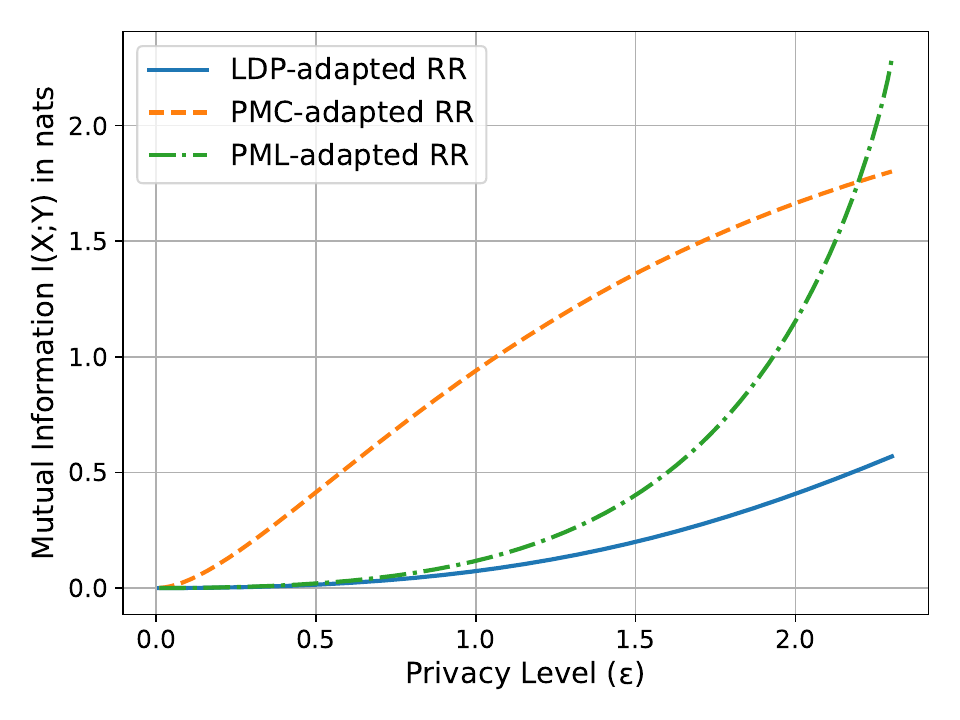}
        \caption{$n=10, P_X$ uniform.}
    \end{subfigure}
    \hfill
    \begin{subfigure}[b]{0.45\textwidth}
        \centering
        \includegraphics[width=0.9\textwidth]{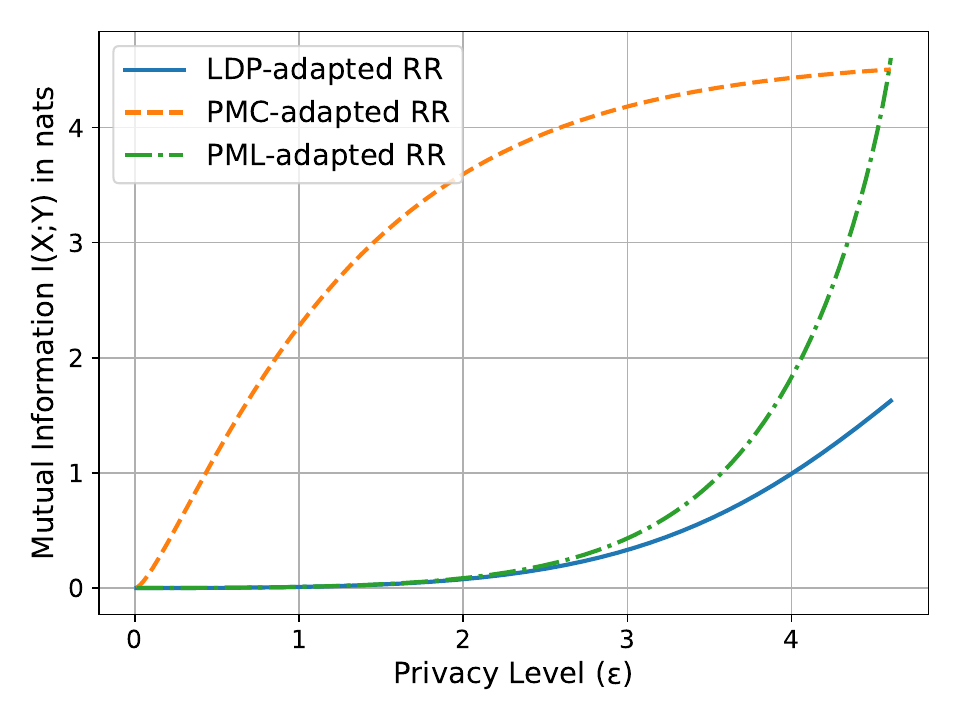}
        \caption{$n=100, P_X$ uniform.}
        \label{fig:five over x}
    \end{subfigure}
    \caption{Privacy–utility tradeoff of the generalized randomized response mechanism. LDP-adapted, PMC-adapted, and PML-adapted mechanisms are compared under different priors, with utility measured by mutual information.}
    \label{fig:rr_put_LDP_vs_PMC}
\end{figure*}

\begin{example}[Generalized Randomized Response~\cite{warnerRRoriginal, extremalmechanismLong}]
\label{ex:randomized_response}
A common mechanism used to achieve local differential privacy is the (generalized) randomized response mechanism. It is applied when the secret $X$ is a finite random variable and produces a privatized version of $X$ while retaining its original alphabet. 

Suppose $\mathcal X = \mathcal Y = [n]$. The randomized response mechanism with parameter $\varepsilon_r \geq 0$ can be expressed as follows:
\begin{equation*}
    P_{Y \mid X=i}(j) = \begin{cases}
    {\displaystyle \frac{e^{\varepsilon_r}}{n-1 + e^{\varepsilon_r}}} & \mathrm{if} \; i=j,\\[0.5em]
    {\displaystyle \frac{1}{n-1 + e^{\varepsilon_r}}} & \mathrm{if} \; i \neq j.
    \end{cases}
\end{equation*}  
It is easy to verify that the randomized response mechanism satisfies $\varepsilon_r$-LDP.

Now, we calculate the PMC of the randomized response mechanism. For all $j \in [n]$, we have $\min_{i \in [n]} P_{Y \mid X=i}(j) = \frac{1}{n-1 + e^{\varepsilon_r}}$ and 
\begin{equation*}
    P_Y(j) = \sum_{i \in [n]} P_{Y \mid X=i}(j) P_X(i) = \frac{1 + P_X(j) (e^{\varepsilon_r} -1)}{n-1 + e^{\varepsilon_r}}.
\end{equation*}
Thus, the PMC associated with outcome $j \in [n]$ of the mechanism is 
\begin{equation*}
    \Lambda(X \!\to j) = \log \max_{i \in [n]} \frac{P_Y(j)}{P_{Y \mid X=i}(j)} = \log \!\Big( 1 + P_X(j) (e^{\varepsilon_r} -1) \Big),
\end{equation*}
implying that the randomized response mechanism satisfies $\varepsilon$-PMC with 
\begin{equation}
\label{eq:rr_pmc}
    \varepsilon = \max_{j \in [n]} \Lambda(X \to j) = \log \; \Big( 1 + p_{\max} \, (e^{\varepsilon_r} -1) \Big),
\end{equation}
where $p_{\max} \coloneqq \max_{j \in [n]} P_X(j)$. Note that $\varepsilon \leq \varepsilon_r$ for all distributions $P_X$, and $\varepsilon \to \varepsilon_r$ as $p_{\max} \to 1$.  

Next, we compare the utility of the generalized randomized response mechanism when its parameter is tuned to satisfy a target $\varepsilon$-PMC, $\varepsilon$-PML, or $\varepsilon$-LDP privacy guarantee. In~\cite{grosse2024extremal}, we showed that the randomized response mechanism with parameter $\varepsilon_r$ satisfies $\log \left( \frac{e^{\varepsilon_r}}{p_{\min} \,(e^{\varepsilon_r} - 1) + 1} \right)$-PML, where $p_{\min} = \min_{j \in [n]} P_X(j)$. Recall that PML is always upper bounded by $\log \frac{1}{p_{\min}}$, whereas both PMC and LDP values can grow arbitrarily large, even for secrets with finite alphabets (see Remark~\ref{rem:pmc_unbounded}). As a result, meaningful comparisons between the three notions can only be made within the range $\varepsilon \in \left[0, \log \frac{1}{p_{\min}}\right]$.

Figure~\ref{fig:rr_put_LDP_vs_PMC} presents the privacy-utility tradeoff curves for the randomized response mechanism across various priors and alphabet sizes, with utility measured via mutual information. In all cases, mechanisms adapted to PMC or PML achieve higher utility than those adapted to LDP, reflecting the advantage of context-aware guarantees over context-free ones. Between PMC and PML, the comparison depends on the prior and the privacy level. When $\varepsilon \leq \log \left(\frac{1 - p_{\max}}{p_{\min}}\right)$, the PMC-adapted mechanism has a larger randomized response parameter ($\varepsilon_r$) than the PML-adapted one, and therefore yields higher utility. This effect is especially prominent for uniform priors over large alphabets. Conversely, when the alphabet is small and the prior is highly skewed, the PML-adapted mechanism offers higher utility.
\end{example}

\begin{example}[PML-extremal mechanism in the high-privacy regime~\cite{grosse2024extremal}]
\label{ex:extreme}
Suppose $X$ is a finite random variable with alphabet $\cX = [n]$. Recall from Section~\ref{ssec:background_pml} that PML-extremal mechanisms constitute a class of optimal mechanisms maximizing sub-convex utility functions~\cite[Def. 5]{grosse2024extremal}. Let $0 \leq \varepsilon_u < \log \, \frac{1}{1 - p_{\mathrm{min}}}$. This range of values for $\varepsilon_u$ corresponds to the \emph{high-privacy regime}. Extremal mechanisms in the high-privacy regime can be expressed as 
\begin{equation}
    P_{Y \mid X=i}(j) = \begin{cases}
        1 - e^{\varepsilon_u}(1-P_X(j)) & \text{if}\; i=j, \\
        e^{\varepsilon_u}P_X(j) & \text{if}\; i \neq j,
    \end{cases}
\end{equation}
where $i,j \in [n]$. It can be verified that these mechanisms satisfy $\varepsilon_u$-PML. 

Now, we calculate the PMC of the extremal mechanisms. We have $\min_{i \in [n]} P_{Y \mid X=i}(j) = 1 - e^{\varepsilon_u}(1-P_X(j))$ and it is easy to check that $P_Y(j) = P_X(j)$ for all $j \in [n]$. Thus, the PMC associated with outcome $j \in [n]$ of the mechanism is 
\begin{align*}
    \Lambda(X \to j) &= \log \; \max_{i \in [n]} \; \frac{P_Y(j)}{P_{Y \mid X=i}(j)}\\
    &= \log \, \frac{P_X(j)}{1 - e^{\varepsilon_u} \Big(1-P_X(j) \Big)}. 
\end{align*}
Since the mapping $p \mapsto \log \frac{p}{1 - e^{\varepsilon_u}(1- p)}$ is decreasing in $p$, the extremal mechanism satisfies $\varepsilon$-PMC with 
\begin{equation}
    \varepsilon = \max_{j \in [n]} \Lambda(X \to j) = \log \; \frac{p_{\mathrm{min}}}{1 - e^{\varepsilon_u}\big(1- p_{\mathrm{min}} \big)}. 
\end{equation} 
\end{example}


Next, we examine the Laplace mechanism~\cite{DPoriginalpaper} which is a central component of many (central) differential privacy algorithms. 

\begin{example}[Laplace mechanism~\cite{DPoriginalpaper,dwork2014algorithmic}]
A central problem in statistics is to estimate the expected value of a random variable based on multiple independently drawn realizations. This problem is also extensively explored under the framework of differential privacy. In particular, when aiming to satisfy pure differential privacy, the Laplace mechanism is often employed which adds noise with Laplace distribution to the sample mean. Importantly, to release the sample mean via the Laplace mechanism with DP guarantees the data domain must be bounded, e.g., an interval. This is because the global (and local) sensitivity of the sample mean is infinite when the data domain is unbounded, e.g., $\bR$~\cite{DPoriginalpaper}.  

Consider an i.i.d database $X^n = (X_1, \ldots, X_n)$, where each $X_i$ is a bounded random variable on the interval $[c,d]$. Suppose $\bE[X_i] = \mu \in [c,d]$, and define $Z_1 \coloneqq X_1 - \mu$. Let $K(t) \coloneqq \log \bE \left[e^{t Z_1} \right]$  denote the cumulant generating function of $Z_1$, where $t \in \bR$. Suppose our objective is to safely disclose the sample mean of $X^n$ by perturbing it with Laplace noise. Let $\mathsf{Lap}(m, b)$ denote Laplace distribution with mean $m \in \bR$ and scale parameter $b >0$. The Laplace mechanism releases an outcome with the conditional distribution $Y \mid X^n = x^n \sim \mathsf{Lap} \left( \frac{\sum_{i=1}^n x_i}{n}, b \right)$. 

Below, we evaluate the PMC from each data point to the released outcome. Due to the symmetry of the problem, the Laplace mechanism leaks the same amount of information about all data points, and for notational convenience, we calculate $\Lambda(X_n \to y)$ with $y \in \bR$.\footnote{While calculating $\Lambda(X_n \to y)$, the randomness of $X^{n-1} = (X_1, \ldots, X_{n-1})$ contributes to the randomness of the mechanism $P_{Y \mid X_n}$. Specifically, we have 
\vspace{-0.5em}
\begin{equation*}
    f_{Y \mid X_n}(y \mid x_n) = \int f_{Y \mid X^n}(y \mid x_1, \ldots, x_n) \; P_{X^{n-1}}(d x_1, \ldots, d x_{n-1}).
\end{equation*}
Furthermore, a bound $\sup_y \Lambda(X_n \to y) \leq \varepsilon$ implies that the mechanism $P_{Y \mid X_n}$ satisfies $\varepsilon$-PMC under the prior $P_{X_n}$.} Our analysis in this example complements a similar analysis performed with the Laplace mechanism and PML in~\cite[Prop. 4.5]{saeidian2023inferential}.   

The largest privacy cost across all possible released outcomes is 
\begin{align}
\label{eq:laplace}
    &\sup_{y \in \bR} \Lambda(X_n \to y) = \log \; \sup_{y \in \bR} \; \frac{f_{Y}(y)}{\inf\limits_{x_n \in [c,d]} f_{Y \mid X_n=x_n}(y)}\\
    &= \log \sup_{y \in \bR} \frac{\displaystyle \frac{1}{2b} \; \bE_{X^n}\left[\exp \left(-\frac{ \lvert y - \frac{\sum_{i=1}^n X_i}{n}\rvert}{b} \right)\right]}{\displaystyle \inf_{x_n \in [c,d]} \frac{1}{2b} \bE_{X^{n-1}}\!\!\left[\exp \!\!\left(\!-\frac{\lvert y - \frac{x_n}{n} - \frac{\sum_{i=1}^{n-1} X_i}{n}\rvert}{b} \!\right)\!\right]}. 
\end{align}
First, we argue that to find the largest PMC across the $y$'s, it suffices to consider $y \geq d$ and $y \leq c$. To see why, note that in the numerator, using the fact that $-\abs{a} \leq \min \{-a,a\}$ for all $a \in \bR$ we have 
\begin{align*}
    &\bE_{X^n}\!\Bigg[\exp \!\Bigg(\!-\frac{ \Big\lvert y - \frac{\sum_{i=1}^n X_i}{n}\Big\rvert}{b} \Bigg)\!\Bigg] \!\leq\\
    &\!\min \!\left\{\bE\Bigg[\!\exp \!\Bigg(\!-\frac{y - \frac{\sum\limits_{i=1}^n X_i}{n}}{b} \!\Bigg)\!\Bigg], \bE\Bigg[\!\exp \!\Bigg(\frac{y - \frac{\sum\limits_{i=1}^n X_i}{n}}{b} \!\Bigg)\!\Bigg]\right\},
\end{align*}
for all $y \in \bR$. Moreover, in the denominator, the mapping
\begin{equation*}
    y \mapsto \bE_{X^{n-1}}\Bigg[\exp \Bigg(-\frac{\lvert y - \frac{x_n}{n} - \frac{\sum_{i=1}^{n-1} X_i}{n}\rvert}{b} \Bigg)\Bigg],
\end{equation*}    
is increasing in  $(-\infty, c]$ and decreasing in $[d, \infty)$ for all $x_n \in [c,d]$. 

Suppose $y \geq d$. The expectation in the numerator of~\eqref{eq:laplace} is
\begin{align*}
    &\bE\!\left[\exp \!\left(-\frac{y - \frac{\sum_{i=1}^n X_i}{n}}{b} \right)\right] \!= \!\exp \!\left(-\frac{y}{b}\right) \bE\!\left[\exp \!\left(\frac{\sum_{i=1}^n X_i}{n b} \right)\right]\\
    &= \exp \left(-\frac{y}{b}\right) \; \prod_{i=1}^n \bE \left[\exp \left(\frac{X_i}{nb} \right)\right]\\
    &= \exp \left(-\frac{y}{b}\right) \; \left( \bE \left[\exp \left(\frac{X_1}{nb} \right)\right] \right)^n,
\end{align*}
and the expectation in the denominator is 
\begin{align*}
    &\bE_{X^{n-1}}\left[\exp \left(-\frac{\displaystyle y - \frac{x_n}{n} - \frac{\sum_{i=1}^{n-1} X_i}{n}}{b} \right)\right]\\
    &=\exp \left(-\frac{y}{b}\right) \cdot \exp \left(\frac{x_n}{n b}\right) \; \bE_{X^{n-1}}\left[\exp \left(\frac{\sum_{i=1}^{n-1} X_i}{n b} \right)\right]\\
    &= \exp \left(-\frac{y}{b}\right) \cdot \exp \left(\frac{x_n}{n b}\right) \; \left(\bE\left[\exp \left(\frac{X_1}{nb} \right)\right] \right)^{n-1}.
\end{align*}
Thus, for $y \geq d$ the PMC is obtained as\footnote{We assume that $\bP \big\{X_n \in [c,c+\delta)\big\} > 0$ for arbitrarily small $\delta >0$ so that $\essinf_{X_n} \exp({\frac{X_n}{nb})} = \exp(\frac{c}{nb})$. Otherwise, the derived expression is an upper bound on the PMC. Similar assumptions are made about intervals of the form $(d-\delta, d]$.} 
\begin{align}
    \Lambda(X_n \to y) &= \log \frac{f_{Y}(y)}{\inf\limits_{x_n \in [c,d]} f_{Y \mid X_n=x_n}(y)}\\
    &= \log \left( \exp \left(-\frac{c}{nb} \right) \cdot \bE\left[\exp \left(\frac{X_1}{nb} \right)\right] \right)\\
    &= - \frac{c}{nb} + \log \bE\left[\exp \left(\frac{X_1}{nb} \right)\right]\\
    &= \frac{\mu - c}{nb} + K(\frac{1}{nb}). \label{eq:laplace_1}
\end{align}
Similarly, when $y \leq c$ the PMC is 
\begin{align}
\label{eq:laplace_2}
    \Lambda(X_n \to y) &= \frac{d - \mu}{nb} + K(-\frac{1}{nb}).
\end{align} 
Combining~\eqref{eq:laplace_1} and \eqref{eq:laplace_2}, we conclude that
\begin{equation*}
    \sup_{y \in \bR} \Lambda(X_n \to y) = \max \!\Bigg\{\!\frac{\mu - c}{nb} + K(\frac{1}{nb}), \frac{d - \mu}{nb} + K(-\frac{1}{nb}) \!\Bigg\}. 
\end{equation*}
In the absence of more assumptions about the distribution of $X_i$'s, we may use $c - \mu \leq Z_1 \leq d - \mu$ to bound $K(\frac{1}{nb})$ and $K(-\frac{1}{nb})$ by
\begin{gather*}
   K(\frac{1}{nb}) = \log \bE \left[e^{ \frac{Z_1}{nb}} \right] \leq \frac{d-\mu}{nb},\\[0.5em]
   K(-\frac{1}{nb}) = \log \bE \left[e^{ -\frac{Z_1}{nb}} \right] \leq \frac{\mu-c}{nb}.
\end{gather*}
This yields
\begin{equation*}
    \sup_{y \in \bR} \; \Lambda(X_n \to y) \leq \frac{d-c}{nb}. 
\end{equation*}
Note that the right-hand side of the above bound corresponds to the differential privacy parameter of the Laplace mechanism for a query with global sensitivity $\frac{d-c}{n}$~\cite{DPoriginalpaper}. 

However, with additional assumptions about the distribution of $X_i$'s, we can derive tighter bounds on the PMC. For example, if $X_i \sim \sfU[c,d]$ then $\mu = \frac{c+d}{2}$ and we have
\begin{align*}
    &K(\frac{1}{nb}) =  K(-\frac{1}{nb})= \log \int_{\frac{c-d}{2}}^{\frac{d-c}{2}} \frac{1}{d-c} \exp (\frac{x}{nb}) \, dx \\
    &= \log \left( \frac{nb}{d-c} \left [\exp(\frac{d-c}{2nb}) - \exp(\frac{c-d}{2nb}) \right]\right)\\
    &= \log \left( \frac{nb}{d-c} \cdot \exp(\frac{c-d}{2nb}) \left [\exp(\frac{d-c}{nb}) - 1 \right]\right).
\end{align*}

In this case, the Laplace mechanism results in the PMC 
\begin{equation*}
    \sup_{y \in \bR} \; \Lambda(X_n \to y) = \log \left( \frac{nb}{d-c} \left [\exp(\frac{d-c}{nb}) - 1 \right]\right), 
\end{equation*}
which is smaller than $\frac{d-c}{nb}$.
\end{example}

\begin{example}[Gaussian noise] 
\label{ex:gaussian_noise}
Let $X$ be a zero mean and bounded random variable with $\abs{X} \leq A$, and let $K(t) \coloneqq \log \bE \left[e^{t X} \right]$ with $t \in \bR$ denote the cumulant generating function of $X$. Suppose our goal is to safely release the realization of $X$ by adding Gaussian noise to it. Thus, we release an outcome of $Y$ with conditional distribution $Y \mid X=x \sim \sfN (x, \sigma^2)$. 

To calculate the PMC $\Lambda(X \to y)$, first suppose $y \geq  0$. We write 
\begin{align*}
    \Lambda(X \to y) &= \log \; \frac{f_{Y}(y)}{\inf\limits_{x \in [-A,A]} f_{Y \mid X=x}(y)}\\
    &= \log \; \frac{\bE_X \left[\exp \left(- \frac{(y-X)^2}{2 \sigma^2}\right) \right]}{\exp \left(- \frac{(y+A)^2}{2 \sigma^2}\right)}\\
    &= \log \; \bE_X \left[\exp \left(\frac{(y+A)^2-(y-X)^2}{2 \sigma^2}\right) \right]\\
    &= \log \; \bE_X \left[\exp \left(\frac{2y(A+X) + A^2 - X^2}{2 \sigma^2}\right) \right].
\end{align*}
Since $0 \leq X^2 \leq A^2$, we obtain the following upper and lower bounds on $\Lambda(X \to y)$:\footnote{When $y$ is small, using $(y-X)^2 \geq 0$ yields a tighter upper bound on the PMC. Here, we are mostly interested in large $y$ since our goal is to characterize the tail of $\Lambda(X \to Y)$.}
\begin{multline*}
    \log \; \bE_X \left[\exp \left(\frac{2y(A+X)}{2 \sigma^2}\right) \right] \leq \Lambda(X \to y)\\
    \leq \log \; \bE_X \left[\exp \left(\frac{2y(A+X) + A^2}{2 \sigma^2}\right) \right],
\end{multline*}
which can be expressed as
\begin{align*}
    \frac{Ay}{\sigma^2} + K(\frac{y}{\sigma^2}) \leq \Lambda(X \to y) \leq \frac{A(A+2y)}{2\sigma^2} + K(\frac{y}{\sigma^2}).
\end{align*}
By Jensen's inequality and using the fact that $\bE[X] =0$, we have $K(\frac{y}{\sigma^2}) \geq 0$. Furthermore, since $X \leq A$, we have the upper bound 
\begin{equation*}
    K(\frac{y}{\sigma^2}) = \log \bE \left[\exp \left(\frac{yX}{\sigma^2} \right) \right] \leq \frac{yA}{\sigma^2}. 
\end{equation*}
Thus, we get 
\begin{equation}
\label{eq:gaussian_pos}
    \frac{Ay}{\sigma^2} \leq \Lambda(X \to y) \leq \frac{A(A+4y)}{2\sigma^2}.
\end{equation}
Similarly, when $y \leq 0$ we have 
\begin{equation*}
    \Lambda(X \to y) = \log \; \bE_X \left[\exp \left(\frac{-2y(A-X) + A^2 - X^2}{2 \sigma^2}\right) \right].
\end{equation*}
This yields the bounds
\begin{align*}
    -\frac{Ay}{\sigma^2} + K(\frac{y}{\sigma^2}) \leq \Lambda(X \to y) \leq \frac{A(A-2y)}{2\sigma^2} + K(\frac{y}{\sigma^2}),
\end{align*}
and 
\begin{equation}
\label{eq:gaussian_neg}
    -\frac{Ay}{\sigma^2} \leq \Lambda(X \to y) \leq \frac{A(A-4y)}{2\sigma^2}.
\end{equation}
Combining~\eqref{eq:gaussian_pos} and \eqref{eq:gaussian_neg} we get the PMC bounds 
\begin{equation}
\label{eq:gaussian_bounds}
    \frac{A \abs{y}}{\sigma^2} \leq \Lambda(X \to y) \leq \frac{A(A+\abs{4y})}{2\sigma^2}, \quad y \in \bR. 
\end{equation}
The lower bound in \eqref{eq:gaussian_bounds} implies that $\Lambda(X \to Y)$ is unbounded. Thus, the Gaussian noise mechanism does not satisfy $\varepsilon$-PMC for any finite value of $\varepsilon$. Nevertheless, we can use the upper bound in \eqref{eq:gaussian_bounds} to characterize the tail of $\Lambda(X \to Y)$. 

Let $\beta \geq 0$ and suppose $A^2 = r \sigma^2$ where $r \geq 0$. Since $X$ is sub-Gaussian with variance proxy $A^2$ and the noise has Gaussian distribution with variance $\sigma^2$, then $Y$ is sub-Gaussian with variance proxy $A^2 + \sigma^2 = (r+1) \sigma^2$. Hence, we can write 
\begin{align}
    \bP \Big\{\Lambda(X \to Y) \geq \beta + &\frac{A^2}{2 \sigma^2} \Big\} \\
    &\leq \bP \left\{\frac{A(A+4\abs{Y})}{2\sigma^2} \geq \beta + \frac{A^2}{2 \sigma^2} \right\}\\
    &= \bP \left\{\abs{Y} \geq \frac{\sigma^2 \beta}{2 A} \right\}\\
    &\leq 2 \exp \left(-\frac{\beta^2 \sigma^4}{8 A^2 (A^2 + \sigma^2)}\right) \label{subeq:gaussian_chernoff}\\
    &= 2 \exp \left(-\frac{\beta^2}{8 (r^2+r)}\right),
\end{align}
where \eqref{subeq:gaussian_chernoff} follows from the Chernoff bound on the tail of a sub-Gaussian random variable. 
Thus, $\Lambda(X \to Y)$ is dominated by a Gaussian random variable with mean $\frac{r}{2}$ and variance $4r(r + 1)$. 
\end{example} 

\section{PMC, PML, and Connections to Other Privacy Measures}
\label{sec:bounds}
In general, the upper and lower bounds on the information density are not independent quantities. These bounds influence each other so that imposing a lower bound on the information density may imply a certain upper bound, and vice versa. In the context of privacy, this naturally raises the question of whether a PMC guarantee inherently provides a certain PML guarantee or whether a PML guarantee implies a certain PMC guarantee. 

In this section, we explore how PMC and PML guarantees interact with each other and with other privacy definitions, namely, ALIP, LIP, and LDP. First, in Section~\ref{ssec:connections_finite}, we assume that $X$ is a finite random variable and discuss how various privacy guarantees interrelate. We also highlight the implications of our results for designing mechanisms with optimal utility. Then, in Section~\ref{ssec:connections_general}, we extend our analysis to secrets on standard Borel spaces.

\subsection{Secrets with Finite Alphabets}
\label{ssec:connections_finite} 
Suppose $X \sim P_X$ is a finite random variable, and recall that $p_\mathrm{min} = \min_{x \in \cX} P_X(x)$. We begin by presenting two results that illustrate how LDP and PMC relate. The first parallels Proposition~\ref{prop:ldp_pml_bound} by showing how LDP constrains the information density through a bound on PMC. The second parallels Theorem~\ref{thm:ldp_pml_sup} and establishes an equivalence between LDP and PMC over all full-support prior distributions. See Appendices~\ref{sec:proof_ldp_pmc} and \ref{sec:proof_ldp_pmc_sup} for proofs.

\begin{proposition}
\label{prop:ldp_pmc}
Suppose $X$ is a finite random variable distributed according to $P_X$. If the mechanism $P_{Y \mid X}$ satisfies $\varepsilon$-LDP, then it satisfies $\varepsilon'$-PMC with
\begin{equation*}
    \varepsilon' = \log \; \Big( e^\varepsilon - p_\mathrm{min} (e^\varepsilon - 1)\Big). 
\end{equation*}
\end{proposition}
Note that similarly to Proposition~\ref{prop:ldp_pml_bound}, $\varepsilon' \to \varepsilon$ as $p_\mathrm{min} \to 0$.

The following result expresses LDP as a PMC constraint and provides a novel interpretation of LDP in terms of risk-averse adversaries. 
\begin{theorem}    
\label{thm:ldp_pmc_sup}
Let $\cP_\cX$ denote the set of distributions with full support on $\cX$. For any $\varepsilon \geq 0$, the mechanism $P_{Y \mid X}$ satisfies $\varepsilon$-LDP, if and only if it satisfies $\varepsilon$-PMC for all $P_X \in \cP_\cX$.
\end{theorem}

We now present the main result of this section. This result demonstrates how PMC and PML guarantees interact with each other, thereby allowing us to relate both measures to other privacy definitions. 
\begin{theorem}
\label{thm:liftasym}
Suppose $X$ is a finite random variable distributed according to $P_X$.
\begin{enumerate}
    \item Let $0 \leq \varepsilon_u < \log \frac{1}{1-p_{\min}}$. If the privacy mechanism $P_{Y \mid X}$ satisfies $\varepsilon_u$-PML, then it also satisfies $ \varepsilon^*_l (\varepsilon_u)$-PMC, where 
    \begin{equation}
    \label{eq:eps_l}
        \varepsilon^*_l (\varepsilon_u) \coloneqq \log \,\frac{p_{\min}}{1-e^{\varepsilon_u}(1-p_{\min})}.
    \end{equation} 

    \item Let $\varepsilon_l \geq 0$. If the privacy mechanism $P_{Y \mid X}$ satisfies $\varepsilon_l$-PMC, then it also satisfies $\varepsilon^*_u (\varepsilon_l)$-PML, where 
    \begin{equation}
    \label{eq:eps_u}
    \varepsilon^*_u (\varepsilon_l) \coloneqq \log \,\frac{1-e^{-\varepsilon_l}(1-p_{\min})}{p_{\min}}.
    \end{equation}
\end{enumerate}
\end{theorem}
Theorem~\ref{thm:liftasym} is proved in Appendix~\ref{sec:proof_liftasym}.

For notational convenience, we let $\varepsilon^*_u \coloneqq \varepsilon^*_u (\varepsilon_l)$ and $\varepsilon^*_l \coloneqq \varepsilon^*_l (\varepsilon_u)$ for fixed values of $\varepsilon_l$ and $\varepsilon_u$. 
\begin{remark}
The condition $\varepsilon_u < \log \frac{1}{1-p_{\min}}$ restricts us to the high-privacy regime of PML and is necessary for $\varepsilon_l^*$ to be well-defined. This is because a mechanism $P_{Y \mid X}$ satisfying $\varepsilon_u$-PML with $\varepsilon_u \geq \log \frac{1}{1-p_{\min}}$ may assign $P_{Y \mid X=x}(y) = 0$ for some $(x,y)$ pair with positive probability. Such a mechanism does not satisfy $\varepsilon_l$-PMC for any finite value of $\varepsilon_l$. This is also why an $\varepsilon_u$-PML guarantee yields useful LIP, ALIP and LDP guarantees only if $\varepsilon_u < \log \frac{1}{1-p_{\min}}$ (see also Corollary~\ref{cor:relations}).     
\end{remark}

\begin{remark}
The translation from PML to PMC in Theorem~\ref{thm:liftasym} in general cannot be improved. In particular, as we saw in Example~\ref{ex:extreme}, the high-privacy extremal mechanism satisfying $\varepsilon_u$-PML also satisfies $\varepsilon_l^*$-PMC. 
\end{remark}

\begin{figure}[!t]
    \begin{subfigure}[b]{.49\textwidth}
        \includegraphics[scale=0.45]{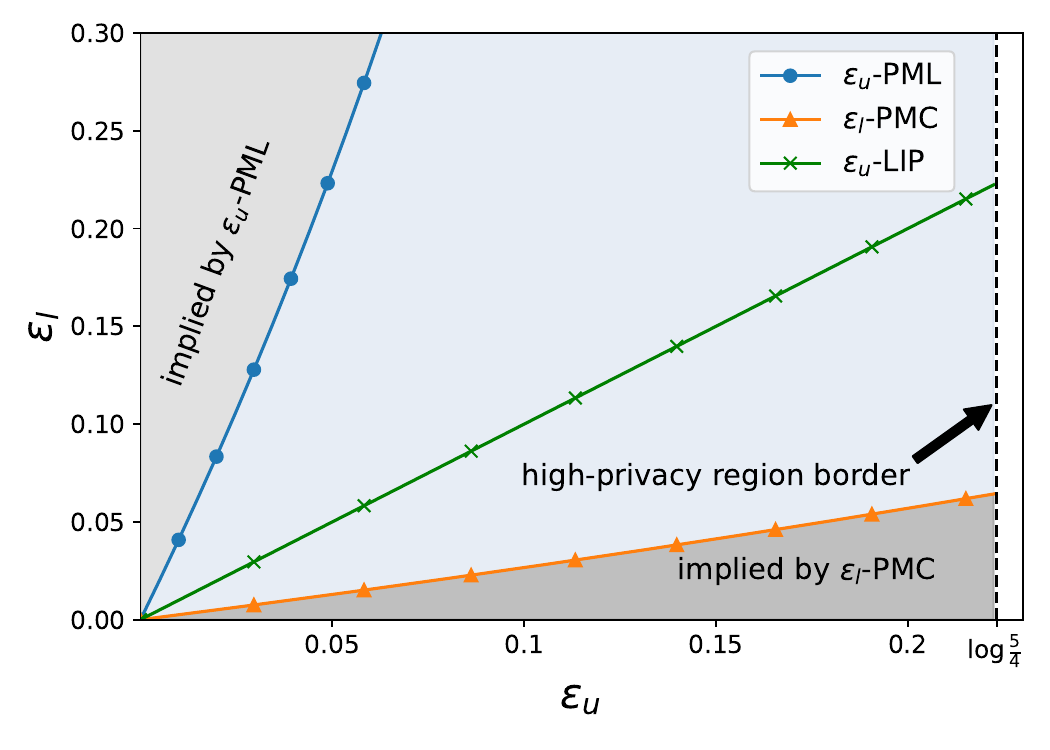}
        \caption{$P_X = (0.3,0.3,0.2,0.2)$}
        \label{fig:eps_u-vs-eps_l:subfig:N4}
    \end{subfigure}
    \begin{subfigure}[b]{.49\textwidth}
        \includegraphics[scale=0.45]{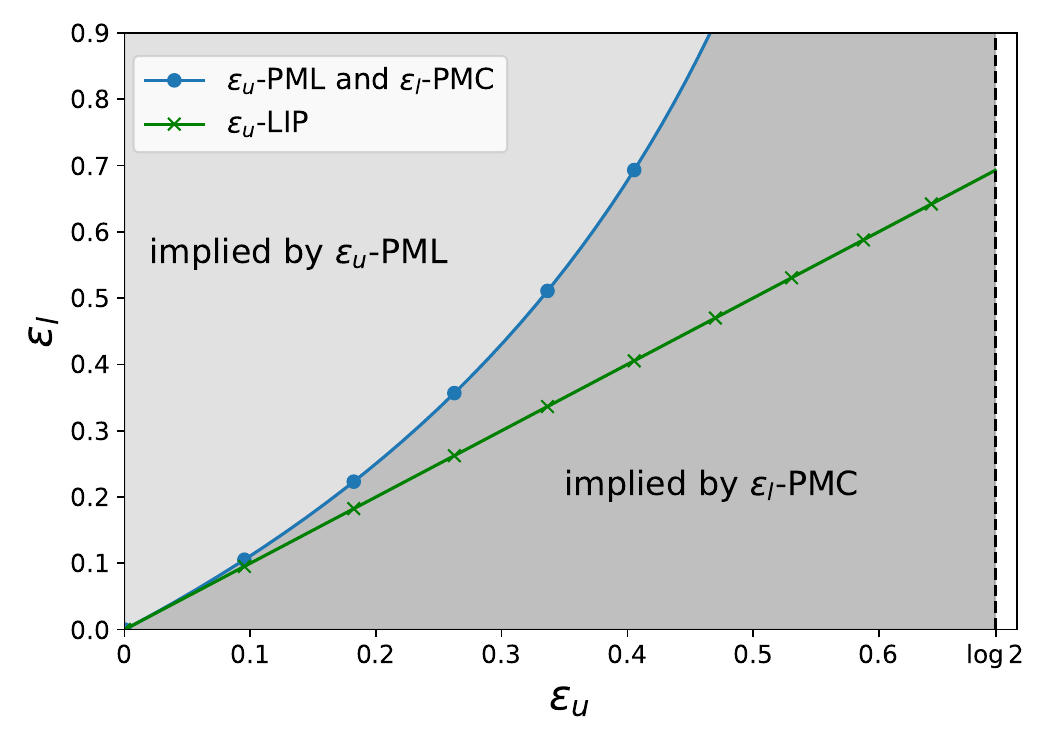}
        \caption{$P_X = (0.5,0.5)$}
        \label{fig:eps_u-vs-eps_l:subfig:N2}
    \end{subfigure}
\centering
\caption{Relationship between the upper and lower bounds on the information density. The blue curve corresponds to $\varepsilon^*_l (\varepsilon_u)$ and the orange curve corresponds to $\varepsilon^*_u (\varepsilon_l)$. When $P_X$ is binary and uniform, the two curves coincide.}  
\label{fig:eps_u-vs-eps_l}
\end{figure}

Figure \ref{fig:eps_u-vs-eps_l} illustrates the bounds presented in Theorem~\ref{thm:liftasym}. Consider a pair $(\varepsilon_u,\varepsilon_l) \geq (0,0)$ with $\varepsilon_u < \log \frac{1}{1-p_\mathrm{min}}$. Theorem~\ref{thm:liftasym} states that if $\varepsilon_l > \varepsilon_l^*$, then $\varepsilon_u$-PML is a strictly stronger privacy guarantee compared to $\varepsilon_l$-PMC. Such pairs constitute the gray area above the blue curve in Figure~\ref{fig:eps_u-vs-eps_l:subfig:N4}. Similarly, if $\varepsilon_u > \varepsilon_u^*$, then $\varepsilon_l$-PMC is a strictly stronger privacy guarantee compared to $\varepsilon_u$-PML. These pairs correspond to the gray area below the orange curve in Figure~\ref{fig:eps_u-vs-eps_l:subfig:N4}. In contrast, the blue region in Figure~\ref{fig:eps_u-vs-eps_l:subfig:N4} represents $(\varepsilon_u,\varepsilon_l)$ pairs that satisfy $\varepsilon_u^* \geq \varepsilon_u$ and $\varepsilon_l^* \geq \varepsilon_l$. In this region, neither the $\varepsilon_u$-PML nor the $\varepsilon_l$-PMC privacy guarantee is stronger than the other. Interestingly, when $X$ is an equiprobable Bernoulli random variable, the gap between the $\varepsilon_u$-PML and $\varepsilon_l$-PMC curves (i.e., the blue region) disappears, as illustrated in Figure~\ref{fig:eps_u-vs-eps_l:subfig:N2}. We formalize this observation in Corollary~\ref{cor:binaryuniform}.

Let us now discuss some of the implications of Theorem~\ref{thm:liftasym}. The primary consequence of this theorem is that both PML and PMC privacy guarantees translate into ALIP, LIP, and LDP guarantees. The ALIP and LIP guarantees simply follow from their definitions, and the LDP guarantees follow from \cite[Prop. 1]{zarrabian2023lift}, where it is shown that $(\varepsilon_l,\varepsilon_u)$-ALIP implies $(\varepsilon_l + \varepsilon_u)$-LDP. 

\begin{corollary}
\label{cor:relations}
Suppose $X$ is a finite random variable distributed according to $P_X$. 
\begin{enumerate}
    \item Let $0 \leq \varepsilon_u < \log \frac{1}{1-p_{\min}}$. If the mechanism $P_{Y \mid X}$ satisfies $\varepsilon_u$-PML, then it also satisfies
    \begin{equation*}
      (\varepsilon_l^*,\varepsilon_u)\text{-ALIP}, \quad \max\{\varepsilon^*_l, \varepsilon_u\}\text{-LIP}, \quad (\varepsilon_l^* + \varepsilon_u)\text{-LDP}.
    \end{equation*}

    \item Let $\varepsilon_l \geq 0$. If the mechanism $P_{Y \mid X}$ satisfies $\varepsilon_l$-PMC, then it also satisfies
    \begin{equation*}
      (\varepsilon_l,\varepsilon_u^*)\text{-ALIP}, \quad \max\{\varepsilon_l, \varepsilon_u^*\}\text{-LIP}, \quad (\varepsilon_l + \varepsilon_u^*)\text{-LDP}.
    \end{equation*}
\end{enumerate}
\end{corollary}

Corollary~\ref{cor:relations} describes implications in one direction: from PML or PMC to ALIP, LIP, and LDP.  Interestingly, when $X$ is an equiprobable Bernoulli random variable the reverse implications also hold. This is because, in this case, we have $(\varepsilon_l^* \circ \varepsilon_u^*) (\varepsilon_l) = \varepsilon_l$ and $(\varepsilon_u^* \circ \varepsilon_l^*) (\varepsilon_u) = \varepsilon_u$. In the following corollary, the LIP statement arises from the fact that when $X$ is an equiprobable Bernoulli random variable, then $\varepsilon_l^* (\varepsilon_u) \geq \varepsilon_u$. The LDP statement follows from Corollary~\ref{cor:relations} and Proposition~\ref{prop:ldp_pml_bound}.   

\begin{corollary}
\label{cor:binaryuniform}
Suppose $\mathcal X = \{0,1\}$ and $P_X(0) = P_X(1) = \frac{1}{2}$. Given $0 \leq \varepsilon_u < \log 2$, a privacy mechanism satisfies $\varepsilon_u$-PML if and only if it satisfies 
\begin{enumerate}
    \item $\varepsilon_l^*$-PMC, or
    \item $(\varepsilon_l^*,\varepsilon_u)$-ALIP, or
    \item $\varepsilon_l^*$-LIP, or
    \item $\varepsilon_l^* + \varepsilon_u$-LDP.
\end{enumerate}
\end{corollary}
Alternatively, we could have fixed $\varepsilon_l \geq 0$ and stated the result for $\varepsilon_l$-PMC and $\varepsilon_u^*$-PML. 

The equivalence between PML and LDP expressed in Corollary~\ref{cor:binaryuniform} also explains why the binary PML-extremal mechanism (Example~\ref{ex:extreme}) for the uniform prior distribution coincides with the (binary) randomized response mechanism (Example~\ref{ex:randomized_response}).\footnote{More precisely, given $\varepsilon \geq 0$, the binary randomized response mechanism with parameter $\varepsilon_r = \varepsilon + \log \frac{2}{2- e^\varepsilon}$ coincides with the binary $\varepsilon$-PML extremal mechanism. See \cite{grosse2024extremal} for more details.} Recall that PML-extremal mechanisms are optimal for maximizing sub-convex utility functions. Similarly, the randomized response mechanism has been shown to be optimal for LDP and the class of sub-convex utility functions~\cite{extremalmechanismLong}.


Finally, we discuss the application of Theorem~\ref{thm:liftasym} to the privacy mechanism design problem. In the following, we use the fact that $(\varepsilon_l, \varepsilon_u)$-ALIP is equivalent to $\varepsilon_u$-PML if $\varepsilon_l \geq \varepsilon_l^*$. 

\begin{corollary}
\label{cor:optimalPML}
 Suppose $X$ is a finite random variable with alphabet $\cX = [n]$. Then, for all $0 \leq \varepsilon_u < \log \frac{1}{1-p_{\min}}$, all $\varepsilon_l \geq \varepsilon_l^*$, and all utility functions, the optimal $(\varepsilon_l,\varepsilon_u)$-ALIP mechanism coincides with the optimal  $\varepsilon_u$-PML mechanism. In particular, the optimal $(\varepsilon_l,\varepsilon_u)$-ALIP mechanism maximizing sub-convex utility functions with $\varepsilon_l \geq \varepsilon_l^*$ is 
\begin{equation}
\label{eq:optPMLsubconvex}
     P_{Y \mid X=i}(j) = \begin{cases}
        1 - e^{\varepsilon_u}(1-P_X(i)) & \text{if}\; i=j, \\
        e^{\varepsilon_u}P_X(j) & \text{if}\; i \neq j,
    \end{cases}
\end{equation}
where $i,j \in [n]$.
\end{corollary}
A similar statement can be made using $\varepsilon_l$-PMC and $\varepsilon_u \geq \varepsilon_u^*$. 


\subsection{Secrets with General Alphabets}
\label{ssec:connections_general}
Now, we consider the more general case where  $(X,Y)$ takes values in the product space $(\mathcal X \times \mathcal Y, \mathcal S_{\mathcal X} \otimes \mathcal S_{\mathcal Y})$. Earlier, we discussed that a mechanism can have finite PML but infinite PMC. This invites the question: Is PMC inherently more restrictive than PML? That is, does bounded PMC imply bounded PML? The following two examples show that, in general, the answer is no.

\begin{example}
\label{ex:bounded_pmc_unbounded_pml_1}
Let $X$ and $Y$ be countably infinite random variables with alphabets $\cX = \cY = \{0, 1, \ldots \}$. Consider the prior 
\begin{equation*}
    P_X(0) = \frac{1}{2}, \quad P_X(n) = 3^{-n} \text{ for } n \geq 1 \\
\end{equation*}
and the mechanism
\begin{align*}
    &P_{Y \mid X = 0}(m) = \begin{cases}
        \frac{3}{4} & \text{if } m=0, \\
        \frac{3^{-m}}{2} & \text{if } m\geq 1, 
    \end{cases}\\
    &P_{Y \mid X = n}(m) = \begin{cases}
        \frac{1}{4} & \text{if } m=0, n\geq 1 \\
        \frac{1}{2} + \frac{3^{-m}}{2} & \text{if } m,n \geq 1, m=n,\\
        \frac{3^{-m}}{2} & \text{if } m,n \geq 1, m \neq n. 
    \end{cases}
\end{align*}
Using the above values, it can be verified that $Y$ has the same marginal distribution as $X$, that is, 
\begin{equation*}
    P_Y(0) = \frac{1}{2}, \quad P_Y(m) = 3^{-m} \text{ for } m \geq 1.
\end{equation*}

We now calculate the PML of the above mechanism at $Y = m \geq 1$. We have 
\begin{align*}
    \ell(X \to m) &= \log \, \frac{\sup_n P_{Y \mid X=n} (m)}{P_Y(m)}\\ 
     &= \log \, \frac{\frac{1}{2} + \frac{3^{-m}}{2}}{3^{-m}}\\[0.4em] 
     &= \log \!\Big( \frac{3^{m}+1}{2} \Big).
\end{align*}
Thus, $\ell(X \to m) \to \infty$ as $m \to \infty$ and the PML is unbounded. 

Next, we calculate the PMC. For $m \geq 1$ we have 
\begin{equation*}
    \Lambda(X \to m) = \log \, \frac{P_{Y}(m)}{\inf_n P_{Y \mid X=n} (m)} = \log \, \frac{3^{-m}}{\frac{3^{-m}}{2}} = \log 2, 
\end{equation*}
and for $m = 0$ we have 
\begin{equation*}  
    \Lambda(X \to 0) = \log \, \frac{P_{Y}(0)}{\inf_n P_{Y \mid X=n} (0)} = \log 2. 
\end{equation*}
Therefore, the mechanism satisfies $\log 2$-PMC.
\end{example}

We also construct an example with absolutely continuous random variables. 
\begin{example}
\label{ex:bounded_pmc_unbounded_pml_2}
Let 
\begin{equation*}
    c_1(x,y) = 2 (xy)^{-2} \Big( \frac{1}{x} + \frac{1}{y} -1\Big)^{-3}, \quad x,y \in (0,1),
\end{equation*}
denote the density of the \emph{Clayton copula} $c_\theta(\cdot, \cdot)$ with parameter $\theta = 1$~\cite{clayton1978model,nelsen2006introduction}. A copula defines a joint distribution with uniform marginals. That is, $X \sim \sfU(0,1)$ and $Y \sim \sfU(0,1)$ since
\begin{gather*}
    \int_{0}^1 c_1(x,y) \, dx = 1 \; \text{ for all } y \in (0,1), \\
    \int_{0}^1 c_1(x,y) \, dy = 1 \; \text{ for all } x \in (0,1).  
\end{gather*}
Observe that $c_1(t,t) \to \infty$ when $t \to 0$, so the density is unbounded. 

Now, fix $\eta \in (0,1)$ and consider the following joint density for $X$ and $Y$:
\begin{equation*}
    f_{XY}(x,y) = \eta + (1 - \eta) \, c_1(x,y), \quad x,y \in (0,1). 
\end{equation*}
This mixture preserves the marginal distributions, so $X \sim \sfU(0,1)$ and $Y \sim \sfU(0,1)$ still hold. Furthermore, the density remains unbounded above, but it is now bounded below by $\eta$.

We now compute both the PML and the PMC. The largest PML over the outcomes $y \in (0,1)$ is  
\begin{align*}
    \sup_{y \in (0,1)} \ell(X \to y) &= \log \, \sup_{x,y \in (0,1)} \, \frac{f_{XY} (x,y)}{f_X(x) f_Y(y)}\\
    &\geq \log \, \sup_{t \in (0,1)} f_{XY} (t,t) = \infty.
\end{align*}
Hence, the PML is unbounded. On the other hand, for the PMC, we have 
\begin{align*}
    \sup_{y \in (0,1)} \Lambda(X \to y) = \log \, \sup_{x,y \in (0,1)} \, \frac{f_X(x) f_Y(y)}{f_{XY} (x,y)} \leq \log \frac{1}{\eta},
\end{align*}
so the mechanism satisfies $\log \frac{1}{\eta}$-PMC. 
\end{example}

Together, these two examples demonstrate that a mechanism can have bounded PMC but unbounded PML. 

Lastly, we show that the connections between LDP, PML, and PMC established in Proposition~\ref{prop:ldp_pml_bound}, Proposition~\ref{prop:ldp_pmc}, and Corollary~\ref{cor:relations} partially extend to general alphabets. The proof is given in Appendix~\ref{sec:proof_general_ldp_pml_pmc}.

\begin{theorem}
\label{thm:general_ldp_pml_pmc}
Let $P_{XY} = P_{Y \mid X} \times P_X$ be the joint distribution of $X$ and $Y$. If the mechanism $P_{Y \mid X}$ satisfies $\varepsilon$-LDP with $\varepsilon \geq 0$, then it satisfies $\varepsilon$-PMC and $\varepsilon$-PML. Conversely, if the mechanism satisfies $\varepsilon_1$-PMC and $\varepsilon_2$-PML, with $\varepsilon_1, \varepsilon_2 \geq 0$, then it satisfies $\varepsilon_1 + \varepsilon_2$-LDP.    
\end{theorem}

The above result demonstrates that, in general, the class of LDP mechanisms coincides with the class of mechanisms that have both bounded PMC and bounded PML. That is, LDP is equivalent to simultaneously restricting the threat imposed by risk-averse and opportunistic adversaries. This can be seen as a general-alphabet extension of~\cite[Prop. 1]{zarrabian2023lift}.

\section{Discussion: PMC As an Information Leakage Measure}
\label{sec:discussion}
While PMC and PML are both operationally meaningful and context-aware privacy measures, they are conceptually different. PML is defined by modeling opportunistic adversaries aiming to maximize gain functions, whereas PMC is obtained by modeling risk-averse adversaries interested in minimizing cost functions. This fundamental difference in the threat models results in substantial qualitative and quantitative differences between the two measures. Here, we discuss their differences along two themes. 

First, in one of the earliest papers on quantitative information flow, \citet{smith2009foundations} posited that a measure of information leakage should capture the difference between the initial uncertainty about a random variable and the remaining uncertainty after observing a correlated quantity. Specifically, he suggested that leakage measures should satisfy
\begin{multline}
\label{eq:smith_axiom}
\text{information leakage} = \\
\text{initial uncertainty} - \text{remaining uncertainty}.
\end{multline}
When $X$ is a finite random variable, the initial uncertainty in $X$ is finite according to all the standard measures of uncertainty, such as the Rényi entropies. Therefore, a direct consequence of~\eqref{eq:smith_axiom} is that information leaked about finite random variables should invariably remain finite, regardless of the mechanism used. This finiteness criterion may be viewed as an axiomatic requirement for information leakage measures and is satisfied by most measures, such as PML, mutual information, and $f$-information, among others. 

However, as noted in Remark~\ref{rem:pmc_unbounded}, PMC can become infinite even when $X$ is finite. As an example, consider a binary random variable $X$ with $P_X(0) = q$ with $q \in (0,1)$ and suppose the posterior distribution satisfies $P_{X \mid Y=y}(0) = 1$ for some $y \in \cY$. Here, according to PMC the information leaked is $\Lambda_{P_{XY}}(X \to y) = \infty$. Consequently, PMC satisfies neither~\eqref{eq:smith_axiom} nor the weaker finiteness criterion. This discrepancy becomes even more counter-intuitive as $q$ approaches 1, where the seemingly incremental knowledge gained about $X$ still yields $\Lambda_{P_{XY}}(X \to y) = \infty$.

A second consequence of the different threat models is that the class of mechanisms satisfying PMC and PML guarantees, while overlapping, are distinct. When the underlying alphabets are finite, PMC is strictly stronger since a
bound on PMC always implies a bound on PML, but the reverse need not hold. In general measurable spaces, the connection is more nuanced. A mechanism can have bounded PMC, bounded PML, both, or neither. Moreover, as long as $P_{XY} \ll P_X \times P_Y$, PML is well-defined even if it is unbounded. However, this is not the case for PMC, which may be undefined altogether. As a concrete example, consider the task of privately estimating the mean of a Gaussian random variable by adding Gaussian noise to the sample mean. This setup is well-defined under PML, since PML permits the addition of Gaussian noise to a Gaussian random variable (see~\cite[Example 4]{saeidian2023pointwisegeneral}). In contrast, PMC  cannot be defined in this setting. As shown in Example~\ref{ex:gaussian_noise} and specifically in~\eqref{eq:gaussian_bounds}, the lower bound on PMC becomes infinite when Gaussian noise is added to an unbounded random variable.

\section{Conclusions}
\label{sec:conclusion}
In this paper, we explored the operational meaning and the implications of guaranteeing privacy by enforcing a lower bound on the information density. We introduced a novel privacy measure, pointwise maximal cost (PMC), and demonstrated that imposing an upper bound on PMC is equivalent to enforcing a lower bound on the information density. PMC quantifies the information leakage about a secret $X$ to risk-averse adversaries who aim to minimize non-negative cost functions after observing an outcome $y$ of a privacy mechanism. When  $X$ is a finite random variable, PMC can also be defined by considering adversaries aiming to minimize the probability of incorrectly guessing randomized functions of $X$. By investigating the operational meaning of the lower bound on the information density, our results have strengthened and increased the relevance of definitions such as LIP and ALIP that previously imposed the lower bound axiomatically. 

This work further examined the connections between upper and lower bounds on the information density by analyzing how PMC and PML relate. When applied to finite random variables, PMC emerges as a more stringent privacy measure, applicable only to mechanisms that satisfy differential privacy-type guarantees.  This characteristic, which sets PMC apart from other measures such as PML and $f$-information, arises from the definition of PMC: If the posterior expected value of a cost function is zero, then PMC is infinite. In general measurable spaces, we showed that the class of mechanisms satisfying LDP is exactly the set of those with bounded PMC and bounded PML. Put differently, LDP is equivalent to simultaneously restricting the threat imposed by risk-averse and opportunistic adversaries.

Overall, our work significantly bridges the gaps in understanding the relationships between various privacy frameworks. Looking ahead, the findings in this paper offer crucial insights for selecting the appropriate framework to ensure privacy in different applications. If the objective is to adopt a conservative approach that prevents any definitive inference about the secret, then LDP, LIP, and ALIP would be suitable choices. When information about the prior distribution is available, LIP and ALIP are preferred as they enable higher utility compared to LDP. ALIP provides additional flexibility by allowing distinct upper and lower bounds, while LIP simplifies the design by imposing symmetric bounds. Conversely, when the prior is not known, then LDP would be a suitable choice since it offers equally strong protection regardless of the prior. 

Finally, we highlight how the results in this work can inform the practical design of privacy mechanisms. A mechanism originally designed to satisfy one privacy guarantee may also fulfill others. Using the relationships between privacy measures developed in this paper, one can assess whether a given mechanism provides additional guarantees beyond its original design. In addition, many of the canonical examples studied here, such as the generalized randomized response and Laplace mechanisms, form the basis for more sophisticated algorithms used in applications like federated learning, data sharing, and privacy-preserving statistics. By introducing a context-aware measure that captures a notion of privacy complementary to existing ones, this work takes a step toward designing high-utility systems with strong and meaningful guarantees.

\appendices
\section{Proof of Theorem~\ref{thm:equivalence}}
\label{sec:proof_thm_equivalence}
Fix $y \in \cY$. First, we argue that given a $U$ satisfying the Markov chain $U-X-Y$ there exists a non-negative cost function $c_U$ such that $\Lambda_{U}(X \to y) = \Lambda_{c}(X \to y)$. Given $U$, let the cost function $c_U: \cX \times \cU \to \bR_+$ be defined as $c_U(x,u) = 1 - P_{U \mid X=x}(u)$ with $x \in \cX$ and $u \in \cU$. Then, we have 
\begin{align*}
    &\exp\Big( \Lambda_{c_U}(X \to y) \Big) = \frac{\inf_{P_{\hat U}} \; \bE \Big[c_U(X,\hat U)\Big]}{\inf_{P_{\tilde U \mid Y}} \; \bE \Big[c_U(X,\tilde U) \mid Y=y \Big]}\\[0.5em] 
    &= \frac{\min_{u \in \mathcal \cU} \bE [c_U(X,u)]}{\min_{u \in \mathcal U} \bE [c_U(X,u) \mid Y=y]}\\[0.5em]
    &= \frac{\min_{u \in \cU} \sum_x c_U(x,u) P_X(x)}{\min_{u \in \cU} \sum_x c_U(x,u) P_{X \mid Y=y}(x)}\\[0.5em]
    &= \frac{\min_{u \in \cU} \sum_x (1 - P_{U \mid X=x}(u)) P_X(x)}{\min_{u \in \cU} \sum_x (1 - P_{U \mid X=x}(u)) P_{X \mid Y=y}(x)}\\[0.5em]
    &= \frac{1 - \max_{u \in \cU} \sum_x P_{U \mid X=x}(u) P_X(x)}{1 - \max_{u \in \cU} \sum_x P_{U \mid X=x}(u) P_{X \mid Y=y}(x)}\\[0.5em]
    &= \exp\Big( \Lambda_{U}(X \to y) \Big).
\end{align*}
Observe that the above construction works even when $\Lambda_{U}(X \to y) = \infty$ in which case we also get $\Lambda_{c_U}(X \to y) = \infty$.

Now, we show that for each non-negative cost function $c$, there exists $U$ satisfying the Markov chain $U-X-Y$ such that $\Lambda_{c}(X \to y) = \Lambda_{U}(X \to y)$. Fix a cost function $c$. Without loss of generality, we assume that $0 \leq c(x,w) \leq 1$ for all $x \in \cX$ and $w \in \cW$. This can be achieved through normalizing $c$ by $\max_{x,w} c(x,w)$. 

First, suppose $\Lambda_c(X \to y) < \infty$. Let $k$ be a large integer. We construct two randomized functions of $X$ denoted by $S$ and $T$ both on the same alphabet $[k+1]$. Let 
\begin{gather*}
    w_S \in \argmin_{w} \sum_x c(x, w) P_{X}(x),\\
    w_T \in \argmin_{w} \sum_x c(x,w) P_{X \mid Y=y}(x),
\end{gather*}
where $w_S$ denotes the adversary's optimal choice prior to observing $y$ and $w_T$ denotes the adversary's optimal choice after observing $y$. For all $x \in \cX$, let
\begin{gather*}
    P_{S \mid X=x}(i) = \frac{c(x,w_S)}{k}, \quad i \in [k],\\ 
    P_{S \mid X=x}(k+1) = 1 - c(x,w_S),\\
    P_{T \mid X=x}(i) = \frac{c(x,w_T)}{k}, \quad i \in [k],\\ 
    P_{T \mid X=x}(k+1) = 1 - c(x,w_T). 
\end{gather*}
Let $0 \leq \delta \leq 1$. We define $U_\delta$ as the mixture of $S$ and $T$ with parameter $\delta$, that is, 
\begin{equation*}
    U_\delta = \begin{cases}
        S & \text{with probability} \; \delta,\\
        T & \text{with probability} \; 1-\delta. 
    \end{cases}
\end{equation*}
Then, $P_{U_\delta \mid X=x}(i) = \delta P_{S \mid X=x}(i) + (1 - \delta) P_{T \mid X=x}(i)$ for $i \in [k+1]$ and we get 
\begin{align*}
    &\inf_{P_{\hat U}}\; \bP [U_\delta \neq \hat{U}] = \min_{u \in [k+1]} \sum_{x} (1-P_{U_\delta \mid X=x}(u)) P_X(x)\\[0.5em]
    &=\min \Bigg \{\sum_{x} \Big( 1 - \delta\frac{c(x,w_S)}{k} - (1 - \delta) \frac{c(x,w_T)}{k}\Big) P_X(x),\\
    &\hspace{6em}\sum_{x} \Big(\delta c(x,w_S) + (1 - \delta) c(x,w_T) \Big) P_X(x) \Bigg\}\\[0.5em]
    &=\sum_{x} \Big(\delta c(x,w_S) + (1 - \delta) c(x,w_T) \Big) P_X(x), 
\end{align*}
where the last equality follows by letting $k \to \infty$. Similarly, we have 
\begin{align*}
    &\inf_{P_{\hat U \mid Y}} \; \bP [U_\delta \neq \hat{U} \mid Y=y]\\
    &= \min_{u \in [k+1]} \sum_{x} (1-P_{U_\delta \mid X=x}(u)) P_{X \mid Y=y}(x)\\[0.5em]
    &=\min \!\Bigg \{\!\sum_{x} \!\Big( \!1 - \delta\frac{c(x,w_S)}{k} - (1 - \delta) \frac{c(x,w_T)}{k}\Big) P_{X \mid Y=y}(x),\\
    &\hspace{5em}\sum_{x} \Big(\delta c(x,w_S) + (1 - \delta) c(x,w_T)\Big) P_{X \mid Y=y}(x) \Bigg\}\\[0.5em]
    &= \sum_{x} \Big(\delta c(x,w_S) + (1 - \delta) c(x,w_T)\Big) P_{X \mid Y=y}(x), 
\end{align*}
and, once again, the last equality follows by letting $k \to \infty$. Therefore, we have
\begin{multline*}
    \exp \Big( \Lambda_{U_\delta} (X \to y) \Big)\\
    = \frac{\sum_{x} \Big(\delta c(x,w_S) + (1 - \delta) c(x,w_T) \Big) P_X(x)}{\sum_{x} \Big(\delta c(x,w_S) + (1 - \delta) c(x,w_T)\Big) P_{X \mid Y=y}(x)}.
\end{multline*}

Next, note that when $\delta = 1$ we have 
\begin{align*}
    \exp\Big(\Lambda_{U_1} (X \to y)\Big) &= \exp\Big(\Lambda_S(X \to y) \Big)\\
    &= \frac{\sum_x c(x,w_S) P_{X}(x)}{\sum_{x} c(x,w_S) P_{X \mid Y=y}(x)}\\[0.7em]
    &\leq \frac{\sum_x c(x,w_S) P_{X}(x)}{\sum_{x} c(x,w_T) P_{X \mid Y=y}(x)}\\
    &= \frac{\min_w \sum_x c(x, w) P_{X}(x)}{\min_w \sum_x c(x, w) P_{X \mid Y=y}(x)}\\[0.7em]
    &= \exp \Big(\Lambda_c(X \to y) \Big), 
\end{align*}
and when $\delta = 0$ we have 
\begin{align*}
    \exp\Big(\Lambda_{U_0} (X \to y)\Big) &= \exp\Big(\Lambda_T(X \to y) \Big)\\
    &= \frac{\sum_x c(x,w_T) P_{X}(x)}{\sum_{x} c(x,w_T) P_{X \mid Y=y}(x)}\\[0.7em]
    &\geq \frac{\sum_x c(x,w_S) P_{X}(x)}{\sum_{x} c(x,w_T) P_{X \mid Y=y}(x)}\\
    &= \frac{\min_w \sum_x c(x, w) P_{X}(x)}{\min_w \sum_x c(x, w) P_{X \mid Y=y}(x)}\\[0.7em]
    &= \exp \Big(\Lambda_c(X \to y) \Big).  
\end{align*}
In other words, we have $\Lambda_{U_1}(X \to y) \leq \Lambda_c(X \to y) \leq \Lambda_{U_0}(X \to y)$. Then, due to the continuity of the mapping $\delta \mapsto \Lambda_{U_\delta}(X \to y)$, there exists $\delta^* \in [0,1]$ such that $\Lambda_{U_{\delta^*}} (X \to y) = \Lambda_c(X \to y)$. 

Finally, we consider the case $\Lambda_c(X \to y) = \infty$. In this case, there exists a proper subset of $\cX$ denoted by $\cE$ such that $P_{X \mid Y=y}(\cE) = 1$ and $c(x,w_T) = 0$ for all $x \in \cE$. Let $U$ be a binary random variable described by $P_{U \mid X=x}(0)  = 1$ for all $x \in \cE$ and $P_{U \mid X=x}(0) = \frac{1}{2}$ for $x \in \cX \setminus \cE$. Then, we have 
\begin{align*}
    \Lambda_U(X \to y) &= \log \frac{1 - \max \{P_U(0), P_U(1) \}}{1 - \max_u \sum_{x \in \cX} P_{U \mid X=x}(u) P_{X \mid Y=y}(x)}\\[0.7em]
    &=\log \frac{1 - \max \{\frac{1}{2} + \frac{P_X(\cE)}{2}, \frac{P_X(\cX \setminus \cE)}{2} \}}{1 - P_{X \mid Y=y}(\cE)}\\
    &= \infty, 
\end{align*}
since $0 < P_X(\cE) < 1$.


\section{Proof of Theorem~\ref{thm:general_leakage}}
\label{sec:proof_thm_general_leakage}
Fix $y \in \cY$. The proof follows along the same lines of the proof of \cite[Thm. 3]{saeidian2023pointwisegeneral}. We begin by noting that since $(\cX, \cS_\cX)$ is standard Borel, then $P_{XY}$ can be disintegrated into the marginal $P_{Y}$ and a conditional distribution $P_{X \mid Y}$ \cite[Thm. IV.2.18]{ccinlar2011probability}. We use this fact to simplify the denominator of~\eqref{eq:general_leakage}. The following lemma is proved in Appendix~\ref{sec:proof_general_leakage_denom}. 

\begin{lemma}
\label{lem:general_leakage_denom}
Let $W$ be a random variable induced by a transition probability kernel $P_{W \mid Y}$ from $(\cY, \cS_\cY)$ into $(\cW, \cS_\cW)$. Suppose $P_{XY}(dx, dy) = P_Y(dy) P_{X \mid Y=y}(dx)$. For all $c \in \cC$, it holds that 
\begin{equation}
\label{eq:leakage_denom}
    \inf_{P_{\hat W \mid Y}} \, \bE \left[c(X, \hat W) \mid Y=y \right] = \inf_{\hat w \in \cW} \int_{\cX} c(x,\hat w) P_{X \mid Y=y}(dx).  
\end{equation}
\end{lemma}

It also follows from Lemma~\ref{lem:general_leakage_denom} that
\begin{equation*}
    \inf_{P_{W}} \, \bE \Big[c(X, W) \Big] = \inf_{w \in \cW} \int_{\cX} c(x, w) P_{X}(dx).  
\end{equation*}

Suppose $P_X \ll P_{X \mid Y=y}$ and let $f \coloneqq \frac{dP_{X}}{dP_{X \mid Y=y}}$ denote the Radon-Nikodym derivative of $P_{X}$ with respect to $P_{X \mid Y=y}$. First, we show that $\Lambda_{P_{XY}}(X \to y) \leq \dinf(P_{X} \Vert P_{X \mid Y=y})$. Fix a measurable space $(\cW, \cS_\cW)$ and a cost function $c \in \cC$. We can write 
\begin{subequations}
\begin{align}
    &\frac{\inf_{P_{W}} \, \bE \Big[c(X, W) \Big]}{ \inf_{P_{\hat W \mid Y}} \bE[c(X, \hat W) \mid Y=y]} = \frac{\inf\limits_{w \in \cW} \int_\cX c(x,w)  P_X(dx)}{\inf\limits_{\hat w \in \cW} \int_{\cX} c(x,\hat w) P_{X \mid Y=y}(dx)} \nonumber\\
    &= \sup_{\hat w \in \cW} \frac{\inf\limits_{w \in \cW} \int_{\cX} c(x,w) \; P_{X}(dx)}{\int_\cX c(x, \hat w) \; P_{X \mid Y=y}(dx)}  \nonumber\\
    &\leq \sup_{\hat w \in \cW} \frac{\int_{\cX} c(x,\hat w) \; P_{X}(dx)}{\int_\cX c(x, \hat w) \; P_{X \mid Y=y}(dx)}  \nonumber\\
    &= \sup_{\hat w \in \cW} \frac{\int_{\cX} c(x,\hat w) \; f(x) \; P_{X \mid Y=y}(dx)}{\int_\cX c(x,\hat w) \; P_{X \mid Y=y}(dx)}  \nonumber\\
    &\leq \esssup_{P_{X \mid Y=y}} f \; \Bigg(\sup_{\hat w \in \cW} \frac{\int_{\cX} c(x,\hat w) \; P_{X \mid Y=y}(dx)}{\int_\cX c(x,\hat w) \; P_{X \mid Y=y}(dx)} \Bigg) \nonumber\\
    &= \esssup_{P_{X}} f \label{subeq:general_up_bound_1} \\
    &= \exp \Big(\dinf(P_{X} \Vert P_{X \mid Y=y}) \Big), \nonumber
\end{align}
\end{subequations}
where \eqref{subeq:general_up_bound_1} is due to the fact that $f$ is zero on all sets $\cA \in \cS_\cX$ such that $P_{X \mid Y=y}(\cA) > 0$ but $P_{X} (\cA) = 0$. Thus, $\Lambda_{P_{XY}}(X \to y) \leq \dinf(P_{X} \Vert P_{X \mid Y=y})$. 

Next, we show that $\Lambda_{P_{XY}}(X \to y) \geq \dinf(P_{X} \Vert P_{X \mid Y=y})$. Let $\cW = \bZ \cup \{-\infty\}$ and $\cS_\cW$ be the discrete $\sigma$-algebra on $\cW$. Fix $\varepsilon > 0$ and consider the following countable partition of $\cX$: 
\begin{equation}
\label{eq:parition}
    \cD_w^\varepsilon = \{x \in \cX : e^{w \varepsilon} \leq f(x) < e^{(w+1) \varepsilon}\},\quad w \in \cW.
\end{equation}
Of course, $\cD_w^\varepsilon \in \cS_\cX$ for all $w \in \cW$. Fix a constant $M >1$ and consider the cost function $c^* : \cX \times \cW \to \bR_+$ defined by
\begin{equation*}
    c^*(x,w) = \begin{cases}
        \frac{1}{P_X(\cD_w^\varepsilon)} \, \ind_{\cD_w^\varepsilon} (x) & \mathrm{if} \; P_X(\cD_w^\varepsilon) > 0,\\
        M & \mathrm{if} \; P_X(\cD_w^\varepsilon) = 0. 
    \end{cases}
\end{equation*}
It is easy to see that $c^* \in (\cS_\cX \otimes \cS_\cW)_+$. Moreover, the absolute continuity $P_X \ll P_{X \mid Y=y}$ ensures that if $P_X(\cD_w^\varepsilon) > 0$ then $P_{X \mid Y=y}(\cD_w^\varepsilon) > 0$. Therefore, we can write
\begin{subequations}
\begin{align}
    &\exp \Big(\Lambda_{P_{XY}}(X \to y) \Big) \geq \frac{\inf\limits_{w \in \cW} \int_\cX c^*(x,w) \; P_X(dx)}{\inf\limits_{w \in \cW} \; \int_{\cX} c^*(x,w) \; P_{X \mid Y=y}(dx)}\\[0.7em]
    &= \frac{\min \left\{M, \inf\limits_{w \in \cW : P_X(\cD_w^\varepsilon) > 0} \; \int_\cX \frac{1}{P_X(\cD_w^\varepsilon)} \, \ind_{\cD_w^\varepsilon} (x) \; P_X(dx) \right\}}{\min \!\left\{\!M, \!\inf\limits_{w \in \cW : P_X(\cD_w^\varepsilon) > 0} \int_\cX \frac{1}{P_X(\cD_w^\varepsilon)}  \ind_{\cD_w^\varepsilon} (x) P_{X \mid Y=y}(dx) \!\right\}}\nonumber \\[0.7em]
    &= \frac{\min \left\{M,\inf\limits_{w \in \cW : P_X(\cD_w^\varepsilon) > 0} \frac{P_X (\cD_w^\varepsilon)}{P_X(\cD_w^\varepsilon)} \right\}}{\min\left\{M,\inf\limits_{w \in \cW : P_X(\cD_w^\varepsilon) > 0} \frac{P_{X \mid Y=y} (\cD_w^\varepsilon)}{P_X(\cD_w^\varepsilon)} \right\}} \nonumber\\[0.7em]
    &=\frac{1}{\min\left\{M,\inf\limits_{w \in \cW : P_X(\cD_w^\varepsilon) > 0} \frac{P_{X \mid Y=y} (\cD_w^\varepsilon)}{P_X(\cD_w^\varepsilon)} \right\}} \label{subeq:achiev_0}\\[0.7em]
    &\geq \frac{1}{\inf\limits_{w \in \cW : P_X(\cD_w^\varepsilon) > 0} \frac{P_{X \mid Y=y} (\cD_w^\varepsilon)}{P_X(\cD_w^\varepsilon)}} \nonumber\\[0.7em]
    &= \sup\limits_{w \in \cW : P_X(\cD_w^\varepsilon) > 0} \frac{P_X (\cD_w^\varepsilon)}{P_{X \mid Y=y}(\cD_w^\varepsilon)} \label{subeq:achiev_1} \\[0.7em]
    &= \esssup_{P_X} \; \bar{f}, \label{subeq:achiev_2}
\end{align}
\end{subequations}
where~\eqref{subeq:achiev_0} follows because $M > 1$, and 
\begin{equation*}
    \bar{f}(x) \coloneqq \sum_{w \in \cW} \frac{P_X(\cD_w^\varepsilon)}{P_{X \mid Y=y}(\cD_w^\varepsilon)} \; \ind_{\cD_w^\varepsilon} (x), \quad x \in \cX.
\end{equation*}
Observe that we have replaced the supremum over $w$ in~\eqref{subeq:achiev_1} with the essential supremum over $x$ in~\eqref{subeq:achiev_2} because $\bar{f}$ is constant on each set $\cD_w^\varepsilon$.\footnote{In fact, $\bar f$ is the conditional expectation of $f$ given $\cF \coloneqq \sigma \{\cD_w^\varepsilon\}$, where $\sigma \{\cD_w^\varepsilon\}$ denotes the $\sigma$-algebra on $\cX$ generated by the collection of sets $\{\cD_w^\varepsilon\}$.} Finally, we have 
\begin{subequations}
\begin{align}
    \Lambda_{P_{XY}}(X \to y) &\geq \log \; \esssup_{P_X} \; \bar{f} \nonumber\\
    &\geq \log \left( \left(\esssup_{P_X} f\right) \; e^{-\varepsilon} \right) \label{subeq:achiev_3}\\
    &= \log \; \esssup_{P_X} f - \varepsilon \nonumber\\
    &= \dinf(P_X \Vert P_{X \mid Y=y}) - \varepsilon,\nonumber
\end{align}
\end{subequations}
where~\eqref{subeq:achiev_3} is because by~\eqref{eq:parition}, $\bar{f}$ never differs from $f$ by more than a factor of $e^\varepsilon$. Then, by letting $\varepsilon \to 0$, we get $\Lambda_{P_{XY}}(X \to y) \geq \dinf(P_{X \mid Y=y} \Vert P_X)$. This completes the proof for the case $P_X \ll P_{X \mid Y=y}$.

All that is left to discuss is the case $P_X \nll P_{X \mid Y=y}$. In this case, there exists a set $\cA_0 \in \cS_\cX$ such that $P_{X \mid Y=y}(\cA_0) = 0$ but $P_X(\cA_0) >0$. Let $(\cW, \cS_\cW)$ be an arbitrary measurable space and consider $c(x,w) = \ind_{\cA_0} (x)$ for all $w \in \cW$. Then, we have $\bE \left[c(X, w) \mid Y=y \right] = P_{X \mid Y=y}(\cA_0) = 0$ whereas $\bE \left[c(X,w) \right] = P_X(\cA_0) > 0$ for all $w \in \cW$. Thus, $\Lambda_{P_{XY}}(X \to y) = \dinf(P_{X} \Vert P_{X \mid Y=y}) = \infty$, as desired.  

\section{Proof of Lemma~\ref{lem:general_leakage_denom}}
\label{sec:proof_general_leakage_denom}
Fix an arbitrary kernel $P_{W \mid Y}$ and $y \in \cY$. We write 
\begin{align*}
    &\bE \left[c(X, W) \mid Y=y \right] = \int_{\cX \times \cW} c(x,w) \; P_{XW \mid Y=y} (dx, dw)\\
    &= \int_{\cW} \; P_{W \mid Y=y}(dw) \int_{\cX} c(x,w)\; P_{X \mid Y=y}(dx) \\ 
    &\geq \int_{\cW} \; P_{W \mid Y=y}(dw) \left(\inf_{w \in \cW} \int_{\cX} c(x,w)\; P_{X \mid Y=y}(dx) \right)\\
    &= \inf_{w \in \cW} \int_{\cX} c(x,w)\; P_{X \mid Y=y}(dx). 
\end{align*}
Taking the infimum over all kernels $P_{W \mid Y}$ we get 
\begin{equation*}
    \inf_{P_{W \mid Y}} \bE \left[c(X, W) \mid Y=y \right] \geq \inf_{w \in \cW} \; \int_{\cX} c(x,w) \; P_{X \mid Y=y}(dx).  
\end{equation*}

To prove the reverse inequality, fix  $a > \inf_{w \in \cW} \int_{\cX} c(x,w)\; P_{X \mid Y=y}(dx)$. By definition, there exists $w' \in \cW$ such that $\int_{\cX} c(x,w')\; P_{X \mid Y=y}(dx) \leq a$. Given $w \in \cW$, let $\delta_w$ denote the Dirac measure on $(\cW, \cS_\cW)$ defined by
\begin{equation*}
    \delta_w (\cA) = \begin{cases}
        1 & \mathrm{if} \; w \in \cA,\\
        0 & \mathrm{if} \; w \notin \cA,\\
    \end{cases} 
\end{equation*}
for each $\cA \in \cS_\cW$. Then, we have
\begin{align*}
    \inf_{P_{W \mid Y}} \bE & \left[c(X, W) \mid Y=y \right]\\
    &\leq \int_{\cX} P_{X \mid Y=y}(dx) \int_{\cW} c(x,w)\; \delta_{w'}(dw) \\ 
    &= \int_{\cX} c(x,w')\; P_{X \mid Y=y}(dx) \leq a. 
\end{align*}
Then, letting $a \to \inf_{w \in \cW} \int_{\cX} c(x,w)\; P_{X \mid Y=y}(dx)$ we obtain
\begin{equation*}
    \inf_{P_{W \mid Y}} \bE \left[c(X, W) \mid Y=y \right] \leq \inf_{w \in \cW} \; \int_{\cX} c(x,w) \; P_{X \mid Y=y}(dx).
\end{equation*}

\section{Proof of Lemma~\ref{lemma:properties}}
\label{sec:proof_lemma_properties}
\begin{enumerate}
    \item Non-negativity of PMC follows from the non-negativity of the \Ren divergence.
    \item If $X$ and $Y$ are independent then $P_{X \mid Y=y} = P_X$ almost surely. Thus, $\Lambda(X \to y) = \dinf(P_X \Vert P_{X \mid Y=y}) = \dinf(P_X \Vert P_X) = 0$ almost surely. 
    \item The assumption implies that $P_{X_1, \ldots, X_k \mid Y_1, \ldots, Y_k} = \prod_{i=1}^k P_{X_i \mid Y_i}$. Therefore, we have 
    \begin{align}
        \Lambda(X_1, \ldots, &X_k \to y_1, \ldots, y_k)\\
        &= \dinf(P_{X_1, \ldots, X_k} \Vert P_{X_1, \ldots, X_k \mid Y_1=y_1, \ldots Y_k=y_k})\\
        &=  \dinf\left(\prod_{i=1}^k P_{X_i} \Big\Vert \prod_{i=1}^k P_{X_i \mid Y_i=y_i}\right) \\
        &= \sum_{i=1}^k \dinf\left(P_{X_i} \Vert P_{X_i \mid Y_i=y_i}\right) \label{subeq:additivity}\\
        &= \sum_{i=1}^k \Lambda(X_i \to y_i),
    \end{align}
    almost surely, where \eqref{subeq:additivity} is due to the additivity property of the \Ren divergence~\cite[Thm. 28]{van2014renyi}.
    \item Fix $y \in \cY$. Let $P_X$ and $Q_X$ denote probability measures on $(\cX, \cS_\cX)$. Define $P_\theta = \theta P_X + (1 - \theta) Q_X$, where $\theta \in [0,1]$. Let $f_{Y \mid X}$ denote the conditional density (defined with respect to $\mu$) of $Y$ given $X$, and let 
    \begin{gather*}
        f_\theta = \frac{dP_\theta}{d\mu}, \quad f_X = \frac{dP_X}{d\mu}, \quad g_X = \frac{dQ_X}{d\mu},\\
        f_Y(y) = \int_{\cX} f_{Y \mid X=x}(y)\,  f_X(x) \, \mu(dx),\\
        g_Y(y) = \int_{\cX} f_{Y \mid X=x}(y)\, g_X(x) \, \mu(dx),
    \end{gather*}
    where $\mu$ is a dominating $\sigma$-finite measure on $\cX$. We have 
    \begin{align}
        &\Lambda_{P_{Y \mid X} \times P_\theta}(X \to y) = \log \; \frac{\int_{\cX} f_{Y \mid X=x}(y) f_\theta(x) \; \mu(dx)}{\essinf\limits_{P_\theta} \; f_{Y \mid X}(y \mid X)}\\
        &= \log \Bigg( \frac{\theta \int f_{Y \mid X=x}(y)  f_X(x) \mu(dx)}{\essinf\limits_{P_\theta} \; f_{Y \mid X}(y \mid X)}\\
        &\hspace{7em} + \frac{(1 - \theta) \int f_{Y \mid X=x}(y)  g_X(x) \mu(dx)}{\essinf\limits_{P_\theta} \; f_{Y \mid X}(y \mid X)}\Bigg) \\
        &= \log \; \Bigg(\frac{\theta f_Y(y)}{\essinf\limits_{P_\theta}  \; f_{Y \mid X}(y \mid X)} + \frac{(1 - \theta) g_Y(y)}{\essinf\limits_{P_\theta}  \; f_{Y \mid X}(y \mid X)}\Bigg)\\
        &\geq \log \Bigg(\theta \frac{f_Y(y) }{\essinf\limits_{P_X} f_{Y \mid X}(y \mid X)}\\
        &\hspace{8em}+ (1 - \theta) \frac{g_Y(y) }{\essinf\limits_{Q_X} f_{Y \mid X}(y \mid X)} \Bigg) \label{subeq:concavity}\\
        &\geq \theta \log \; \frac{f_Y(y) }{\essinf\limits_{P_X} \; f_{Y \mid X}(y \mid X)}\\
        &\hspace{7em}+ (1 - \theta) \log \; \frac{g_Y(y) }{\essinf\limits_{Q_X} \; f_{Y \mid X}(y \mid X)}\\
        &= \theta \Lambda_{P_{Y \mid X} \times P_X}(X \to y) + (1 - \theta) \Lambda_{P_{Y \mid X} \times Q_X}(X \to y),
    \end{align}
    where \eqref{subeq:concavity} follows because $P_X, Q_X \ll P_\theta$, implying that $\essinf\limits_{P_\theta}  \; f_{Y \mid X}(y \mid X) \leq \essinf\limits_{P_X}  \; f_{Y \mid X}(y \mid X)$ and $\essinf\limits_{P_\theta}  \; f_{Y \mid X}(y \mid X) \leq \essinf\limits_{Q_X}  \; f_{Y \mid X}(y \mid X)$. 
    \item Fix $y \in \cY$. Consider the Markov chain $Z - X - Y$. The kernel $P_{Z \mid X}$ induces distributions $P_Z = P_{Z \mid X} \circ P_{X}$ and $P_{Z \mid Y=y} = P_{Z \mid X} \circ P_{X \mid Y=y}$. Therefore, we can write
    \begin{align*}
        \Lambda(Z \to y) &= \dinf(P_Z \Vert P_{Z \mid Y=y})\\
        &= \dinf(P_{Z \mid X} \circ P_X \Vert P_{Z \mid X} \circ P_{X \mid Y=y})\\
        &\leq \dinf(P_X \Vert P_{X \mid Y=y}) = \Lambda(X \to y),
    \end{align*}
    where the inequality is due to the data-processing inequality for the \Ren divergence~\cite[Thm. 9]{van2014renyi}. 
    
    \item Consider the Markov chain $X - Y - Z$. The kernel $P_{Z \mid Y}$ induces distributions $P_{XZ} = P_{Z \mid Y} \circ P_{XY}$ and $P_Z = P_{Z \mid Y} \circ P_{Y}$. Hence, we can write 
    \begin{align*}
        \esssup_{P_Z} \Lambda(&X \to Z) = \dinf(P_X \times P_Z \Vert P_{XZ})\\
        &=  \dinf(P_X \times (P_{Z\mid Y} \circ P_Y) \Vert P_{Z \mid Y} \circ P_{XY})\\
        &=  \dinf(P_{Z\mid Y} \circ (P_X \times P_Y) \Vert P_{Z \mid Y} \circ P_{XY})\\
        &\leq \dinf (P_X \times P_Y \Vert P_{XY})\\
        &= \esssup_{P_Y} \Lambda(X \to Y),
    \end{align*}
    where the inequality is due to the data-processing inequality for the \Ren divergence~\cite[Thm. 9]{van2014renyi}. 

    \item Fix $y_1 \in \cY_1$ and $y_2 \in \cY_2$. First, suppose $P_{X} \ll P_{X \mid Y_1=y_1}$. We can express the PMC using densities: 
    \begin{align}
        &\Lambda(X \to y_1, y_2) = \log \; \esssup_{P_X} \; \frac{f_{Y_1, Y_2}(y_1, y_2)}{f_{Y_1, Y_2 \mid X}(y_1, y_2 \mid X)}\\
        &= \log \; \esssup_{P_X} \; \frac{f_{Y_1}(y_1) \; f_{Y_2 \mid Y_1}(y_2 \mid y_1)}{f_{Y_1\mid X}(y_1 \mid X) \; f_{Y_2 \mid Y_1, X}(y_2 \mid y_1, X)}\\
        &= \esssup_{P_X} \!\left(\!\log \frac{f_{Y_1}(y_1)}{f_{Y_1\mid X}(y_1 \!\!\mid \!\!X)} + \!\log \frac{f_{Y_2 \mid Y_1}(y_2 \!\mid\! y_1)}{f_{Y_2 \mid Y_1, X}(y_2 \! \mid \! y_1, X)}\!\right)\\
        &\leq \log \; \esssup_{P_X} \; \frac{f_{Y_1}(y_1)}{f_{Y_1\mid X}(y_1 \mid X)}\\
        &\hspace{5em}+ \log \; \esssup_{P_X} \; \frac{f_{Y_2 \mid Y_1}(y_2 \mid y_1)}{f_{Y_2 \mid Y_1 , X}(y_2 \mid y_1, X)}\\
        &\leq \log \; \esssup_{P_X} \; \frac{f_{Y_1}(y_1)}{f_{Y_1\!\mid \!X}(y_1 \!\mid \!X)}\\
        &\hspace{5em}+ \log \; \esssup_{P_{X \mid Y_1 = y_1}} \; \frac{f_{Y_2 \mid Y_1}(y_2 \mid y_1)}{f_{Y_2 \mid Y_1, X}(y_2 \mid y_1, X)}, \label{subeq:bigger_support}\\
        &= \Lambda(X \to y_1) +  \Lambda(X \to y_2 \mid y_1). 
    \end{align}
where \eqref{subeq:bigger_support} is due to the assumption that $P_{X} \ll P_{X \mid Y_1=y_1}$. Next, suppose $P_{X} \nll P_{X \mid Y_1=y_1}$. Then, $\Lambda(X \to y_1) = \dinf(P_{X} \Vert P_{X \mid Y_1=y_1}) = \infty$, which results in the trivial inequality $\Lambda(X \to y_1, y_2) \leq \infty$. 
\end{enumerate}

\section{Proof of Proposition~\ref{prop:ldp_pmc}}
\label{sec:proof_ldp_pmc}
Let $f_{Y \mid X} = \frac{d P_{Y \mid X}}{d \nu}$ and $f_{Y} = \frac{d P_Y}{d \nu}$, where $\nu$ is a dominating $\sigma$-finite measure on $(\cY, \cS_\cY)$. If $P_{Y \mid X}$ satisfies $\varepsilon$-LDP, then for all pairs $x,x' \in \cX$ there exists a measurable set $\cA_{x,x'} \in \cY$ with probability $P_Y(\cA_{x,x'}) = 1$ such that 
\begin{equation*}
    \frac{f_{Y \mid X=x}(y)}{f_{Y \mid X=x'}(y)} \leq e^\varepsilon, \quad y \in \cA_{x,x'}. 
\end{equation*}
Since $\cX$ is finite, we can take $\cA = \bigcap_{x,x' \in \cX} \cA_{x,x'}$ which satisfies $P_Y(\cA) = 1$. On this set, we have 
\begin{equation*}
    \frac{f_{Y \mid X=x}(y)}{f_{Y \mid X=x'}(y)} \leq e^\varepsilon, \quad x,x' \in \cX, \; y \in \cA. 
\end{equation*}
Fix $x \in \cX$ and $y \in \cA$. We can write  
\begin{align*}
    &\frac{f_Y(y)}{f_{Y \mid X=x}(y)} = \frac{\sum_{x' \in \cX} f_{Y \mid X=x'}(y) P_X(x')}{f_{Y \mid X=x}(y)}\\
    &\leq \frac{f_{Y \mid X=x}(y) P_X(x) + e^\varepsilon f_{Y \mid X=x}(y) (1 - P_X(x))}{f_{Y \mid X=x}(y)}\\
    &= e^\varepsilon - P_X(x) (e^\varepsilon - 1). 
\end{align*}
Maximizing over $x \in \cX$ on both sides and taking the logarithm, we get 
\begin{align*}
    \Lambda(X \to y) &= \log \; \max_{x \in \cX} \frac{f_Y(y)}{f_{Y \mid X=x}(y)}\\
    &\leq \log \; \Big( e^\varepsilon - p_\mathrm{min} (e^\varepsilon - 1)\Big), \quad y \in \cA.  
\end{align*}
Therefore, the mechanism satisfies $\log \big( e^\varepsilon - p_\mathrm{min} (e^\varepsilon - 1)\big)$-PMC. 

\section{Proof of Theorem~\ref{thm:ldp_pmc_sup}}
\label{sec:proof_ldp_pmc_sup}

Let $f_{Y \mid X} = \frac{d P_{Y \mid X}}{d \nu}$ and $f_{Y} = \frac{d P_Y}{d \nu}$. Fix $y \in \cY$. Since $\cX$ is finite, we have
\begin{align*}
    &\sup_{P_X \in \cP_\cX} \Lambda(X \to y) 
    = \sup_{P_X \in \cP_\cX} \log \, \max_{x \in \cX} \frac{f_Y(y)}{f_{Y \mid X=x}(y)} \\[0.5em]
    &= \log \, \max_{x \in \cX} \frac{\sup\limits_{P_X \in \cP_\cX} \sum_{x' \in \cX} f_{Y \mid X=x'}(y) P_X(x')}{f_{Y \mid X=x}(y)} \\[0.5em]
    &= \log \, \max_{x \in \cX} \frac{\max_{x' \in \cX} f_{Y \mid X=x'}(y)}{f_{Y \mid X=x}(y)} \\[0.5em]
    &= \log \, \max_{x,x' \in \cX} \frac{f_{Y \mid X=x'}(y)}{f_{Y \mid X=x}(y)}.
\end{align*}

Therefore, for all $\varepsilon \geq 0$ and all $y \in \cY$, we have
\begin{equation*}
    \sup_{P_X \in \cP_\cX} \Lambda(X \to y) \leq \varepsilon 
    \iff 
    \log \, \max_{x,x' \in \cX} \frac{f_{Y \mid X=x'}(y)}{f_{Y \mid X=x}(y)} \leq \varepsilon.
\end{equation*}

\section{Proof of Theorem~\ref{thm:liftasym}}
\label{sec:proof_liftasym}
\begin{enumerate}
    \item Fix some $y \in \mathcal Y$ and $x^* \in \cX$. Since $P_{Y \mid X}$ satisfies $\varepsilon_u$-PML we have $i(x;y) \leq \varepsilon_u$ for all $x \in \cX$. Thus, we can write  
    \begin{align*}
        1 &= \sum_{x \in \mathcal X} P_{X \mid Y=y}(x)\\
        &= P_{X \mid Y=y}(x^*) + \sum_{x \in \cX \setminus\{x^*\}} P_{X \mid Y=y}(x)\\
        &\leq P_{X \mid Y=y}(x^*) + \sum_{x \in \cX \setminus\{x^*\}} e^{\varepsilon_u} P_X(x)\\
        &= P_{X \mid Y=y}(x^*) + e^{\varepsilon_u} \big(1 - P_X(x^*) \big). 
    \end{align*}
    Rearranging, we get 
    \begin{equation}
        \frac{P_X(x^*)}{P_{X \mid Y=y}(x^*)} \leq \frac{P_X(x^*)}{1 - e^{\varepsilon_u} (1 - P_X(x^*))}. 
    \end{equation}
    This yields the following upper bound on the PMC: 
    \begin{align}
    \begin{split}
    \label{eq:liftlimit}
        \Lambda(X \to y) &= \log \; \max_{x \in \mathcal X}\frac{P_X(x)}{P_{X \mid Y=y}(x)}\\
        &\leq \log \; \max_{x\in \mathcal X} \frac{P_X(x)}{1 - e^{\varepsilon_u} (1 - P_X(x))}\\
        &= \log \frac{p_\mathrm{min}}{1 - e^{\varepsilon_u} (1 - p_\mathrm{min})},
    \end{split}
    \end{align}
    where the last equality is due to the fact that the mapping $p \mapsto \frac{p}{1-e^{\varepsilon_u}(1-p)}$ is decreasing for $0 \leq \varepsilon_u < \log \frac{1}{1-p_{\min}}$. 

    \item Fix some $y \in \cY$ and $x^* \in \cX$. Since $P_{Y \mid X}$ satisfies $\varepsilon_l$-PMC we have $i(x;y) \geq - \varepsilon_l$ for all $x \in \cX$. Thus, we can write  
    \begin{align*}
        1 &= \sum_{x \in \mathcal X} P_{X \mid Y=y}(x)\\
        &= P_{X \mid Y=y}(x^*) + \sum_{x \in \cX \setminus\{x^*\}} P_{X \mid Y=y}(x)\\
        &\geq P_{X \mid Y=y}(x^*) + \sum_{x \in \cX \setminus\{x^*\}} e^{-\varepsilon_l} P_X(x)\\
        &= P_{X \mid Y=y}(x^*) + e^{-\varepsilon_l} \big(1 - P_X(x^*) \big). 
    \end{align*}
    Rearranging, we get 
    \begin{equation}
        \frac{P_{X \mid Y=y}(x^*)}{P_{X}(x^*)} \leq \frac{1 - e^{-\varepsilon_l} (1 - P_X(x^*))}{P_X(x^*)}. 
    \end{equation}
    This yields the following upper bound on the PML: 
    \begin{align*}
        \ell(X \to y) &= \log \; \max_{x \in \mathcal X}\frac{P_{X \mid Y=y}(x)}{P_{X}(x)}\\
        &\leq \log \; \max_{x\in \mathcal X} \frac{1 - e^{-\varepsilon_l} (1 - P_X(x))}{P_X(x)}\\
        &= \log \frac{1 - e^{-\varepsilon_l} (1 - p_\mathrm{min})}{p_\mathrm{min}},
    \end{align*}
    where the last equality is due to the fact that the mapping $p \mapsto \frac{1-e^{-\varepsilon_l}(1-p)}{p}$ is decreasing for $\varepsilon_l \geq 0$. 
\end{enumerate}

\section{Proof of Theorem~\ref{thm:general_ldp_pml_pmc}}
\label{sec:proof_general_ldp_pml_pmc}

Let $f_{Y \mid X} = \frac{d P_{Y \mid X}}{d \nu}$ and $f_{Y} = \frac{d P_Y}{d \nu}$, where $\nu$ is a dominating $\sigma$-finite measure on $(\cY, \cS_\cY)$. Recall that by Definition~\ref{def:ldp}, a mechanism satisfies $\varepsilon$-LDP if 
\begin{align*}
    \esssup_{P_X \times P_X} \; \dinf \Big(P_{Y \mid X} (\cdot \mid X) \big \Vert P_{Y \mid X} (\cdot \mid X') \Big)\\
    = \esssup_{P_X \times P_X \times P_Y} \, \log \, \frac{f_{Y \mid X}(Y \mid X)}{f_{Y \mid X}(Y \mid X')} \leq \varepsilon.
\end{align*}
Note that in the expression for the \Ren divergence, we have taken the essential supremum with respect to $P_Y$ since our assumptions in Section~\ref{ssec:notation} ensure that $P_{Y \mid X=x} \ll P_Y$ almost surely. Equivalently, there exists a set $\cE \in \cS_\cX \otimes \cS_\cX \otimes \cS_\cY$ with probability one such that 
\begin{equation*}
    \frac{f_{Y \mid X}(y \mid x)}{f_{Y \mid X}(y \mid x')} \leq e^{\varepsilon}, \quad \text{ for all } (x,x',y) \in \cE.
\end{equation*}
For each $(x,y)$ define the section $\cE_{x,y} = \{x' \in \cX : (x,x',y) \in \cE \}$ and note that $P_X(\cE_{x,y}) =1$ for $P_X \times P_Y$-a.s. $(x,y)$. Fix such a pair $(x,y)$. Then, we have   
\begin{align*}
    f_{Y}(y) &= \int_\cX f_{Y \mid X} (y \mid x') \, P_X(dx')\\
    &\leq \int_{\cE_{x,y}} f_{Y \mid X} (y \mid x') \, P_X(dx')\\
    &\leq e^\varepsilon f_{Y \mid X} (y \mid x) \int_{\cE_{x,y}} P_X(dx')\\
    &= e^\varepsilon f_{Y \mid X} (y \mid x) P_X(\cE_{x,y}) = e^\varepsilon f_{Y \mid X} (y \mid x).
\end{align*}
Hence, 
\begin{equation*}
    \frac{f_Y (y)}{f_{Y \mid X} (y \mid x)} \leq e^\varepsilon, \quad P_X \times P_Y\text{-a.s.},  
\end{equation*}
i.e., the mechanism satisfies $\varepsilon$-PMC. Applying the same argument, we get $f_Y (y) \geq e^{-\varepsilon} f_{Y \mid X} (y \mid x)$ almost surely, so the mechanism satisfies $\varepsilon$-PML. 

Conversely, suppose the mechanism satisfies $\varepsilon_1$-PMC and $\varepsilon_2$-PML. Let  
\begin{gather*}
    \cA_1 = \Big\{(x,y) : \frac{f_{Y}(y)}{f_{Y \mid X}(y \mid x)} \leq e^{\varepsilon_1} \Big\}, \\
    \cA_2 = \Big\{(x',y) : \frac{f_{Y \mid X}(y \mid x')}{f_{Y}(y)} \leq e^{\varepsilon_2} \Big\},  
\end{gather*}
and note that $(P_X \times P_Y) (\cA_1) = (P_X \times P_Y) (\cA_2) = 1$. Define
\begin{equation*}
    \cA = \Big \{(x,x',y) : (x,y) \in \cA_1, \text{ and } (x',y) \in \cA_2 \Big\},
\end{equation*}
so $\cA$ has probability one under the probability measure $P_X \times P_X \times P_Y$ (since the complements of both $\cA_1$ and $\cA_2$ have probability zero). Moreover, for all $(x,x',y) \in \cA$, we have
\begin{align*}
    \frac{f_Y(y)}{f_{Y \mid X}(y \mid x)} \leq e^{\varepsilon_1}, \quad \text{and} \quad  \frac{f_{Y \mid X}(y \mid x')}{f_{Y}(y)} \leq e^{\varepsilon_2}, 
\end{align*}
and multiplying these inequalities yields
\begin{equation*}
     \frac{f_{Y \mid X}(y \mid x')}{f_{Y \mid X}(y \mid x)} \leq e^{\varepsilon_1 + \varepsilon_2}, \quad \text{ for all } (x,x',y) \in \cA.
\end{equation*}
Therefore, the mechanism satisfies $\varepsilon_1 + \varepsilon_2$-LDP.

\bibliographystyle{IEEEtranN}
\footnotesize
\bibliography{main}

\begin{IEEEbiographynophoto}{Sara Saeidian}
(M’20) received the B.Sc. degree in Electrical Engineering from the University of Tehran, Iran in 2017 and the M.Sc. degree in Information and Network Engineering from KTH Royal Institute of Technology, Sweden in 2019. She received the Ph.D. degree from the Division of Information Science and Engineering at KTH in 2024. In the same year, she was awarded an international postdoctoral grant from the Swedish Research Council and is currently a postdoctoral researcher at Inria Saclay, France. Her research interests include trustworthy machine learning, with a focus on privacy and algorithmic fairness, as well as information theory.
\end{IEEEbiographynophoto}

\begin{IEEEbiographynophoto}{Leonhard Grosse}
(S’23) received the B.S. degree in electrical engineering and information technology from the University of Stuttgart, Germany in 2021 and the M.S. degree in information and network engineering from the KTH Royal Institute of Technology in 2023, where he is currently pursuing the Ph.D. degree with the Division of Information Science and Engineering. His research interests include information theory with applications to privacy, statistics and learning.
\end{IEEEbiographynophoto}


\begin{IEEEbiographynophoto}{Parastoo~Sadeghi}
received the bachelor’s and master’s degrees in electrical engineering from the Sharif University of Technology, Tehran, Iran, in 1995 and 1997, respectively, and the Ph.D. degree in electrical engineering from the University of New South Wales (UNSW), Sydney, in 2006. She is currently a Professor with the School of Engineering and Technology, UNSW Canberra. She has co-authored over 220 refereed journal articles and conference papers. Her research interests include information theory, data privacy, index coding, and network coding. From 2016 to 2019 and since April 2025, she has served as an Associate Editor for the IEEE Transactions on Information Theory. In 2022, she was selected as a Distinguished Lecturer of the IEEE Information Theory Society. She is currently serving as a member of the Board of Governors of the IEEE Information Theory Society.
\end{IEEEbiographynophoto}

\begin{IEEEbiographynophoto}{Mikael~Skoglund}
(S'93-M'97-SM'04-F'19) received the Ph.D. degree in
1997 from Chalmers University of Technology, Sweden.  In 1997, he
joined the Royal Institute of Technology (KTH), Stockholm, Sweden,
where he was appointed to the Chair in Communication Theory in 2003.
At KTH he heads the Division of Information Science and Engineering.

He has contributed to source-channel coding, coding and transmission
for wireless communications, Shannon theory, information-theoretic
security, information theory for statistics and learning, information
and control, quantum information, and signal processing. In these
research fields, he has supervised more than 30 PhD students to
completion

Mikael Skoglund is a Fellow of the IEEE. During 2003--08 he was an
associate editor for the IEEE Transactions on Communications.  In the
interval 2008--12 he was on the editorial board for the IEEE
Transactions on Information Theory and starting in the Fall of 2021
he joined it once again. He has served on numerous technical program
committees for IEEE sponsored conferences, he was general co-chair for
IEEE ITW 2019 and TPC co-chair for IEEE ISIT 2022. He is a member of
the ELLIS Society, and an elected member of the IEEE Information
Theory Society Board of Governors.

\end{IEEEbiographynophoto}

\begin{IEEEbiographynophoto}{Tobias~J.~Oechtering}
(S’01-M’08-SM’12) received his Dipl-Ing degree in Electrical Engineering and Information Technology in 2002 from RWTH Aachen University, Germany, his Dr-Ing degree in Electrical Engineering in 2007 from the Technische Universit\"at Berlin, Germany. In 2008 he joined KTH Royal Institute of Technology, Stockholm, Sweden and has been a Professor since 2018. In 2009, he received the ``F\"orderpreis 2009” from the Vodafone Foundation.

Dr. Oechtering was Senior Editor of IEEE Transactions on Information Forensic and Security during 2020-2025 and served previously as Associate Editor for the same journal since 2016, and IEEE Communications Letters during 2012-2015. He has served on numerous technical program committees for IEEE-sponsored conferences, and he was general co-chair for IEEE ITW 2019. His research interests include privacy and physical layer security, statistical learning and signal processing, communication and information theory, as well as communications for networked control. 
\end{IEEEbiographynophoto}

\end{document}